\documentclass[aoas,preprint]{imsart2}

\usepackage{amsmath,amssymb,amsthm}
\usepackage[final]{graphicx}
\usepackage{float}
\usepackage[round,authoryear]{natbib}
\usepackage{times}

\RequirePackage[OT1]{fontenc}
\RequirePackage[colorlinks,citecolor=blue,urlcolor=blue]{hyperref}
\RequirePackage{hypernat}

\usepackage[dvipsnames]{xcolor}
\usepackage{xy}\xyoption{all} \xyoption{poly} \xyoption{knot}

\startlocaldefs

\def\IW{\mbox{IW}}
\def\Sig#1{\Sigma_{#1}}     %\def\SigallT{\Sigma_{1:T}}
\def\Om#1{\Omega_{#1}}     %\def\OmallT{\Omega_{1:T}}
\def\Ph#1{\Phi_{#1}}    \def\La#1{\Lambda_{#1}}
\def\Ps#1{\Psi_{#1}}      \def\Ga#1{\Gamma_{#1}}
\def\Up#1{\Upsilon_{#1}}  \def\De#1{\Delta_{#1}}
\def\beq{\begin{equation}}\def\eeq{\end{equation}}
\def\bpmat{\begin{pmatrix}}\def\epmat{\end{pmatrix}}
\def\bbmat{\begin{bmatrix}}\def\ebmat{\end{bmatrix}}
\def\tr{\mbox{\rm tr}}\def\diag{\mbox{\rm diag}}

\newcommand{\figdir}{figs}
\def\postcap{\vspace{-0.275in}}

\numberwithin{equation}{section}

\newtheorem{theorem}{Theorem}[section]

\endlocaldefs

\begin{document}
	
\begin{frontmatter}

% "Title of the paper"
\title{Autoregressive Models for Variance Matrices:\break Stationary Inverse Wishart Processes}
\runtitle{Inverse Wishart Processes}

\begin{aug}
\author{\fnms{Emily~B.} \snm{Fox}\corref{}\thanksref{t1}\ead[label=e1]{(fox,mw)@stat.duke.edu}}
\and
\author{\fnms{Mike} \snm{West}\thanksref{t2}\ead[label=u1,url]{http://www.stat.duke.edu}}

\thankstext{t1}{Research partially supported by a Mathematical Sciences Postdoctoral Research Fellowship
                from the National Science Foundation}
\thankstext{t2}{Research partially supported by the National Science Foundation under grants
		 DMS-0342172 and DMS-1106516.
                Any opinions, findings and conclusions or recommendations expressed in this work
                are those of the authors and do not necessarily reflect the views of the NSF. }
\runauthor{E. Fox and M. West}
\affiliation{Duke University, Durham NC, USA}
\address{Emily Fox and Mike West\\ Department of Statistical Science\\ Duke Box \#90251\\ Durham NC 27708-0251\\
\printead{u1}\\ \printead{e1}}
\end{aug}

\begin{abstract}

	We introduce and explore a new class of stationary time series models for variance matrices based
on a constructive definition exploiting inverse Wishart distribution theory.  The main class of
models explored is a novel class of stationary, first-order autoregressive (AR) processes on the
cone of positive semi-definite matrices.  Aspects of the theory and structure of these new models for multivariate ``volatility'' processes are described in detail and exemplified.  We then develop approaches to model fitting via Bayesian simulation-based computations, creating a custom filtering method that relies on an efficient innovations sampler. An example is then provided in analysis of a multivariate electroencephalogram (EEG) time series in neurological studies. We conclude by
discussing potential further developments of higher-order AR models and a number
of connections with prior approaches.

\end{abstract}

\begin{keyword}[class=AMS]
\kwd[Primary ]{62M10}
\kwd[; secondary ]{37M10,62H99}
\end{keyword}

\begin{keyword}
\kwd{Autoregressive models}
\kwd{Bayesian forecasting}
\kwd{Innovations sampling}
\kwd{Matrix-Variate Autoregressions}
\kwd{Multiple time series analysis}
\kwd{Multivariate stochastic volatility}
\kwd{Time-varying variance matrix}
\end{keyword}

\end{frontmatter}
	
\section{Introduction}

Modeling the temporal dependence structure in a sequence of variance matrices is of increasing interest  in multi- and matrix-variate time series analysis, with motivating applications in fields as diverse as econometrics, neuroscience, epidemiology and 
spatial-temporal modeling.
Some key interests and needs are in defining:
(i) classes of stationary stochastic process models
on the cone of symmetric, non-negative definite matrices that offer flexibility to model differing degrees of
dependence structures as well as short-term predictive ability;
(ii) models that are open to theoretical study and interpretation; and
(iii) models generating some degree of analytic and computational tractability for model fitting and exploitation in  applied work.

The context is a sequence of $q\times q$ variance matrices (i.e., symmetric, non-negative definite matrices) $\Sig{t}$ in discrete time $t=0,1,\ldots,$ typically the variance matrices of components of
more elaborate state-space models for an observable time series.
In econometrics and finance, variants of ``observation-driven'' multivariate ARCH~\citep{Engle:02}
models and state-space or ``parameter-driven''  multivariate stochastic volatility models~\citep{QuintanaWest1987,Harvey:94} are widely used.   While the former directly specify
``volatility'' matrices $\Sig{t}$ as  functions of lagged values and past data, state-space approaches use
formal stochastic process models that offer cleaner interpretation, access to theoretical understanding as 
well as potential to scale more easily with dimension; see~\cite{Chib:09} for a survey of such approaches. The class of state-space models based on Bayesian discount methods~\citep{QuintanaWest1987,Quintana92,Uhlig94,WestHarrison1997,Quintana03,Quintana2010},
are also widely used in financial applications for local volatility estimation and smoothing.
These methods are, however,  restricted to local estimation due to the underlying non-stationary random-walk style model for $\Sig{t};$ see~\cite{PradoWest2010} for recent review and additional developments.

%These approaches have also been successfully applied and used in vector and matrix time series models and with dynamic graphical model structuring; see  \cite{Carvalho2007} and \cite{Wang2009}...\cite{CarvalhoWest:07,CarvalhoReesonWang:10}

Two recent contributions explore constructions of AR(1) style models based on conditional Wishart transition distributions~\citep{PhilipovGlickman:06,PhilipovGlickman:06b,Gourieroux:09}.  These aim to
provide flexibility in modeling  one-step dependencies balanced with parsimony in parameterization
through properties of the Wishart distribution.  These models tend to be
rather intractable theoretically, hence somewhat difficult to understand and interpret, while
model fitting is challenging and there are open questions of how
useful  potential higher-order variants might be. We discuss these approaches and
issues further in Section~\ref{sec:related}.

The centrality of inverse Wishart theory to current Bayesian state-space approaches
underlies the ideas for new model classes explored in this paper.  We introduce a class of stationary, non-linear autoregressive (AR) models for variance matrices by exploiting the structure of
conditional and marginal distributions in the inverse Wishart family.  We denote the resulting models by
AR or IW-AR for definiteness, and use AR(1) or IW-AR(1) to be more specific about first-order models
when needed; most of the development of this paper is for first-order models.  The new IW-AR models are open to some useful theoretical analysis of margins, stationarity, reversibility, and conditional moments, among other properties.  Exploiting the state-space nature of the IW-AR(1) process, we develop an MCMC sampler based on forward filtering backward sampling (FFBS) proposals that results in tractable Bayesian computations. This operates locally on a matrix innovations process to ameliorate issues arising from global accept-rejects of the variance matrix process (e.g., exponential decrease in acceptance rates with increasing sequence length) albeit at  increased computational cost.

Section~\ref{sec:model} introduces the new models and some aspects of the theoretical structure are explored in Section~\ref{sec:theory}.  Posterior computations are developed in Section~\ref{sec:posterior}, building on a data augmentation idea discussed in Section~\ref{sec:data}.  An example in EEG time series analysis is given in Section~\ref{sec:eeg} and Section~\ref{sec:higherOrder} discusses extensions to higher-order AR dependencies.
Section~\ref{sec:related} discusses connections with other approaches, Section~\ref{sec:final} provides summary
comments and supporting technical material is appended. 

For time ranges we use the concise notation $s:t$ to denote the sequence of time indices $s,s+1,\ldots,t;$ e.g., $\Sig{0:T} = \{ \Sig{0},\ldots,\Sig{T} \}$ and $\Sig{t-1:t} = \{\Sig{t-1},\Sig{t}\}.$

\section{First-Order Inverse Wishart Autoregressive Processes}
\label{sec:model}

\subsection{Construction}
\label{sec:IWARconstruction}

As context, suppose we are to observe a series of $q\times 1$ vector observations $x_t$ with
\begin{align}
	x_t | \Sig{t} \sim N(0,\Sigma_t), \hspace{0.25in} t=1:T,
	\label{eqn:obsModel}
\end{align}
where $x_t$ is independent of $\{ x_s, \Sig{s}; \, s<t\}$ conditional on $\Sig{t}.$
We aim to capture the volatility dynamics with a stationary, first-order Markov model for the
$\Sig{t}$ sequence. The joint density for matrices over an arbitrary time period $t=0:T$ is
\begin{align}
	p(\Sigma_{0:T}) =  \frac{\prod_{t=1}^T p(\Sigma_{t-1},\Sigma_t)}{\prod_{t=2}^T p(\Sigma_{t-1})}
	\label{eqn:jointDist}
\end{align}
for some time invariant joint density $p(\Sig{t-1},\Sig{t})$ for consecutive matrices in the numerator terms; this joint density has common
margins given by the time invariant $p(\Sig{t})$ appearing in the denominator terms.

We take the defining joint density $p(\Sig{t-1},\Sig{t})$ as arising from an
inverse Wishart on an augmented state-space.
 Specifically, introduce random matrices $\phi_t$ such that
\begin{align}
	\left(\begin{array}{cc}
		\Sigma_{t-1} & \phi_t'\\ \phi_t & \Sigma_t
	\end{array}\right)
	\sim \mbox{IW}_{2q}\left(n+2,n
	\left(\begin{array}{cc}
	S & SF'\\ FS & S
	\end{array}\right)\right)
	\label{eqn:SigmaJoint}
\end{align}
for some degree of freedom parameter $n>0,$ a $q\times q$ variance matrix parameter $S$
and a $q\times q$ matrix parameter $F$ such that the $2q\times 2q$ parameter matrix
parameter of the distribution above is non-negative definite.
This inverse Wishart has common margin for the diagonal blocks; for each $t,$
\begin{align}
	\Sig{t} \sim \IW_q(n+2,nS)
	\label{eqn:SigmaMarg}
\end{align}
with  $E[\Sig{t}] = S.$   It is now clear that eqn.~\eqref{eqn:jointDist} defines stationary first-order
process with eqn.~\eqref{eqn:SigmaMarg} as the stationary (marginal) distribution. Transitions
are governed by the conditional density $p(\Sig{t}|\Sig{t-1})$ implicitly defined by
eqn.~\eqref{eqn:SigmaJoint}. This has no closed analytic form but is now
explored theoretically.
\subsection{Innovations Process}
\label{sec:innovations}
The joint distribution of $\{\Sigma_{t-1},\Sigma_t,\phi_t\}$ defined in eqn.~\eqref{eqn:SigmaJoint} can be reformulated in terms of $\{\Sigma_{t-1},\Upsilon_t,\Psi_t\}$ where $\{\Upsilon_t,\Psi_t\}$ are marginally  matrix normal, inverse Wishart distributed and independent of $\Sigma_{t-1}$.  Specifically,
standard inverse Wishart theory  \citep[e.g.][]{Carvalho:07b} implies that
\begin{align}
	\Sigma_t = \Psi_t + \Upsilon_t \Sigma_{t-1} \Upsilon_t'.
	\label{eqn:SigmaAR}
\end{align}
and $\phi_t = \Upsilon_t\Sigma_{t-1}$ where the $q\times q$ matrices $\{\Upsilon_t,\Psi_t\}$
follow
\begin{equation}
\begin{aligned}
	\Psi_t &\sim \IW_q(n+q+2,nV)\\
	\Upsilon_t \mid \Psi_t &\sim N(F,\Psi_t,(nS)^{-1})
	\label{eqn:MNIWprior}
\end{aligned}
\end{equation}
with $V = S - F S F'$ and where $\{\Upsilon_t,\Psi_t\}$ are conditionally independent of
$\Sig{t-1}\sim \IW_q(n+2,nS).$   Eqn.~\eqref{eqn:SigmaAR} is an explicit AR(1) equation in which
$\Upsilon_t$ acts as a random autoregressive coefficient matrix and $\Psi_t$ an additive random disturbance.  Since $\{\Upsilon_t,\Psi_t\}$ are independent at each $t$ and drive the dynamics of this IW-AR process, we refer to them as latent \emph{innovations}.
\subsection{Special Case of $q=1$}
\label{sec:q1}
When $q=1,$ $\Sigma_t\equiv\sigma_t>0$ and
the IW-AR process reduces to an inverse gamma autoregressive process.  Now $S\equiv s>0$
and $F \equiv f \in (-1,1)$ and the joint density of eqn.~\eqref{eqn:SigmaJoint} is
\begin{align}
	\left(\begin{array}{cc}
		\sigma_{t-1} & \phi_t\\ \phi_t & \sigma_t
	\end{array}\right)
	\sim \mbox{IW}_{2}\left(n+2,n
	s\left(\begin{array}{cc}
	1 & f\\ f & 1
	\end{array}\right)\right),
	\label{eqn:SigmaJoint_q1}
\end{align}
such that
\begin{align}
	\sigma_t \sim IG\left(\frac{n+2}{2},\frac{ns}{2}\right).
	\label{eqn:SigmaMarg_q1}
\end{align}
Equivalently, with scalar innovations $\{\Upsilon_t,\Psi_t\} \equiv \{ \upsilon_t,\psi_t\},$
\begin{align}
\sigma_t = \psi_t + \upsilon_t^2\sigma_{t-1}
\label{eqn:SigmaAR_q1}
\end{align}
where
\begin{equation}
\psi_t \sim IG\left(\frac{n+3}{2},\frac{ns(1-f^2)}{2}\right)\quad\textrm{and}\quad
\upsilon_t \mid \psi_t \sim N\left(f,\frac{\psi_t}{ns}\right).
\label{eqn:MNIW_q1}
\end{equation}
We can see immediate analogies with the standard linear, Gaussian AR(1) process with a random AR coefficient.
The marginal mean of $\upsilon_t^2$ is $(nf^2+1)/(n+1)$ which plays the role of an average linear 
autoregressive coefficient. For $|f|$ close to 1, the model approaches the stationary/non-stationary 
boundary, and when  $n$ is large, the mean AR(1) coefficient is close to $f^2.$   
Also, $E[\psi_t] = ns(1-f^2)/(n+1)$ so that for fixed $n$ and $s$ the additive innovation 
noise tends to be smaller as $|f|$ approaches unity.  Parameters $(n,f)$ also control dispersion of
the additive innovations through, for example, $V(\psi_t) = 2[ns(1-f^2)]^2/[(n+1)^2(n-1)]$. 
Section~\ref{sec:theory} further explores this in the general multivariate setting as well as this  special case of $q=1$.

This inverse gamma autoregressive process is related to the formulation of~\cite{Pitt:02}.  In that work, the authors construct a stationary autoregressive process $\{\sigma_t\}$ with inverse gamma marginals by harnessing a conditionally gamma distributed latent process $\{z_t\}$.  The sequence $\{\sigma_t,z_t\}$ obtained by generating $\sigma_t \mid z_{t-1}$ and $z_t \mid \sigma_t$ from the respective closed form conditional distributions leads to a marginal process $\{\sigma_t\}$ with the desired autoregressive structure.  Extensions to Bayesian nonparametric transition kernels is considered in~\cite{Mena:05} and to state-space volatility processes in~\cite{Pitt:05}.  Although related in spirit to this work, the proposed IW-AR process represents a novel construction.  One attribute of the IW-AR approach, as explored in Section~\ref{sec:theory}, is that our process need not be reversible depending upon the parameterization specified by $f$ and $s$.  Furthermore, our formulation allows straightforward higher-order extensions, discussed in Section~\ref{sec:higherOrder}.  Additional discussion and other related approaches appears in Section~\ref{sec:related}. 

\section{Theoretical Properties}
\label{sec:theory}

\subsection{Marginal Processes for Submatrices and Univariate Elements}
\label{sec:blockevolution}

Consider any partition of $\Sig{t}$ into blocks $\Sig{t,ij},$ $(i=1:I, j=1:J),$ where $i,j$ represent 
consecutive blocks of consecutive sets of row and column indices, respectively.  As a special case this also 
defines scalar elements. Then the
evolution of each submatrix $\Sig{t,ij}$  depends upon every element of  $\Sigma_{t-1}$ as follows:
\begin{align}
	\Sig{t,ij} = \Psi_{t,ij} + \begin{bmatrix}\Upsilon_{t,i1} & \dots & \Upsilon_{t,iJ}\end{bmatrix} \Sigma_{t-1} \begin{bmatrix} \Upsilon_{t,1j}'\\ \vdots \\ \Upsilon_{t,Ij}'\end{bmatrix}.
\end{align}
Here $\Psi_{t,ij} \sim \mbox{IW}(n+q+2,nV_{ij})$ and $\begin{bmatrix}\Upsilon_{t,i1} & \dots & \Upsilon_{t,iJ}\end{bmatrix}$ has a 
conditional matrix normal distribution induced from the joint distribution of eqn.~\eqref{eqn:MNIWprior}.  

\subsection{Stationarity}
\label{sec:stationarity}
\begin{theorem}
	The process defined  via eqn.~\eqref{eqn:SigmaJoint} is strictly stationary when the parameterization of the inverse Wishart of  yields a valid distribution:  that is, when $S$ and $S-FSF'$ are positive definite. 
	\begin{proof}
		This follows directly from the constructive definition using eqn.~\eqref{eqn:SigmaJoint}.
		For a valid model, the scale matrix must be positive definite.  Equivalently, via Sylvester's criterion and the Schur complement, $S$ and $S-FSF'$ must be positive definite.
	\end{proof}
\end{theorem}

 We note extensions to non-negative definite cases when the
		resulting $\Sig{t}$ matrices are singular with singular inverse Wishart distributions, although
		 these are of limited practical interest so we focus on non-singular cases throughout. 
		 
In simple cases of $F= \rho I_q$ the stationarity condition reduces to $|S|>0$ and $|\rho|<1$.  Other special cases are those in which $F,S$ share eigenvectors with
eigen-decompositions $F=ERE'$ and $S=EQE'$. Stationarity is assured when
$||Q||_0 > 0$ and $||R||_0 < 1$.

\subsection{Reversibility}

\begin{theorem}
	The process is time-reversible if and only if $FS = SF'.$

	\begin{proof}

The reverse-time process on the $\Sigma_t$ is as follows.  Eqn.~\eqref{eqn:SigmaJoint} implies that
\begin{align}
	\left(\begin{array}{cc}
		\Sigma_{t} & \tilde{\phi}'_t\\ \tilde{\phi}_t & \Sigma_{t-1}
	\end{array}\right)
	\sim \mbox{IW}_{2q}\left(n+2,n
	\left(\begin{array}{cc}
	S & FS\\ SF' & S
	\end{array}\right)\right),
\end{align}
for some latent process $\tilde{\phi}_t$.  Then, as in Section~\ref{sec:model}, we have 
\begin{align}
	\Sigma_{t-1} = \tilde{\Psi}_t + \tilde{\Upsilon}_t \Sigma_{t} \tilde{\Upsilon}'_t
\end{align}
where
\begin{equation}
	\tilde{\Psi}_t \sim \mbox{IW}_q(n+q+2,n\tilde{V})\quad\textrm{and}\quad
	\tilde{\Upsilon}_t \mid \tilde{\Psi}_t \sim N(\tilde{F},\tilde{\Psi}_t,(nS)^{-1})
\label{eqn:tildeInnovations_prior}
\end{equation}
with $\tilde{V} = S-SF'S^{-1}FS$ and $\tilde{F}=SF'S^{-1}$.  If $FS = SF'$, then $\tilde{V} = V$ and $\tilde{F}=F$ and the reverse-time process follows the same model as the forward-time process.  Conversely, assume a reversible process (i.e., $\tilde{F} = F$ and $\tilde{V} = V$) with $FS \neq SF'$. Since $\tilde{F}=SF'S^{-1}$, a contradiction immediately arises.
\end{proof}
\label{thm:reversibility}
\end{theorem}

Examples of reversible IW-AR processes include cases 
when $F= \rho I_q$ or when $F=ERE'$ and $S=EQE'$.  Note, however, that the process 
is irreversible when $F = \mbox{diag}(\rho_1,\dots,\rho_q)$ with distinct elements and $S$ is
non-diagonal.

\subsection{Conditional Mean}
\label{sec:moments}
The IW-AR yields a simple form for the conditional expectation of $\Sigma_t$ given $\Sigma_{t-1}$.
\begin{theorem}
\begin{align}
	E[\Sigma_t \mid \Sigma_{t-1}] = F\Sigma_{t-1}F' + c_{n,q}\left[ 1 + \tr\{\Sigma_{t-1}(nS)^{-1}\}\right]V
	\label{eqn:SigmaCondMean}
\end{align}	with $c_{n,q}= n/(n+q).$ 
\begin{proof}
	See Appendix. 
\end{proof}
\label{thm:condMean}
\end{theorem}
Theorem~\ref{thm:condMean} illuminates the inherent matrix linearity of the model and the interpretation of $F$ as a \lq\lq square root" 
AR parameter matrix. The conditional mean regression form is $F\Sig{t-1}F'$ corrected by a term that
reflects the skewness of the conditional distribution. For large $n,$ the underlying inverse Wishart distributions 
are less skewed and  this latter term is small; indeed
\begin{align}
	\lim_{n\rightarrow \infty}E[\Sigma_t \mid \Sigma_{t-1}] &= S +  F(\Sigma_{t-1}-S)F'.
\end{align}
%
% In the illuminating special
%	case of $q=1$ with $F=\rho$ (referring to the formulation of Section~\ref{sec:q1}) this reduces to 
%	%
%	$$
%		E[\sigma_t\mid \sigma_{t-1}] = \rho^2\sigma_{t-1} + (1+\sigma_t/(ns))V/(n+1).
%	$$

%
\subsection{Principal Component IW-AR Processes}
\label{sec:principalIWAR}
Assume that $F$ and $S$ share eigenvectors so that the IW-AR model is reversible. There exists 
a principal component IW-AR process, as follows. 

\begin{theorem}
Suppose that $F=ERE'$ and $S=EQE'$ where $E$ is orthogonal and 
$R = \diag(\rho_1,\dots,\rho_q)$ and $Q = \diag(\xi_1,\dots,\xi_q)$ with
positive elements. 
Then the sequence of matrices defined by $\hat{\Sigma}_t = E'\Sigma_t E$
follows an IW-AR(1) model with degrees of freedom $n,$ scale matrix $Q$ and AR matrix $R$. Specifically, 
$$
		\hat{\Sigma}_t = \hat{\Psi}_t + \hat{\Upsilon}_t \hat{\Sigma}_{t-1} \hat{\Upsilon}_t'
$$
	where $\hat{\Psi}_t = E'\Psi_t E$ and $\hat{\Upsilon}_t = E'\Upsilon_t E$ are such that 
$$
		\hat{\Psi}_t \sim \mbox{IW}_q(n+q+2,nQ(I-R^2))\quad\textrm{and}\quad
		\hat{\Upsilon}_t \mid \hat{\Psi}_t \sim N(R,\hat{\Psi}_t,(nQ)^{-1})
$$
and with marginal distribution $\hat{\Sigma}_t\sim \IW_q(n+2,nQ).$ 
	
	The conditional moment is given by
	\begin{align}
		E[\hat{\Sigma}_t \mid \hat{\Sigma}_{t-1}] = R\hat{\Sigma}_{t-1}R + c_{n,q}\left[1 + \sum_i (n\xi_i)^{-1}\hat{\Sigma}_{t-1,ii}\right]Q(I-R^2) .
		\label{eqn:principalCondMean}
	\end{align}
	
	eqn.~\ref{eqn:principalCondMean} implies that there exists a zero-mean noise process $v_{t,j}$ such that
	\begin{align}
		\hat{\Sigma}_{t,jj} = \rho_j^2\hat{\Sigma}_{t-1,jj} + c_{n,q}\left[1 + \sum_i (n\xi_i)^{-1}\hat{\Sigma}_{t-1,ii}\right](1-\rho_j^2)\xi_j  + v_{t,j}.
	\end{align}
	Autoregressive processes for the other terms of $\hat{\Sigma}_t$ are similarly defined.
\label{thm:principalIWAR}
\begin{proof} See Appendix. \end{proof}
\end{theorem}

\subsection{Exponential Forgetting}
It is also interesting to examine properties of the IW-AR process as a function of the parameters $F$, $S$, and $n$ and an initial value $\Sigma_0$. The mean of the IW-AR process forgets its initial condition exponentially fast under a wide range of conditions on the parameterization.  %That is, $\lim_{t\rightarrow \infty} E[\Sigma_t \mid \Sigma_0] = E[\Sigma_t] = S$.

\begin{theorem}
	Assuming that $||S||_{\infty}$, $||S^{-1}||_{\infty}$, and $||F||_{\infty}$ are each bounded by some finite $\lambda \in \mathbb{R}$, then
	\begin{align}
		E[\Sigma_{t}\mid \Sigma_0] = F^{t}\Sigma_0 F^{'t} + c_{n,q}(S- F^{t} S F^{'t}) + O(n^{-1}1\cdot 1')
	\end{align}
	where $c_{n,q}=n/(n+q)$ and  $1$ denotes a column vector of ones.  Further, 
	\begin{align}
		\lim_{n\rightarrow \infty} E[\Sigma_{t}\mid \Sigma_0] = S + F^{t}(\Sigma_0-S) F^{'t}.
		\label{eqn:SigmaCondMeanLimit}
	\end{align}
	\label{thm:SigmaCondMean}
	\begin{proof} See Appendix. \end{proof}
\end{theorem}
From eqn.~\eqref{eqn:SigmaCondMeanLimit}, based on fixed $S$ and in the limit as $n \rightarrow \infty$, we can directly analyze the effects of the elements of $F$ on the conditional mean.

\paragraph{Unitary Bounded Spectral Radius of $F$}
If we assume that $F$ has spectral radius $\rho(F)< 1$ (i.e., the magnitude of the largest eigenvalue of $F$ is less than 1), Theorem~\ref{thm:SigmaCondMean} implies that for $n\rightarrow \infty$ the conditional mean goes exponentially fast to the marginal mean $S$, with a rate proportional $\rho(F)$.

\paragraph{Univariate Process ($q=1$)}
In the univariate case we can analytically examine the conditional mean $E[\sigma_t \mid \sigma_0]$ without relying on the limit of $n \rightarrow \infty$.
%%
%\begin{theorem}
%For $q=1$, and denoting $F=f \in \mathbb{R}$,
%%
%\begin{align}
%	E[\sigma_t \mid \sigma_0] = s + \left(\frac{nf^2+1}{n+1}\right)^t(\sigma_0-s)
%	\label{eqn:SigmaCondMean_scalar}
%\end{align}
%%
%whenever $|f|<1$.
%%
%\end{proof}
%\end{theorem}
Using the notation of Section~\ref{sec:q1} and recursing on the form of $E[\sigma_t \mid \sigma_{t-1}]$ specified in Theorem~\ref{thm:condMean},
	\begin{align}
		E[\sigma_t \mid \sigma_0] = \left(\frac{nf^2+1}{n+1}\right)^t\sigma_0 + \frac{ns(1-f^2)}{n+1} \sum_{\tau = 0}^{t-1} \left(\frac{nf^2+1}{n+1}\right)^{\tau}.
	\end{align}
	Since we assume $|f|<1$ this becomes
	\begin{align}
		E[\sigma_t \mid \sigma_0] %&= \left(\frac{1+nf^2}{n+1}\right)^t\Sigma_0 + \frac{1-\left(\frac{1+nf^2}{n+1}\right)^t}{1-f^2}V\\
		%&= \left(\frac{nf^2+1}{n+1}\right)^t\sigma_0 + \left(1-\left(\frac{nf^2+1}{n+1}\right)^t\right)s\\
		&= s + \left(\frac{nf^2+1}{n+1}\right)^t(\sigma_0-s).
		\label{eqn:SigmaCondMean_scalar}
	\end{align}
This has the form of a linear AR(1) model with AR parameter $(nf^2+1)/(n+1)$. 
Then  $\lim_{t\rightarrow \infty} E[\sigma_t \mid \sigma_0] = s$ when $|f|<1$, and does so exponentially fast regardless of $n$.  However, the overall rate of this exponential forgetting is governed by the AR parameter $(nf^2+1)/(n+1)$. 

\paragraph{Shared Eigenvectors Between $F$, $S$ and $\Sigma_0$}
In Theorem~\ref{thm:principalIWAR}, we examined a principal component IW-AR process based on the shared eigenvectors of $F$ and $S$.  If we further assume that $\Sigma_0$ shares eigenvectors with $F$ and $S$, the following theorem shows that $\hat{\Sigma}_t = E'\Sigma_t E$ remains diagonal in expectation and has a closed form mean recursion.  

\begin{theorem}
Under the conditions of Theorem~\ref{thm:principalIWAR} assume
$\Sigma_0 = E\Theta_0E'$ where $\Theta_0$ is diagonal, $\Theta_0=\diag(\theta_0)$ 
for some $q-$vector $\theta_0$ with positive elements. 
Let $\theta_{t|0}$ denote the $q-$vector of the eigenvalues of $E[\Sigma_t \mid \Sigma_0],$
$\xi = \diag(Q)$ and $\xi_{-1} = \diag\left(Q^{-1}\right)$. Then $E[\Sigma_t \mid \Sigma_0]$ has the form
of a first-order, non-diagonal autoregression on $\theta_{t|0}$
\begin{align}
	\theta_{t|0} =c_{n,q}(I-R^2)\xi + \left[n^{-1}c_{n,q}(I-R^2)\xi\xi_{-1}' + R^2\right]\theta_{t-1|0},
\end{align}
or
\begin{align}
	\theta_{t|0} &= B^t\theta_0 + \sum_{\tau =0}^{t-1}B^\tau \alpha
\end{align}
where $\alpha = c_{n,q}(I-R^2)\xi$ and $B = n^{-1}c_{n,q}(I-R^2)\xi\xi_{-1}' + R^2$.  Assuming a stationary process such that $||R||_0 < 1$,
\begin{align}
	\theta_{t|0} &= B^t\theta_0 + (I-B)^{-1}(I-B^t)\alpha,
\end{align}
and
\begin{align}
	\lim_{t\rightarrow \infty} \theta_{t|0} &= \xi.
\end{align}
That is, the eigenvalues of the limiting conditional mean are exactly those of the marginal mean $S$.
\label{thm:eigCondMean}
	\begin{proof} See Appendix. \end{proof}
\end{theorem}

Recalling that $\theta_{t|0}$ fully determines $E[\Sigma_t \mid \Sigma_0]$ in the case of shared eigenvectors, we once again conclude that the conditional mean of the process forgets the initial condition $\Sigma_0$ exponentially fast---this occurs irregardless of the value of $n$.

\section{Data Augmentation}
\label{sec:data}

\begin{figure}[t!] \centering
	\includegraphics[height = 2in]{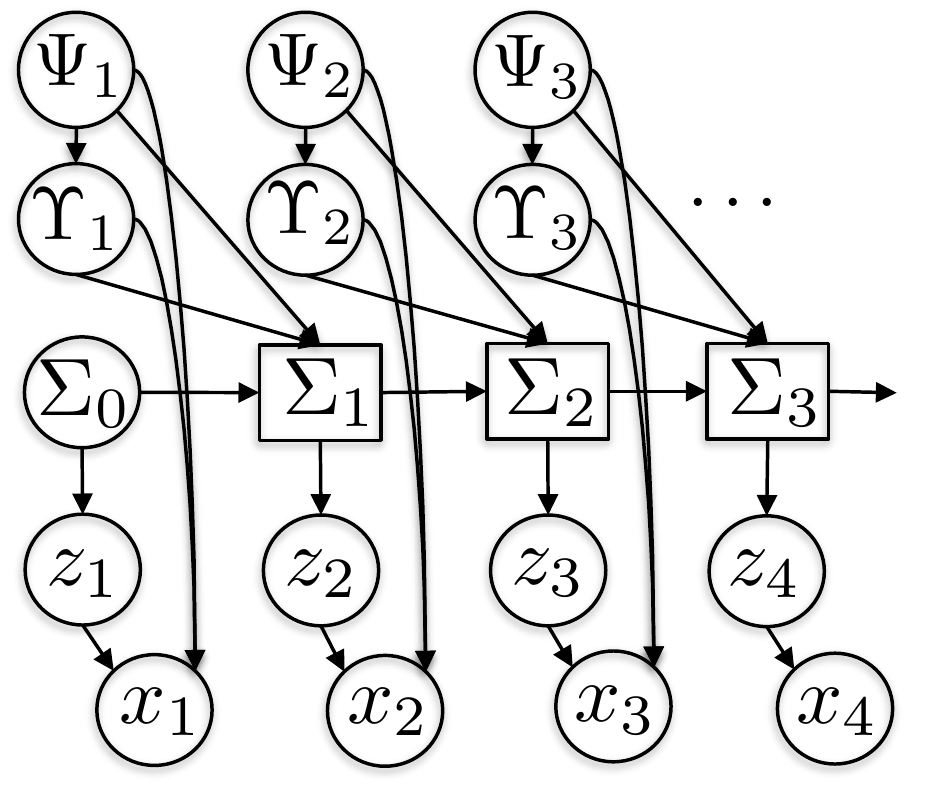}
	\caption{Representation of the graphical model of the IW-AR(1) process under augmentation by the latent  $z_t.$  Circles indicate random variables, arrows imply probabilistic conditional relationships while squares represent 
	quantities that are deterministic based on an instantiation of the variables in their parents nodes.} \label{fig:IWARfull} \postcap \vspace{0.1in}
\end{figure}

Augmentation of the observation model eqn.~\eqref{eqn:obsModel} provides interpretation of the latent innovations process $\{\Upsilon_t,\Psi_t\}$ as well as forming central and critical theoretical development for posterior 
computations  as detailed in Section~\ref{sec:posterior}.  Conditional on $\Sig{0}$ and the innovations 
sequence, the observation model can be regarded as arising by marginalization over an inherent latent $q-$vector
process $z_t,$ $(t=1,\ldots),$ where 
\begin{equation}
	x_t\mid z_t \sim N(\Upsilon_tz_t,\Psi_t)\quad\textrm{and}\quad 
	z_t \sim N(0,\Sigma_{t-1})
\label{eqn:xzreg}
\end{equation} 
independently over time. 
That is, the observations $x_t$ are from a conditionally linear model with latent covariate vectors
$z_t$ and regression parameters $\{\Upsilon_t,\Psi_t\}$.  The normal-inverse Wishart prior for $\{\Upsilon_t,\Psi_t\}$ provides a conjugate prior in this standard multivariate regression framework.  See Figure~\ref{fig:IWARfull} for a graphical model representation of this process.

Let $y_t = [z_t' \,\, x_t']'$ and $\Delta_t = \{\Sigma_{t-1},\Upsilon_t,\Psi_t\}$. Then  
\begin{equation}
\begin{aligned}
	p(y_{1:T}\mid \Delta_{1:T}) = \prod_{t=1}^T N(x_t \mid \Upsilon_t z_t,\Psi_t)N(z_t\mid 0,\Sigma_{t-1})\\
	p(\Delta_{1:T}) = p(\Sigma_0\mid n,S) \prod_{t=1}^T NIW_q(\Upsilon_t,\Psi_t\mid F,n,S)
\end{aligned}
\end{equation}
where $NIW$ denotes the matrix normal, inverse Wishart prior on $\{\Upsilon_t,\Psi_t\}$ of  eqn.~\eqref{eqn:MNIWprior}.  We omit the dependency of the left hand side on the hyperparameters $n$, $F$, and $S$ for notational simplicity. Figure~\ref{fig:IWARdelta} displays the resulting graphical model, clearly 
illustrating the simplified conditional independence structure that enables computation as developed below.  Note that $\Delta_t$ plays the
role of an augmented state and the evolution to time $t$ defines $\Sig{t}$ as a deterministic function of this state.

\section{Model Fitting via MCMC}
\label{sec:posterior}
\begin{figure}[t!] \centering
	\includegraphics[height = 1in]{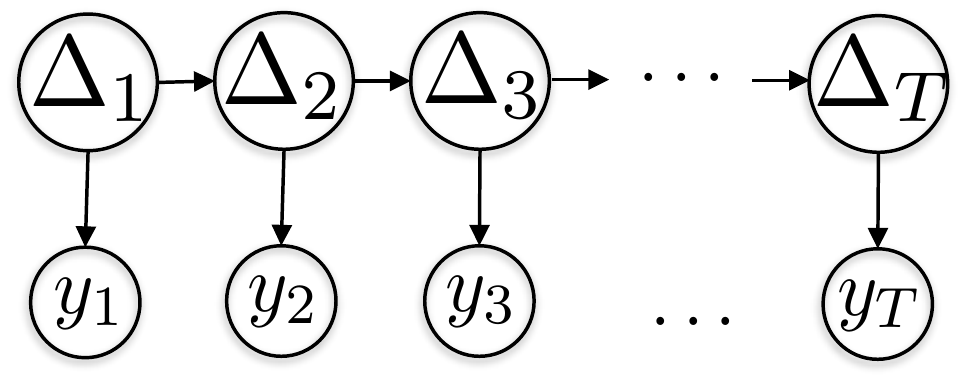}
	\caption{Representation of the graphical model for the augmented data $y_t = [z_t' \,\, x_t']'$ and 
	 latent states $\Delta_t = \{\Sigma_{t-1},\Upsilon_t,\Psi_t\}$.  }
	\label{fig:IWARdelta} \postcap \vspace{0.1in}
\end{figure}
For model fitting, we develop a Markov chain Monte Carlo (MCMC) sampler that harnesses the simplified state-space structure of the 
augmented model comprised of Gaussian observations with an IW-AR process.  This structure (Figure~\ref{fig:IWARdelta}) 
immediately suggests a natural MCMC sampler that iterates between the following steps:

\paragraph{Step 1} Impute the latent process by sampling each $z_t$ from 
\begin{align}
p(z_t \mid x_t, \Upsilon_t,\Psi_t,\Sigma_{t-1}) &\propto N(z_t\mid 0,\Sigma_{t-1}) N(x_t\mid \Upsilon_tz_t,\Psi_t) \nonumber\\
&\hspace{-0.5in}= N(z_t\mid (\Sigma_{t-1}^{-1} + \Upsilon_t'\Psi_t^{-1}\Upsilon_t)^{-1}\Upsilon_t'\Psi_t^{-1}x_t, (\Sigma_{t-1}^{-1} + \Upsilon_t'\Psi_t^{-1}\Upsilon_t)^{-1}).
\end{align}

\paragraph{Step 2} Update the hyperparameters $S$ and $F$ conditioned on $x_{1:T}$ and $z_{1:T}$ by sampling steps defined in Section~\ref{sec:hypers} below. 

\paragraph{Step 3} Impute the augmented variance matrix states using a Metropolis-Hastings approach targeting the
 conditional posterior 
\begin{align}
	p(\Delta_{1:T} \mid x_{1:T}, z_{1:T}, F, S).
	\label{eqn:DeltaPost}
\end{align}
We do this using an approximate forward filtering, backward sampling (FFBS) algorithm to define proposal 
distributions; see Section~\ref{sec:FFBS} below. 

Note that \emph{Step 2} and \emph{Step 3} comprise a block-sampling of the IW-AR hyperparameters $\{F,S\}$ and the augmented process $\Delta_{1:T}$ conditioned on $x_{1:T}$ and $z_{1:T}$.  This greatly improves efficiency relative to a sampler that iterates between (i) sampling $\Delta_{1:T}$ given $F$, $S$, $x_{1:T}$. and $z_{1:T}$ and (ii) sampling $\{F,S\}$ given $\Delta_{1:T}$ (which is then conditionally independent of $x_{1:T}$. and $z_{1:T}$).

%In Section~\ref{sec:SMC} we instead examine a sequential Monte Carlo (SMC) algorithm that provides an online approach to model fitting.  Here, however, we assume that we are working with a batch of data and are interested in retrospective analysis.

\subsection{Forward Filtering, Backward Sampling}
\label{sec:FFBS}

We utilize the fact that there is a deterministic mapping from $\Delta_t$ to the augmented matrix
\begin{align}
	\Delta_t \rightarrow \left(\begin{array}{cc}
		\Sigma_{t-1} & \Sigma_{t-1}\Upsilon'_t\\
		 \Upsilon_t\Sigma_{t-1} & \Sigma_t
	\end{array}\right),
\end{align}
and thus use the two interchangeably.  Our goal is to develop an approximate forward filtering algorithm that produces an approximation to $p(\Delta_{1:T} \mid y_{1:T})$, which can then be used in backward-sampling a posterior sequence $\Sigma_{0:T}$.  We examine the  filtering and sampling stages in turn. 

\paragraph{Approximate Forward Filtering}
An exact forward filtering would involve recursively \emph{updating} $p(\Delta_t \mid y_{1:t-1})$ to $p(\Delta_t \mid y_{1:t})$ and \emph{propagating} $p(\Delta_t \mid y_{1:t})$ to $p(\Delta_{t+1} \mid y_{1:t})$.  However, as examined in the Appendix, this filter is analytically intractable for the IW-AR so we use an approximate filtering procedure based on moment-matching in order to maintain inverse Wishart approximations to each propagate and update step.  Specifically, let $g_{t-1|t-1}(\Delta_{t-1} \mid y_{1:t-1})$ denote the approximation to the posterior $p(\Delta_{t-1} \mid y_{1:t-1})$ at time $t-1$.  We then approximate the predictive distribution $p(\Delta_t \mid y_{1:t-1})$ by
\begin{align}
	g_{t|t-1}(\Delta_t \mid y_{1:t-1}) &= \mbox{IW}\left(r_t,(r_t-2)E_{g_{t-1|t-1}}[\Delta_t\mid y_{1:t-1}] \right).
\end{align}
Here, $r_t$ is a specified degree of freedom to use in the approximation at time $t$.  The subsequent update step of incorporating observation $y_t$ is exact based on the approximations made so far.  Namely,
\begin{align}
	g_{t|t}(\Delta_t \mid y_{1:t}) &= \mbox{IW}\left(r_t+1,(r_t-2)E_{g_{t-1|t-1}}[\Delta_t\mid y_{1:t-1}] + y_ty_t'\right).
	\label{eqn:updatedProposal}
\end{align}
The required expectation here is easily seen to be 
\begin{multline}
	E_{g_{t-1|t-1}}[\Delta_t\mid y_{1:t-1}]\\ = \left(\begin{array}{cc}
			S_{t-1} & S_{t-1}F'\\ FS_{t-1} & FS_{t-1}F' + c_{n,q}(1+\tr(S_{t-1}(nS)^{-1})V)
		\end{array}\right)
\end{multline}
where the $S_t$ sequence is updated using the identity
\begin{align}
	(r_t-1)S_t &= (r_t-2)\left\{FS_{t-1}F' + c_{n,q} (1+\tr(S_{t-1}(nS)^{-1}))V\right\} + x_tx_t'
	\label{eqn:FF_recursion}
\end{align}
as further detailed in the Appendix. 

In summary, the approximate forward filtering is defined by a recursion in which the inverse Wishart distribution for $\Delta_{t-1}$ is converted into a predictive distribution for $\Delta_t$.  This predictive distribution is also taken to be inverse Wishart degree of freedom $r_t$ and mean $E_{g_{t-1|t-1}}[\Delta_t\mid y_{1:t-1}]$ -- i.e., the predictive mean under the distribution $g_{t-1|t-1}$ and the dynamics specified by the IW-AR prior.  The inverse Wishart predictive distribution is then directly updated to the 
resulting inverse Wishart distribution for $\Delta_t$.

\paragraph{Backward Sampling}
\begin{figure}[t!] \centering
	\includegraphics[height = 2in]{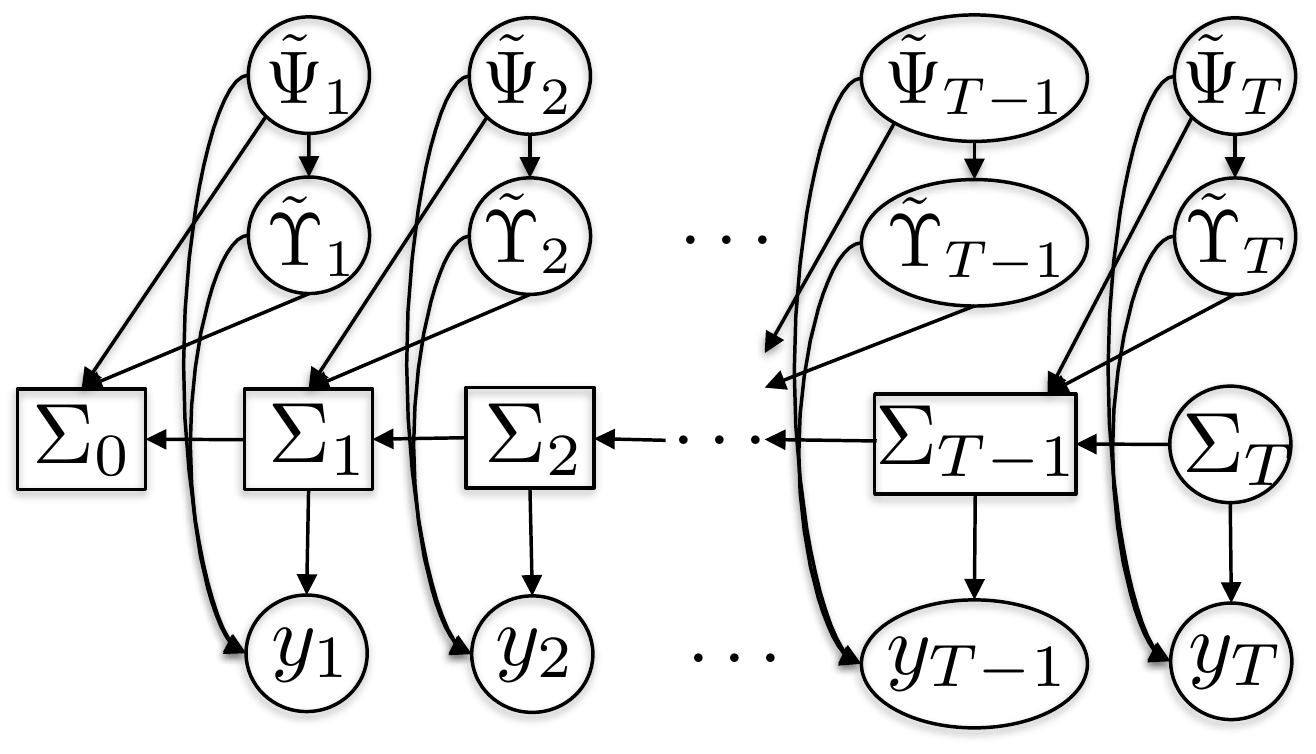}
	\caption{Graphical model of the reverse time inverse Wishart autoregressive process where  $\tilde{\Psi}_t \sim IW(n+q+2,n\tilde{V})$ and $\tilde{\Upsilon}_t \mid \tilde{\Psi}_t \sim N(\tilde{F},\tilde{\Psi}_t,(nS)^{-1})$ with $\tilde{F} = SF'S^{-1}$ and $\tilde{V}=S-SF'S^{-1}FS$. The reverse time transitions are defined by $\Sigma_{t-1} = \tilde{\Psi}_t + \tilde{\Upsilon}_t\Sigma_{t}\tilde{\Upsilon}_t'$.  Since one can deterministically compute $\{\Upsilon_t,\Psi_t\}$ from $\tilde{\Delta}_t = \{\tilde{\Upsilon}_t,\tilde{\Psi}_t,\Sigma_t\}$ (see eqn.~\eqref{eqn:tilde2reg}), the augmented observation $y_t$ are conditionally independent given $\tilde{\Delta}_{1:T}$.}
	\label{fig:IWARreverse} \postcap \vspace{0.1in}
\end{figure}
Running these forward filtering computations to time $t=T$ yields $g_{T|T}(\Delta_T\mid y_{1:T})$ approximating the true posterior $p(\Delta_T\mid y_{1:T})$.  We use the sequence of approximations $g_{t|t}(\Delta_t\mid y_{1:t})$ in deriving a backwards sampling stage, which we show is exact based on the approximations made in the forward filtering. At time $t=T$, we sample $\Sigma_T$ from the implied approximate posterior margin
\begin{align}
\Sigma_T \sim g_{T|T}(\Sigma_T | y_{1:T}) = \mbox{IW}(r_T+1,(r_T-2)E_{g_{T-1|T-1}}[\Sigma_T | y_{1:T-1}] + x_Tx_T').
\end{align}
We then harness the reverse time process, depicted in Figure~\ref{fig:IWARreverse}.  As we iterate backwards in time, we condition on the previously sampled $\Sigma_t$ to sample $\Sigma_{t-1}$ as follows.  We first sample
\begin{equation}
\begin{aligned}		
\tilde{\Psi}_t &\sim \mbox{IW}(r_t+1+q,G_t^{11} - G_t^{21'}(G_t^{22})^{-1}G_t^{21})\\
\tilde{\Upsilon}_t &\mid \tilde{\Psi}_t \sim N(G_t^{21'}(G_t^{22})^{-1},\tilde{\Psi}_t,(G_t^{22})^{-1}),
\end{aligned}
\label{eqn:tildeInnovations}
\end{equation}
and then set
\begin{align}
	\Sigma_{t-1} = \tilde{\Psi}_t + \tilde{\Upsilon}_t\Sigma_t\tilde{\Upsilon}_t'.
\end{align}
Here, $G_t = (r_t-2)E_{g_{t-1|t-1}}[\Delta_t | y_{1:t-1}] + y_ty_t'$, with $G_t^{11}$, $G_t^{21}$, $G_t^{22}$ denoting the three unique $q\times q$ sub-blocks ($G_t^{12} = G_t^{21'}$).  These terms, which can be regarded as sufficient statistics of the forward filtering procedure, can be written as
\begin{equation}
\begin{aligned}
	G_t^{11} &= (r_t-2)S_{t-1} + z_tz_t'\\
	G_t^{21} &= (r_t-2)FS_{t-1} + x_tz_t'\\
	G_t^{22} &= (r_t-1)S_t,
\end{aligned}
\label{eqn:Gt}
\end{equation}
with $S_t$ recursively defined as in~\eqref{eqn:FF_recursion}.  In practice, conditioned on $\{n,S,F\}$, the sequence $S_{1:T}$ is precomputed and simply accessed in the backward sampling of $\Sigma_{0:T}$.

Note that if we wish to impute $\{\Upsilon_t,\Psi_t\}$, we can deterministically compute them based on the sampled $\{\Sigma_{t-1},\Sigma_t,\tilde{\Upsilon}_t,\tilde{\Psi}_t\}$; that is, 
\begin{equation}
\Upsilon_t = \Sigma_t\tilde{\Upsilon}_t'\Sigma_{t-1}^{-1}\quad\textrm{and}\quad
\Psi_t = \Sigma_t - \Upsilon_t\Sigma_{t-1}\Upsilon_t'.
\label{eqn:tilde2reg}
\end{equation}

\paragraph{Accept-Reject Calculation} We use the proposed approximate FFBS scheme as a proposal distribution for a Metropolis Hastings stage.  Let $q(\cdot\mid \cdot)$ represent the proposal distribution for $\{\Sigma_T,\tilde{\Upsilon}_{1:T},\tilde{\Psi}_{1:T}\}$ implied by the sequence of forward filtering approximations $g_{t|t}(\Delta_t\mid y_{1:t})$.  For every proposed $\Delta^*_{1:T}$, we compare the ratio
\begin{align}
	r(\Delta^*_{1:T}) = \frac{p(\Sigma^*_T,\tilde{\Upsilon}^*_{1:T},\tilde{\Psi}^*_{1:T}\mid y_{1:T})}{q(\Sigma^*_T,\tilde{\Upsilon}^*_{1:T},\tilde{\Psi}^*_{1:T}\mid y_{1:T})}
	\label{eqn:acceptReject}
\end{align}
to $r(\Delta_{1:T})$, where $\Delta_{1:T}$ is the previous sample of the augmented sequence.  If $r(\Delta^*_{1:T}) > r(\Delta_{1:T})$, we accept the proposed sequence.  Otherwise, we accept the sequence with probability $r(\Delta^*_{1:T})/r(\Delta_{1:T})$.

The accept-reject ratio is calculated as follows.  Noting that there is a one-to-one mapping between $\{\Sigma_T,\tilde{\Upsilon}_{1:T},\tilde{\Psi}_{1:T}\}$ and $\{\Sigma_0,\Upsilon_{1:T},\Psi_{1:T}\}$,
\begin{align}
	\frac{p(\Sigma_T,\tilde{\Upsilon}_{1:T},\tilde{\Psi}_{1:T}\mid y_{1:T})}{q(\Sigma_T,\tilde{\Upsilon}_{1:T},\tilde{\Psi}_{1:T}\mid y_{1:T})} \propto
	\frac{p(y_{1:T}\mid \Sigma_0,\Upsilon_{1:T},\Psi_{1:T})p(\Sigma_T,\tilde{\Upsilon}_{1:T},\tilde{\Psi}_{1:T})}{q(\Sigma_T,\tilde{\Upsilon}_{1:T},\tilde{\Psi}_{1:T}\mid y_{1:T})}.
\end{align}
The augmented data likelihood is given by
\begin{align}
	p(y_{1:T}\mid \Sigma_0,\Upsilon_{1:T},\Psi_{1:T}) = \prod_{t=1}^T N(x_t\mid  \Upsilon_tz_t,\Psi_t)N(z_t\mid  0,\Sigma_{t-1}).
\end{align}
As specified in eqn.~\eqref{eqn:tildeInnovations_prior}, the prior of the reverse time process is given by
\begin{multline}
	p(\Sigma_T,\tilde{\Upsilon}_{1:T},\tilde{\Psi}_{1:T}) = \mbox{IW}(\Sigma_T\mid n+2,nS)\\
	\times \prod_{t=1}^T \mbox{IW}(\tilde{\Psi}_t\mid n+q+2,n\tilde{V})N(\tilde{\Upsilon}_t\mid \tilde{F}, \tilde{\Psi}_t,(nS)^{-1}).
	\label{eqn:backwardsInnovationsPrior}
\end{multline}
Similarly, the proposal density decomposes as
\begin{equation}
	q(\Sigma_T,\tilde{\Upsilon}_{1:T},\tilde{\Psi}_{1:T}\mid y_{1:T}) =
	\mbox{IW}(\Sigma_T\mid r_T+1,G_T^{22}) 
	 \prod_{t=1}^T  f_t(\tilde{\Upsilon}_t) h_t(\tilde{\Psi}_t \mid \tilde{\Upsilon}_t) 
	\label{eqn:backwardsInnovationsProp}
\end{equation}
where 
$$ f_t(\tilde{\Upsilon}_t) = \mbox{IW}(\tilde{\Psi}_t\mid r_t+q+1,G_t^{11}-G_t^{21'}(G_t^{22})^{-1}G_t^{21})$$
and 
$$ h_t(\tilde{\Psi}_t \mid \tilde{\Upsilon}_t) =N(\tilde{\Upsilon}_t\mid G_t^{21'}(G_t^{22})^{-1},\tilde{\Psi}_t,(G_t^{22})^{-1}). $$

\paragraph{FFBS Computations} One important note is that the acceptance rate decreases exponentially fast with the length of the time series, as with all Metropolis-based samplers for sequences of states in hidden Markov models.  Recall that the proposed $\Sigma^{*}_{0:T}$ sequence is based on a sample $\Sigma^{*}_T$ from $g_{T|T}$ and a collection of $T$ independent samples $\{\tilde{\Upsilon}^{*}_t,\tilde{\Psi}^{*}_t\}$ from distributions based on $g_{t|t}$.  If the final approximate filtered distribution $g_{T|T}$ is a poor approximation to the true distribution, then the collection of $T$ independent proposed innovations $\{\tilde{\Upsilon}^{*}_t,\tilde{\Psi}^{*}_t\}$ are unlikely to result in a $\Sigma^{*}_{0:T}$ that explains the data well.  The accuracy of the approximation $g_{T|T}$ decreases with $T$.  Furthermore, even if $g_{T|T}$ is a good approximation to the true posterior, a single poor innovations sample $\{\tilde{\Upsilon}^{*}_t,\tilde{\Psi}^{*}_t\}$ can be detrimental since the effects propagate in defining $\Sigma^{*}_{0:t-1}$.  The chance of obtaining an unlikely sample $\{\tilde{\Upsilon}^{*}_t,\tilde{\Psi}^{*}_t\}$ for some $t$ increases with $T$.

Since the distributions contributing to the accept-reject ratio $r(\Delta_{1:T})$ factor over $t$, one can sequentially compute and monitor this ratio based on the samples of $\Sigma^{*}_T$ and $\{\tilde{\Upsilon}^{*}_{t:T},\tilde{\Psi}^{*}_{t:T}\}$.  One can then imagine harnessing ideas from randomness recycling~\citep{Huber:01} to improve efficiency by rejecting locally instead of rejecting an entire sample path from $t=0,\dots,T$.  Additionally, one could develop adaptive methods in which samples $\{\tilde{\Upsilon}^{*}_{t},\tilde{\Psi}^{*}_{t}\}$ leading to drastic declines in the acceptance-ratio-to-$t$ were rejected and $\{\tilde{\Upsilon}^{*}_{t},\tilde{\Psi}^{*}_{t}\}$ was then resampled, but only for some finite period of adaptation.  These ideas all focus on the ability to accept or reject entire sub-sequences $\{\Sigma^*_T,\tilde{\Upsilon}^{*}_{t:T},\tilde{\Psi}^{*}_{t:T}\}$, and require theoretical analysis to justify convergence to the correct stationary distribution.  Alternatively, we develop below an {\em innovations-based} sampling
approach in which we fix $\{\Sigma_T,\tilde{\Upsilon}_{1:t-1,t+1:T},\tilde{\Psi}_{1:t-1,t+1:T}\}$ and simply consider accepting or rejecting $\{\tilde{\Upsilon}^{*}_t,\tilde{\Psi}^{*}_t\}$ at a single time step $t$.

\paragraph{An Innovations-Based Sampler} We propose an \emph{innovations} sampler that accepts or rejects $\{\tilde{\Upsilon}^{*}_t,\tilde{\Psi}^{*}_t\}$ for every $t$ instead of accepting or rejecting the entire chain $\Sigma^*_{0:T}$ induced from the collection of these backwards innovations samples (and $\Sigma^*_T$).  Specifically, let $\theta_t = \{\tilde{\Upsilon}_t,\tilde{\Psi}_t\}$ and
$	\mathcal{S}_t = \{\theta_1,\theta_2,\dots,\theta_T,\Sigma_T\}.$
For each $t$ we propose
\begin{align}
	\mathcal{S}^*_t = \{\theta_1,\theta_2,\dots,\theta_{t-1},\theta^*_t,\theta_{t+1},\dots,\theta_T,\Sigma_T\},
\end{align}
with $\theta^*_t=\{\tilde{\Upsilon}^*_t,\tilde{\Psi}^*_t\}$.  The accept-reject ratio based on eqn.~\eqref{eqn:acceptReject} simplifies to
\begin{equation}
\begin{aligned}
	\frac{r(\Delta^*_{1:T})}{r(\Delta_{1:T})} &= \frac{p(\mathcal{S}^*_t \mid y_{1:T})}{q(\mathcal{S}^*_t\mid y_{1:T})}\frac{q(\mathcal{S}_t \mid y_{1:T})}{p(\mathcal{S}_t\mid y_{1:T})}\\
	&= \frac{p(y_{1:T}\mid \Sigma^*_0,\Upsilon^*_{1:t},\Psi^*_{1:t},\Upsilon_{t+1:T},\Psi_{t+1:T})}{p(y_{1:T}\mid \Sigma_0,\Upsilon_{1:T},\Psi_{1:T})}\frac{p(\mathcal{S}^*_t)}{p(\mathcal{S}_t)}\frac{q(\mathcal{S}_t\mid y_{1:T})}{q(\mathcal{S}^*_t\mid y_{1:T})}\\
	&= \frac{\prod_{\tau=1}^t N(x_{\tau}\mid \Upsilon^*_{\tau} z_{\tau},\Psi^*_{\tau})N(z_{\tau}\mid \Sigma^*_{\tau-1})}{\prod_{\tau=1}^t N(x_{\tau}\mid \Upsilon_{\tau} z_{\tau},\Psi_{\tau})N(z_{\tau}\mid \Sigma_{\tau-1})}
	\frac{p_t(\theta^*_t)}{p_t(\theta_t)}\frac{q_t(\theta_t \mid y_{1:T})}{q_t(\theta^*_t \mid y_{1:T})}.
\end{aligned}
\end{equation}
Here $p_t(\theta_t)$ denotes the matrix normal, inverse Wishart prior on the backwards innovations $\theta_t=\{\tilde{\Upsilon}_t,\tilde{\Psi}_t\}$ and $q_t(\theta_t\mid y_{1:T})$ the corresponding distribution under the forward-filtering based proposal.  That is, $p_t(\cdot)$ and $q_t(\cdot\mid y_{1:T})$ represent the time $t$ components of eqns.~\eqref{eqn:backwardsInnovationsPrior} and~\eqref{eqn:backwardsInnovationsProp}, respectively.  We utilize the fact that the prior, proposal and likelihood terms all factor over $t$.  For the prior and proposal, the only terms that differ between the proposed sequence $\mathcal{S}^*_t$ and the previous $\mathcal{S}_t$ are the backwards innovations at time $t$ (i.e, $\theta_t$).  For the likelihood, the effects of the change in \emph{backwards} innovations at time $t$ propagate to the \emph{forward} parameters $\{\Sigma^*_0,\Upsilon^*_{1:t},\Psi^*_{1:t}\}$ while leaving $\{\Upsilon_{t+1:T},\Psi_{t+1:T}\}$ unchanged.  See Figure~\ref{fig:IWARreverse}.

Note that the proposed innovations sampler is quite computationally intensive since the accept-reject ratio calculation for each proposed $\{\tilde{\Upsilon}^{*}_t,\tilde{\Psi}^{*}_t\}$ requires recomputing $\Sigma^*_{0:t-1}$.  In practice, we employ an approximate sampler that harnesses the fact that the effects of a given $\{\tilde{\Upsilon}^{*}_t,\tilde{\Psi}^{*}_t\}$ on $\Sigma^{*}_{t-\tau}$ decreases in expectation as $\tau$ increases.  That is, $\{\tilde{\Upsilon}^{*}_t,\tilde{\Psi}^{*}_t\}$ represents a stochastic input whose effect is propagated through a stable dynamical system (assuming the $F$ has spectral norm less than 1).  In particular, we only calculate $\Sigma^{*}_{t-\tau}$ for $\tau$ increasing until the Frobenius norm $||\Sigma^{*}_{t-\tau}-\Sigma_{t-\tau}||_2<\epsilon$ for some pre-specified, small value $\epsilon >0$.  That is, we propagate the effects of $\{\tilde{\Upsilon}^{*}_t,\tilde{\Psi}^{*}_t\}$ until the value of $\Sigma^{*}_{t-\tau}$ becomes nearly indistinguishable numerically from the previous $\Sigma_{t-\tau}$.  Alternatively, in order to maintain a constant per-sample computational complexity, one can specify a fixed lag $\tau$ based on $F$ and $n$ since these hyperparameters determine (in expectation) the rate at which the effects of the proposed $\{\tilde{\Upsilon}^{*}_t,\tilde{\Psi}^{*}_t\}$ decay.

In contrast to sequential sampling of $\{\tilde{\Upsilon}_t,\tilde{\Psi}_t\}$ for $t=T,\dots,1$, one can imagine focusing on regions where the current $\Sigma_t$ is ``poor'', where ``poor'' is determined by some specified metric (e.g., the likelihood function $N(x_t\mid 0,\Sigma_t)$).  As long as there is still positive probability of considering any $t \in \{1,\dots,T\}$, the resulting sampler will converge to the correct stationary distribution.

Finally, instead of always running an approximate forward filter and performing backward sampling (where the ``backward sampling'' can occur in any order based on the innovations representation), one could run a \emph{backward filter} and perform \emph{forward sampling} (BFFS), exploiting the theory of the reverse-time IW-AR process.  By interchanging FFBS with BFFS, the errors aggregated during filtering and the uncertainty inherent at the filter's starting point alternate from $t=0$ to $t=T$, thus producing samples closer to the values that would be obtained if smoothing were analytically feasible.

\subsection{Hyperparameters}
\label{sec:hypers}

In sampling the IW-AR hyperparameters $F$ and $S$, we need to ensure that $V=S-FSF'$ remains positive definite.  Section~\ref{sec:stationarity} explored two cases in which simple constraints on $F$ imply $V$ positive definite for $S$ positive definite: (i) $F=\rho I_q$ or (ii) $F = ERE'$ with $E$ the eigenvectors of $S$.  A simple framework for sampling the hyperparameters in this case is to propose $S$ from a Wishart distribution, thus ensuring its positive-definiteness, and the eigenvalues of $F$ from a beta proposal, thus ensuring spectral radius bounded by 1.  The induced $V$ will then be positive definite.  One can also assume Wishart and beta priors.

Both of the above specifications of $F$ and $S$ lead to reversible IW-AR processes.  For a non-reversible IW-AR process (assuming $S$ non-diagonal), we can take $F = \diag(\rho)$, implying $V = (1\cdot1' - \rho\cdot\rho')\circ S$  where $\circ$ denotes the Hadamard product.  Note that even with $|\rho_i|<1$ and $S$ positive definite, $V$ need not be positive definite.  However, for $V$ positive definite and $|\rho_i|<1$, $S$ will be positive definite and has elements simply defined by $S_{ij} = V_{ij}/(1-\rho_i\rho_j)$.  Thus, in the case of $F$ diagonal we sample $F$ and $V$ and then compute $S$ from these values.  Once again, we employ a beta prior on $\rho_i$ and a Wishart prior now on $V$.  The details of the posterior computations for the case of $F$ diagonal are outlined below.  The case of $F = ERE'$ and $S = EQE'$ follows similarly.

Let $\mbox{W}(v_0,V_0)$ denote a Wishart prior for $V$ and $\mbox{Beta}(c\rho_{0,i},c(1-\rho_{0,i}))$ a beta prior for $\rho_i$.  We use an independence chain sampler in which $V^*$ is proposed from a Wishart proposal $\mbox{W}(v_1,V_1)$ and $\rho^*_i$ from a beta proposal $\mbox{Beta}(d\rho_{1,i},d(1-\rho_{1,i}))$. The accept-reject ratio is then calculated based on the ratio
\begin{equation}
\begin{aligned}
	\frac{r(V^*,F^*)}{r(V,F)} %&= \frac{p(V^*,F^* \mid x_{1:T},z_{1:T})q(V)q(F)}{p(V,F \mid, x_{1:T},z_{1:T})q(V^*)q(F^*)}\\
	 					&= \frac{p(x_{1:T}\mid z_{1:T},F^*,V^*)p(z_{1:T}\mid F^*,V^*)p(V^*)p(F^*)q(V)q(F)}
								{p(x_{1:T}\mid z_{1:T},F,V)p(z_{1:T}\mid F,V)p(V)p(F)q(V^*)q(F^*)}.
\end{aligned}
\end{equation}
Here, $p(\cdot)$ and $q(\cdot)$ denote the prior and proposal for the specified argument, respectively.  The conditional likelihood $p(x_{1:T}\mid z_{1:T},F,V)$ and marginal likelihood $p(z_{1:T}\mid F,V)$ are derived below.  We interchange $\{F,V\}$ and $\{F,S\}$ since there is a bijective mapping between the two when $F$ is diagonal.

\paragraph{Conditional Likelihood $p(x_{1:T} \mid z_{1:T},F,S)$}
Since $x_t \sim N(\Upsilon_tz_t,\Psi_t)$ with $\Upsilon_t \mid \Psi_t \sim N(F,\Psi_t,(nS)^{-1})$ and $\Psi_t \sim IW(n+q+2,nV)$, marginalizing $\Upsilon_t$ and $\Psi_t$ yields a multivariate t distribution as detailed in the
Appendix. This results in   
\begin{align}	
	p(x_{1:T} \mid z_{1:T},F,S) = \prod_{t=1}^T t_{n+q+2}\left(x_t\mid Fz_t,c_{n,q+2}\{1+z_t'(nS)^{-1}z_t\}V \right).
\end{align}

\paragraph{Marginal Likelihood $p(z_{1:T} \mid F,S)$}
Computing the marginal likelihood requires the evaluation of the analytically intractable integral
\begin{align}
	p(z_{1:T} \mid F,S) = \int \prod_t p(z_t \mid \Sigma_{t-1}) p(\Sigma_{0:T}\mid F,S) d\Sigma_{0:T}.
\end{align}
However, we can approximate the marginal likelihood by employing an approximate filter in a 
manner analogous to that of Section~\ref{sec:FFBS}.  In particular, if we had an exact filter that produced the \emph{predicted} distribution $p(\Sigma_{t-1} \mid z_{1:t-1})$ and the \emph{updated} distribution $p(\Sigma_{t-1} \mid z_{1:t})$,  we could recursively compute the marginal likelihood as 
\begin{align}
	p(z_{1:t}) = \frac{p(z_t \mid \Sigma_{t-1})p(\Sigma_{t-1}\mid z_{1:t-1})}{p(\Sigma_{t-1}\mid z_{1:t})}p(z_{1:t-1})
	\label{eqn:zmarg}
\end{align}
(here $F$ and $S$ are omitted for notational simplicity).
Recall that $p(z_t \mid \Sigma_{t-1}) = N(z_t\mid 0,\Sigma_{t-1})$.  Since exact filtering is not possible, we propose an approximate moment-matched filter using ideas parallel to those used for the FFBS approximation to 
$p(\Delta_t\mid y_{1:t})$. Specifically,  
\begin{align}
	g_{t|t}(\Sigma_{t} \mid z_{1:t}) &= \mbox{IW}\left(r_{t},(r_{t}-2)\bar{\Sigma}_{t|t} \right)\\
	g_{t|t+1}(\Sigma_{t} \mid z_{1:t+1}) &= \mbox{IW}\left(r_{t}+1,(r_{t}-2)\bar{\Sigma}_{t|t}+ z_{t+1}z_{t+1}'\right),
\end{align}
with $\bar{\Sigma}_{t|t} = E_{g_{t-1|t}}[\Sigma_t \mid z_{1:t}]$.  The matched-means are recursively computed 
using
\begin{align}
	(r_{t-1}-1)\bar{\Sigma}_{t-1|t} &= (r_{t-1}-2)\bar{\Sigma}_{t-1|t-1} + z_tz_t' \\
	\bar{\Sigma}_{t|t} &= F\bar{\Sigma}_{t-1|t}F' + c_{n,q}\left[ 1+\tr\{\bar{\Sigma}_{t-1|t}(nS)^{-1}\}\right],
\end{align}
with initial condition $\bar{\Sigma}_{0|0} = S$.

Using this approximate filter in eqn.~\eqref{eqn:zmarg} for the marginal likelihood and canceling terms yields
\begin{align}
	p(z_{1:t}\mid F,S) = \frac{1}{\pi^{Tq/2}}\prod_{t=1}^T \frac{|(r_{t-1}-2)\bar{\Sigma}_{t-1|t-1}|^{(r_{t-1}+q-1)/2}}{|(r_{t-1}-2)\bar{\Sigma}_{t-1|t-1} + z_tz_t'|^{(r_{t-1}+q)/2}}\frac{\Gamma\left(\frac{r_{t-1}+q}{2}\right)}{\Gamma\left(\frac{r_{t-1}}{2}\right)}.
\end{align}
Note that we could have employed our filter for $\Delta_{t}$ based on observations $y_t = [z_t' \,\, x_t']'$ to produce an approximation to $p(x_{1:T},z_{1:T}\mid F,S)$.  However, since we have an exact form for $p(x_{1:T} \mid z_{1:T}, F,S)$ we choose to reduce the impact of our approximation by simply using it to compute $p(z_{1:T} \mid F, S)$.

We note further that it is straightforward to analyze $p(F,S \mid \Sigma_0,\Upsilon_{1:T},\Psi_{1:T})$, suggesting that the MCMC use Gibbs sampler components for $F,S$ that would avoid 
 approximations.  However, in practice we found use of this leads to extremely slow mixing rates
 relative to our proposed strategy above.

\section{Stochastic Volatility in Time Series}
\label{sec:eeg}

In this section, we consider a full analysis in which actual \emph{observations} $\xi_t$ are from a 
VAR$(r)$ model whose \emph{innovations} $x_t$ have IW-AR(1) volatility matrices.  
That is, we observe $q-$vector
data $\xi_t$ such that
\begin{align}
	\xi_t = \sum_{i=1}^r A_i \xi_{t-i} + x_t, \hspace{0.25in} x_t \sim N(0,\Sigma_t),
	\label{eqn:ARobs}
\end{align}
where $A_i$ is the $q\times q$ autoregressive parameter matrix at lag $i$ and we now assume that 
 $\Sigma_t$ is an IW-AR(1) process. Define
$A = \begin{bmatrix} A_1 & \cdots & A_r \end{bmatrix}.$ 

We modify the MCMC of Section~\ref{sec:posterior} as follows.  In place of \emph{Step 1} that previously sampled $z_{1:T}$ given $\Delta_{1:T}$ and $x_{1:T}$, we now block sample $\{A,z_{1:T}\}$ given $\Delta_{1:T}$ and $\xi_{1-r:T}$.  That is, we first sample $A$ given $\Delta_{1:T}$ and $\xi_{1-r:T}$ and then $z_{1:T}$ given $A$, $\Delta_{1:T}$ and $\xi_{1-r:T}$.  Noting that $x_{1:T}$ is a deterministic function of $\xi_{1-r:T}$ and the autoregressive matrix $A$, the latter step follows exactly as before.  \emph{Step 2} and \emph{Step 3} remain unchanged.  Thus, the only modification to the sampler is the insertion of a \emph{Step 0} to sample $A$ given $\Delta_{1:T}$ and $\xi_{1-r:T}$.  See the Appendix for further details.

\subsection{Example in Analysis of EEG Time Series}

In multi-channel electroencephalogram (EEG)  studies, multiple probes on the scalp of a patient 
undergoing an induced brain seizure generate electrical potential fluctuation signals that represent 
the spatially localized read-outs of the underlying signal~\citep{Krystal1999b}.  
Much of prior work with clinically relevant data sets has
been on the evaluation of time:frequency structure in such series~\citep{Krystal2000,Ombao:05,Freyermuth:10}
and time-varying parameter vector autoregressions are key tools in this
applied context, as in others~\citep{Prado1997,West1999,PradoWest2010}. 
Existing models represent some aspects of cross-series structure in this inherently spatially distributed multiple 
time series context~\citep{Prado:99,Prado2001a},
but past studies have shown substantial residual dependencies among estimated innovations processes 
across EEG probe locations and the implications for estimation of such
structure in models that ignore significant patterns of time-varying cross-series correlations are 
largely unexplored.   Hence it is of interest to explore models that use IW-AR models for 
multivariate volatility processes of innovations driving vector autoregressions as an obvious first-step.

We explore one initial example using the model of eqn.~\eqref{eqn:ARobs}.  Define 
$\nu_t = \Sigma_t^{-1/2}\xi_t$ so that 
\begin{align}
	\nu_t = \sum_{i=1}^r A_{t,i}\nu_{t-i} + w_t, \hspace{0.25in} w_t \sim N(0,I)
\end{align}
where the $A_{t,i} = \Sigma_t^{-1/2}A_i \Sigma_{t-i}$ are structured, time-varying AR parameter matrices 
for the transformed process. We can fit this model in the original form of eqn.~\eqref{eqn:ARobs} 
and this transformed series is then of
interest as defining underlying independent component series. 

An example data analysis uses $q=10$  channels of a sub-sampled series of 1000 time points, taken from the
larger data set of \cite{West1999}. The original series were collected at a rate of 256/second and these are 
down-sampled by a factor of 2 here to yield $T=1000$ observations over roughly 8 seconds.  
The data were first standardized over a significantly longer time window,
and the selected 8 second section of data corresponds to a recording period containing abnormal neuronal 
activity and thus increased changes in volatility.
The example sets $r=8$ and uses underlying  diagonal autoregressive matrices $A_i = \diag(a_i)$
with independent and relatively diffuse priors $a_i\sim N(0,10I_q).$  

For the IW-AR model component, we assume the rather general,
irreversible IW-AR process with a diagonal $F=\diag(\rho)$  and set $n=6.$  We specify priors on $\rho$ and 
$S$ based on an exploratory analysis of an earlier held-out section of the time series, $\xi^0_{1:T}$, also of length 1000. 
Specifically, this was based on estimating innovations $x^{0}_{1:T}$ from $q$ separate, univariate TVAR models
as in \cite{Prado:99}.  Treating these constructed zero-mean series as raw data, the standard variance matrix 
discounting method~\citep{PradoWest2010}  was applied using an initial 20 degrees of freedom and a discount 
factor $\beta = 0.95$ to generate 100 independent posterior samples of the series of $q\times q$ variance 
matrices, say $U_{0:T},$ across this
prior, hold-out period.  We then applied individual univariate IW-AR(1) models -- the inverse gamma processes 
of Section~\ref{sec:q1} --  to each of the diagonal data sets $U_{ii,t}.$  From these, we extracted summary information 
on which to base the priors for the real data analysis, as follows. First, we take 
 $\rho_i \sim \mbox{Beta}(100\rho_{0,i},100(1-\rho_{0,i})),$ independently, where $\rho_{0,i}^2 = (a_i(n+1)-1)/n$
and   $a_i$ is the approximate posterior mean of the IW-AR autoregressive parameter from the hold-out data 
 analysis of $U_{ii,t};$ second, we set $\nu_0 = q+2$ and $V_0 = (1\cdot 1' - \rho_0\cdot \rho_0') \circ S_0$, where $S_0$ is the 
 sample mean of all of the the $U_{1:T}.$ 

Although centered around a held-out-data-informed mean, the chosen Wishart prior for $S$ is quite diffuse
and the beta priors for the $\rho_i$ are weakly informative relative to the number of observations $T=1000.$ 
Our use of initial hold-out data to specify priors is coherent and consistent with common practice in other
areas of Bayesian forecasting and dynamic modeling such as in using factor models;~\cite{AguilarWest00}, for 
example, adopt such an approach and give useful discussion  of the importance of centering hyperprior support 
around ``reasonable'' values for these types of dynamic models.

From an identical analysis on the batch of test data, we infer values $\rho_{1}$ and $V_{1}$ that are used in specifying the $\mbox{Beta}(d\rho_{1,i},d(1-\rho_{1,i}))$ proposal for $\rho_i$ and $W(v_1,V_{1})$ proposal for $V$ used in our MCMC algorithm.  After some experimentation, this used tuning parameters 
$d=750$ and $v_1 = 40$.  The FFBS proposals also rely on defining the moment-matched IW degree of freedom parameters $r_{0:T}$ for which we set $r_0 = n+2$, which matches the prior specification, and then discount as $r_t = 0.98 r_{t-1} + 1$.  Also, in employing the approximate innovations-based sampler described in Section~\ref{sec:FFBS},  the 
analysis monitors based on $||\Sigma^{*}_{t-\tau}-\Sigma_{t-\tau}||_2<\epsilon$ and uses $\epsilon = 1e-4$.

Some summaries of analysis are based on running 5 separate MCMC chains for 5000 iterations, discarding the first 
1000 samples of each and thinning by examining every 10th MCMC iteration.  Note that we count one full sweep of side-by-side innovations based FFBS of $\Sigma_{0:T}$ as one step in an iteration.  The sampler was initialized with $F$ and $S$ based on the mean of their respective proposal distributions and the residuals $x_{1:T}$ computed from $q$ separate
univariate TVAR analyses.  The sequences $\Sigma_{0:T}$, $\Upsilon_{1:T}$, and $\Psi_{1:T}$ were initialized by 
directly accepting the first proposal from one step of the FFBS algorithm. 

\begin{figure}[htbp!]
	\centering
	\begin{tabular}{ccc}
		\hspace{-0.2in}
		\includegraphics[width = 1.65in]{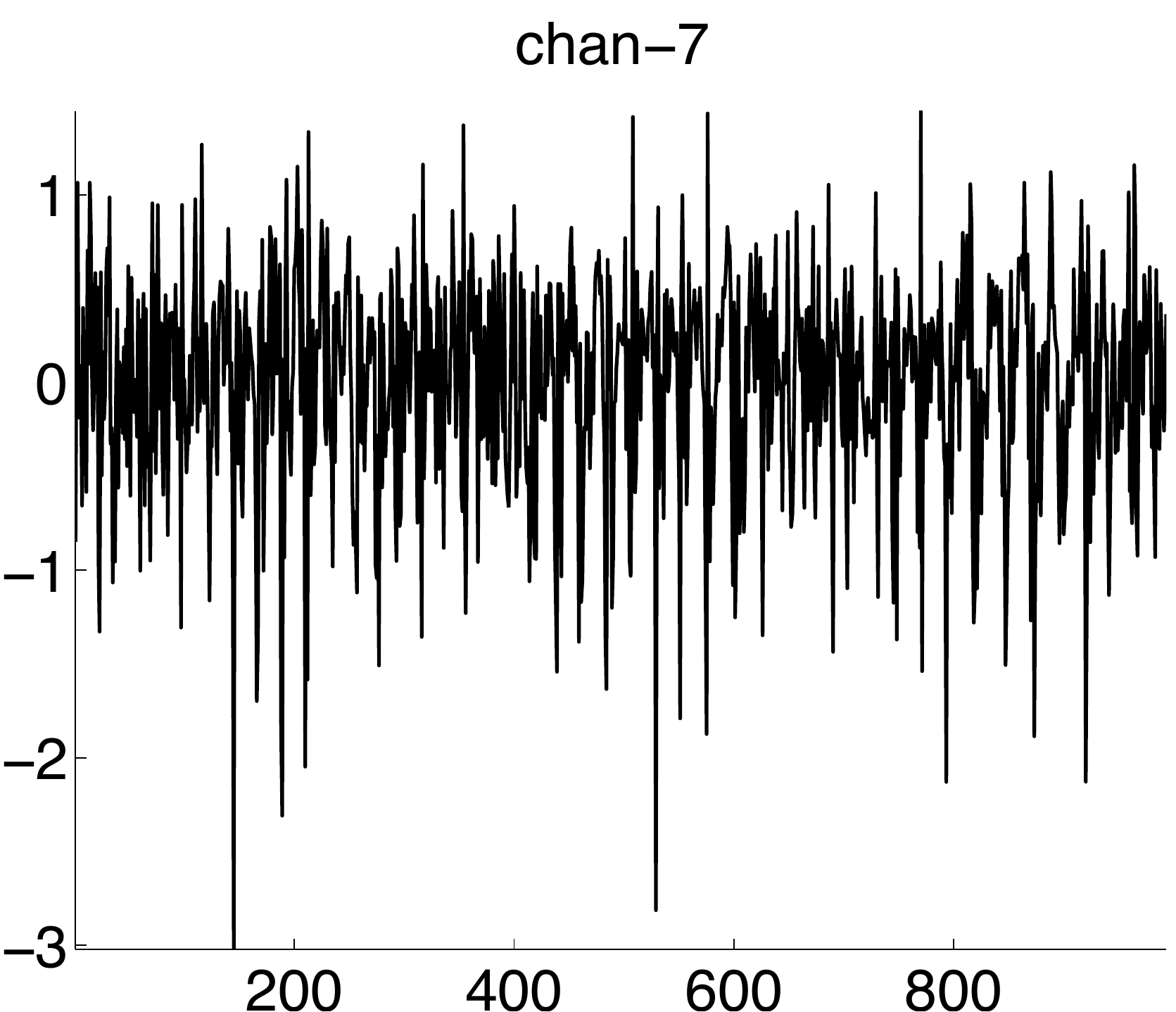} & \hspace{-0.1in}
		\includegraphics[width = 1.65in]{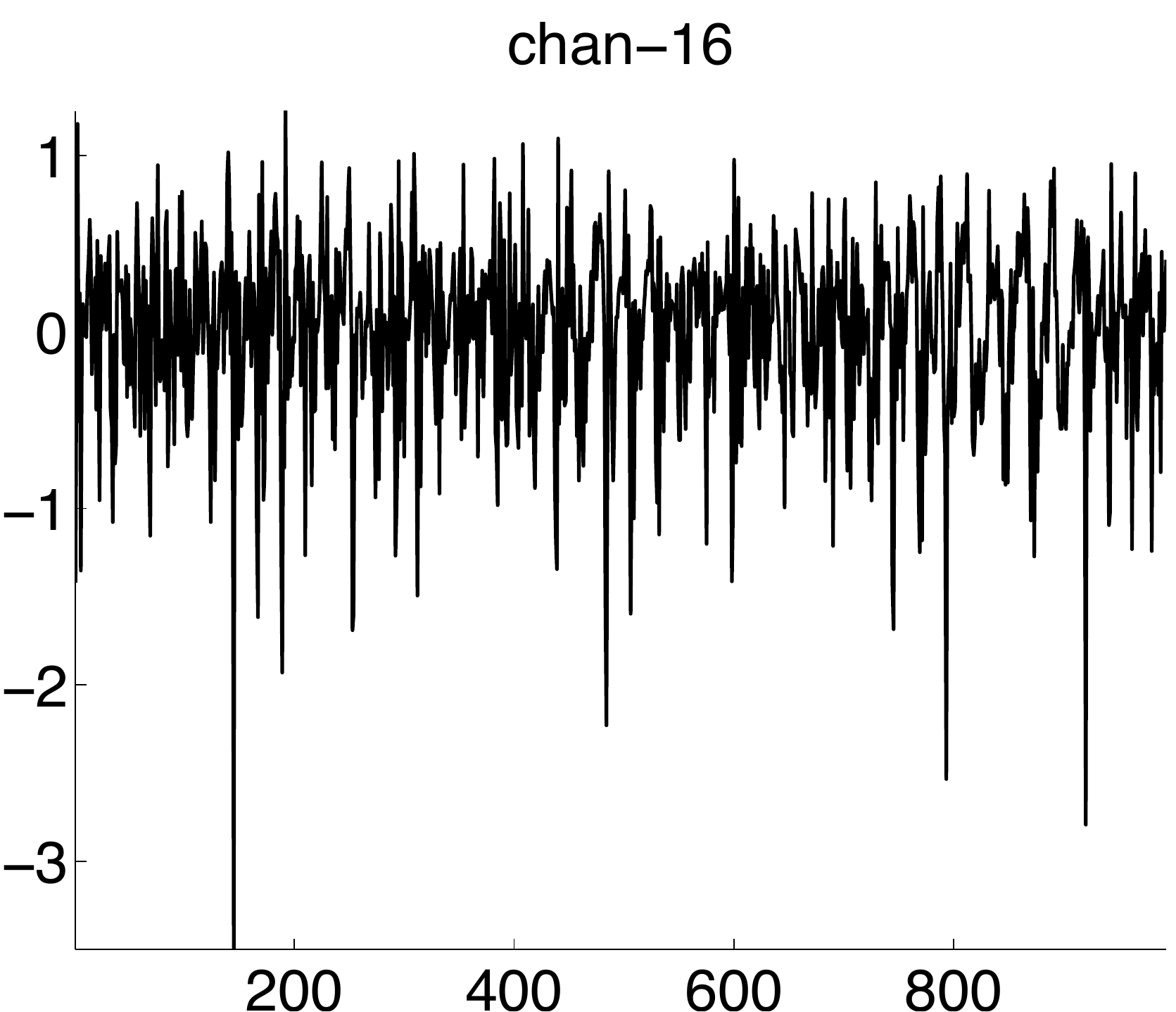} & \hspace{-0.1in}
		\includegraphics[width = 1.65in]{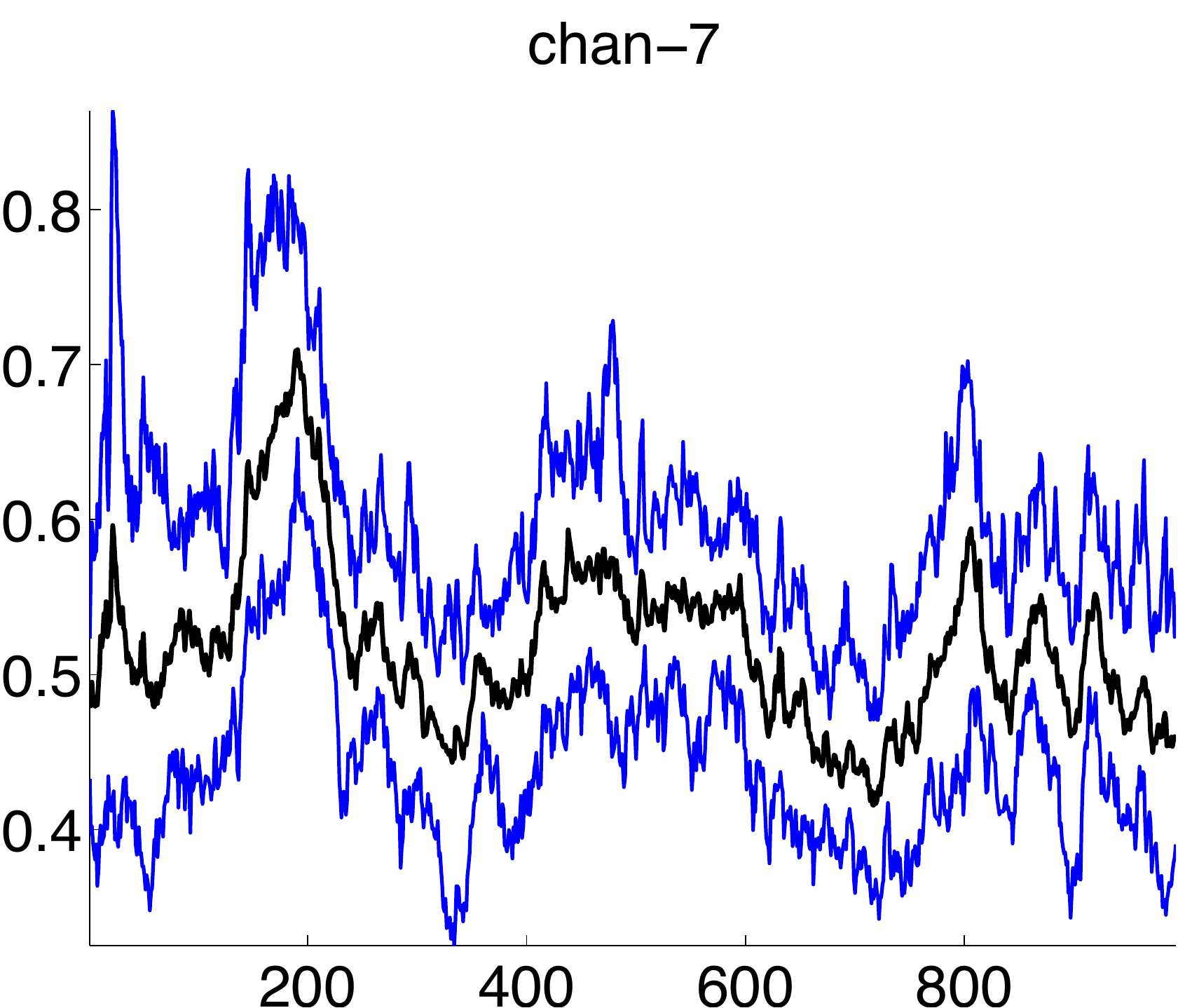} \\
		\hspace{-0.2in}
		\includegraphics[width = 1.65in]{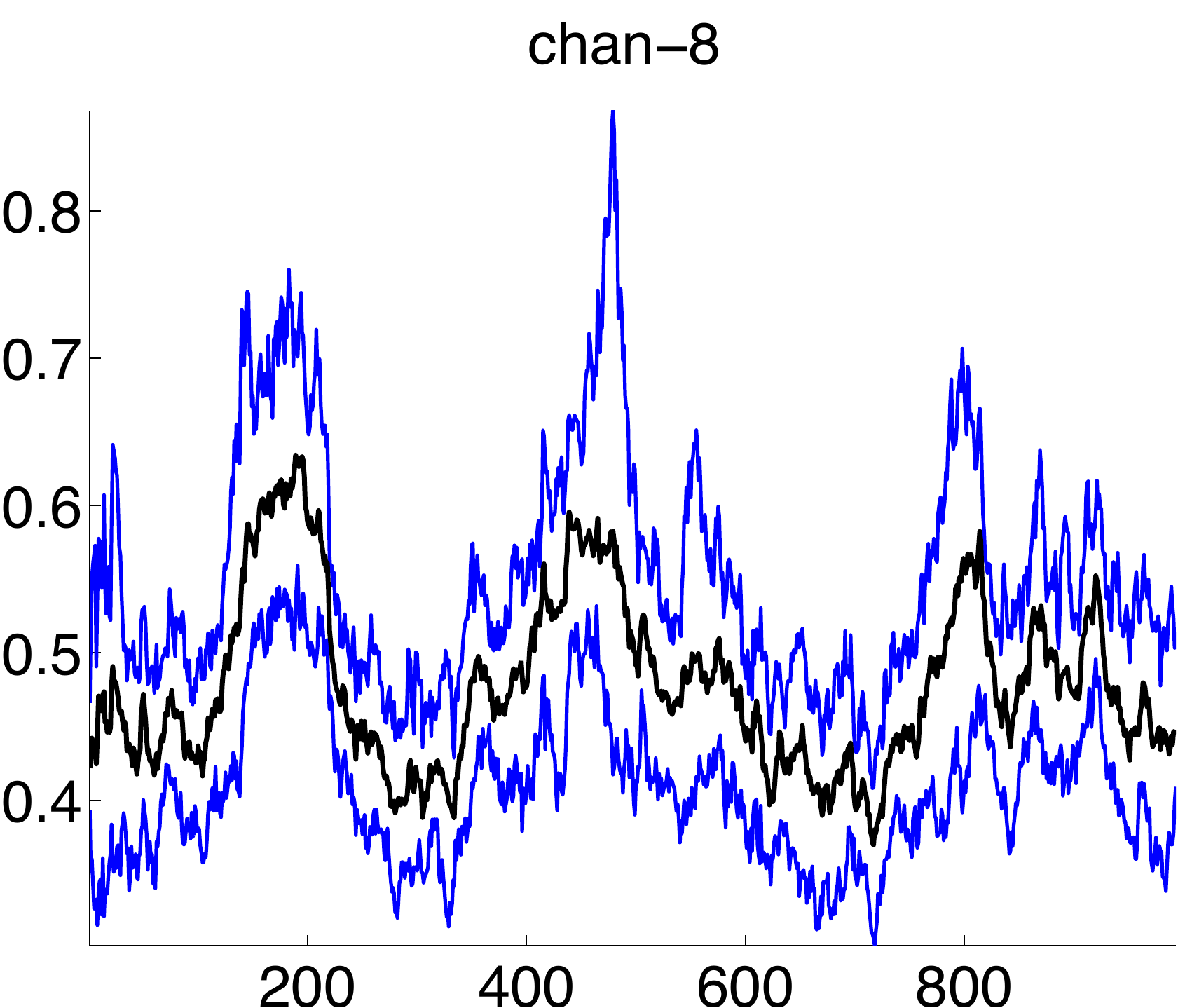} & \hspace{-0.1in}
		\includegraphics[width = 1.65in]{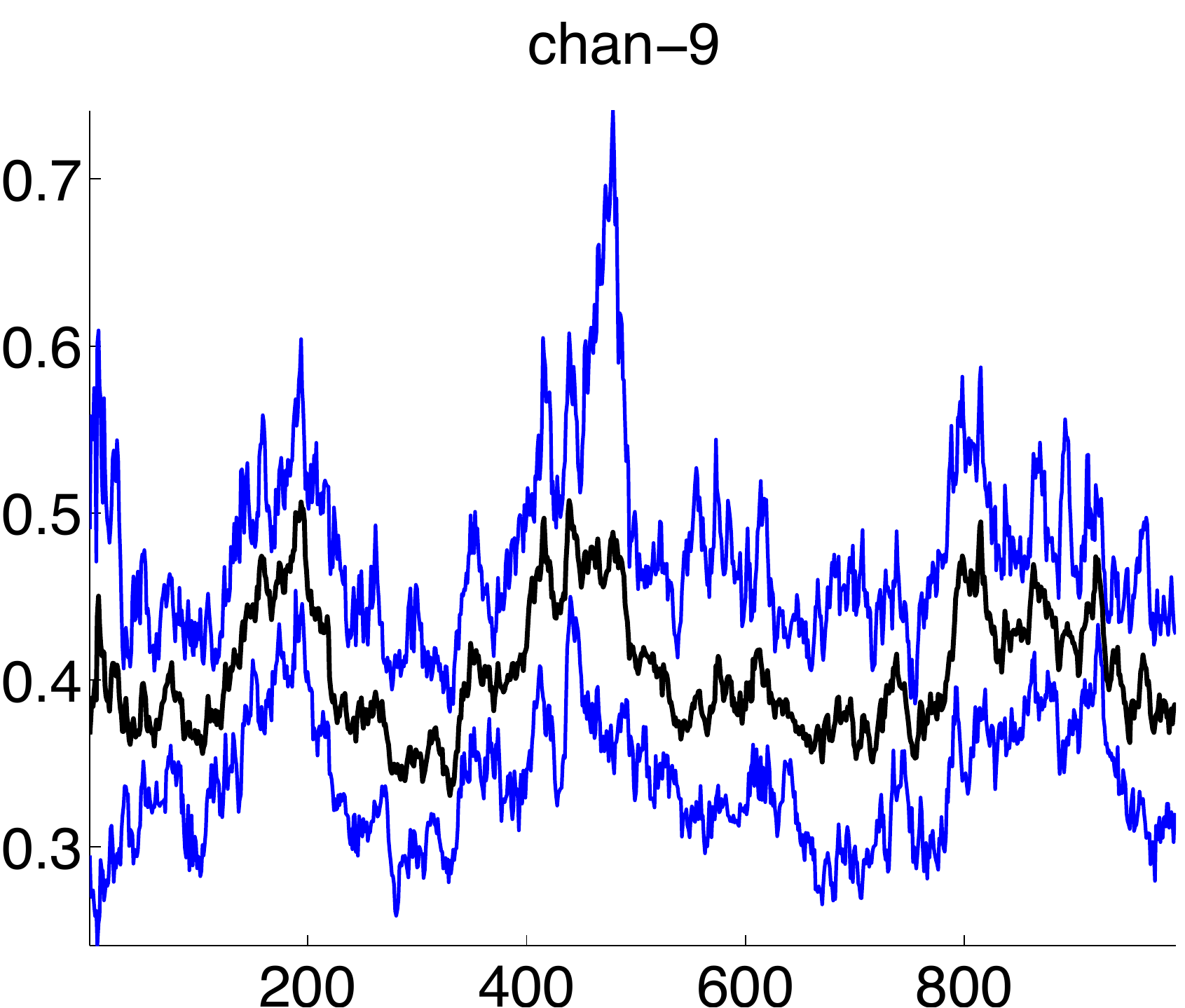} & \hspace{-0.1in}
		\includegraphics[width = 1.65in]{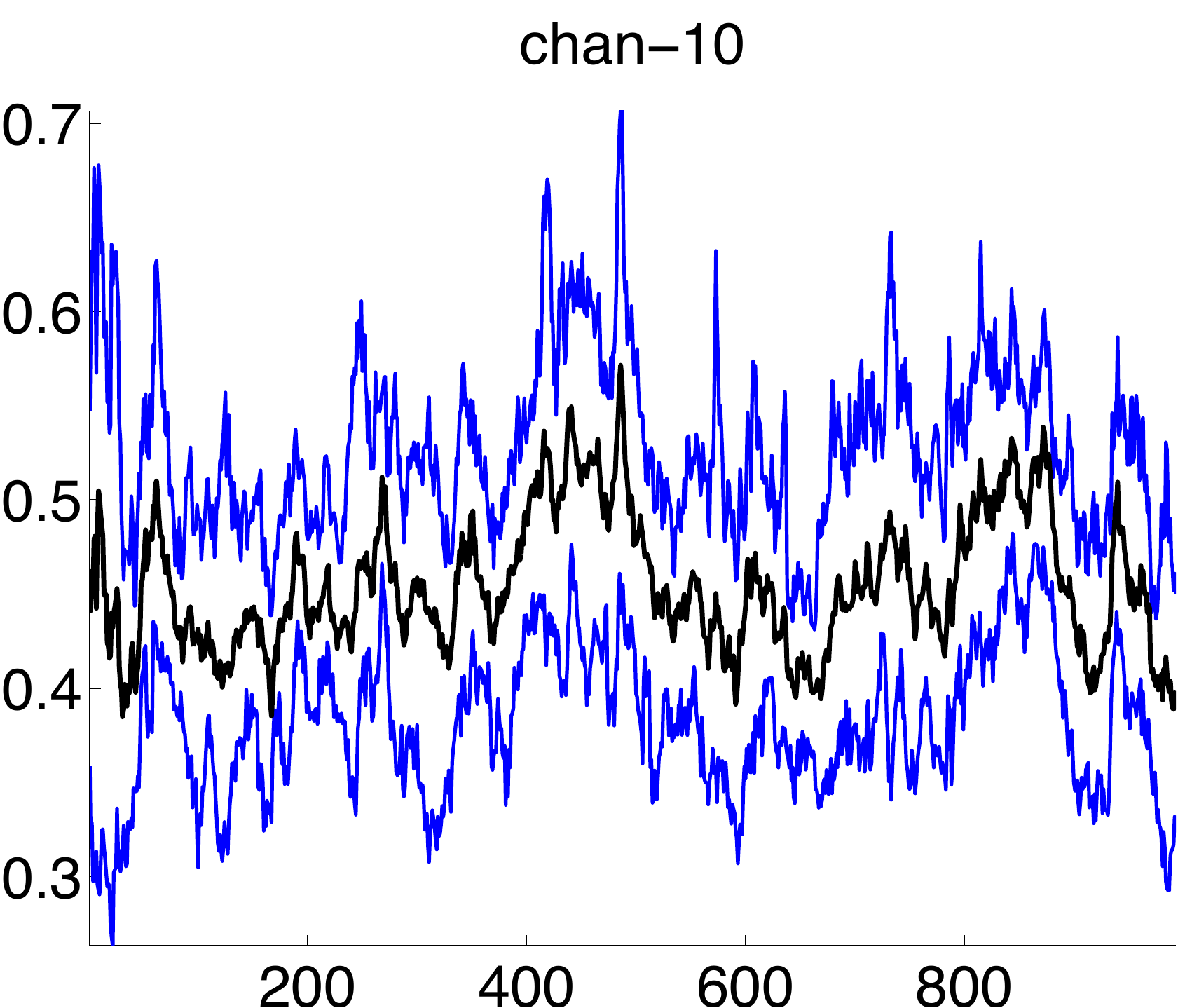} \\
		\hspace{-0.2in}
		\includegraphics[width = 1.65in]{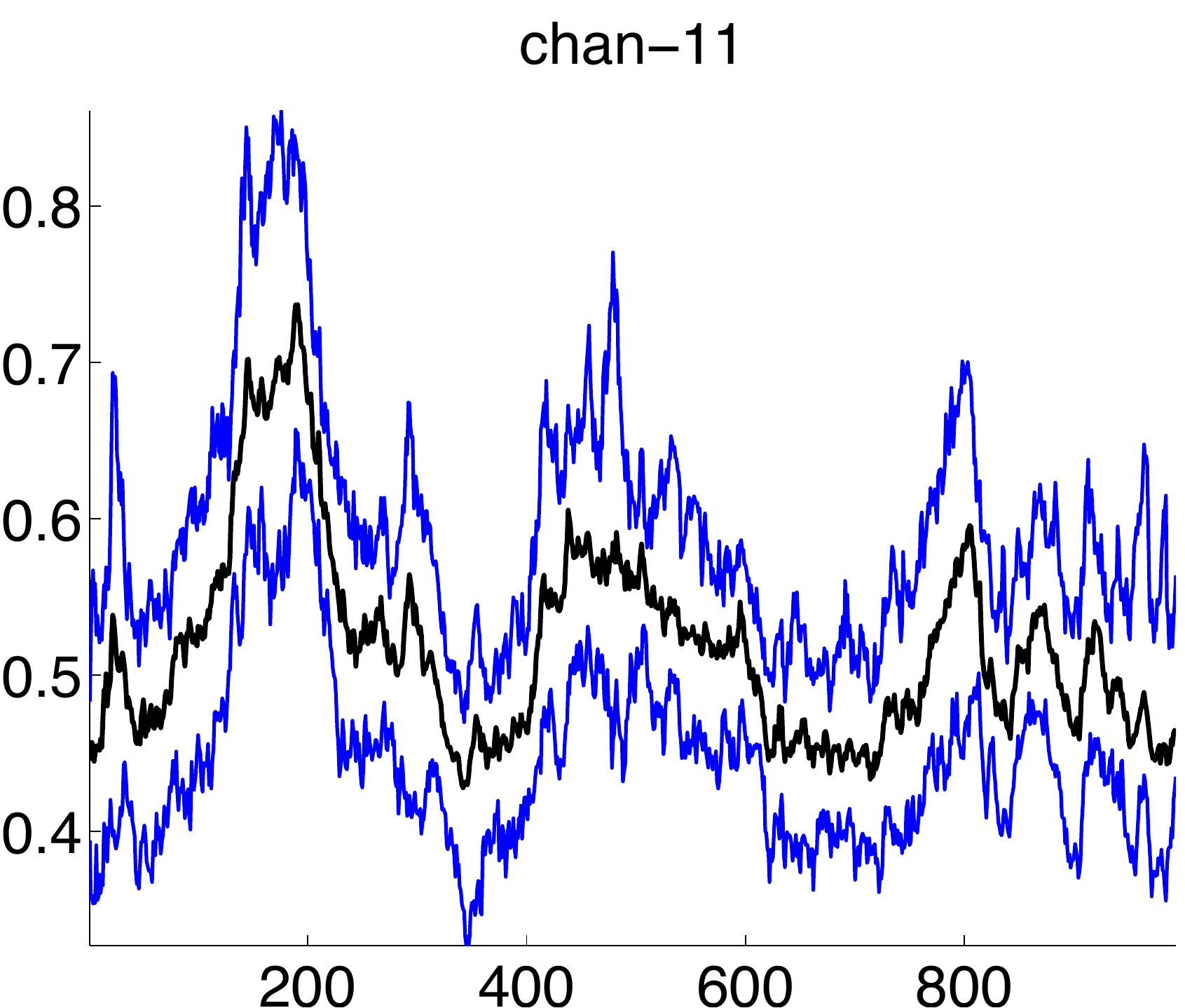} & \hspace{-0.1in}
		\includegraphics[width = 1.65in]{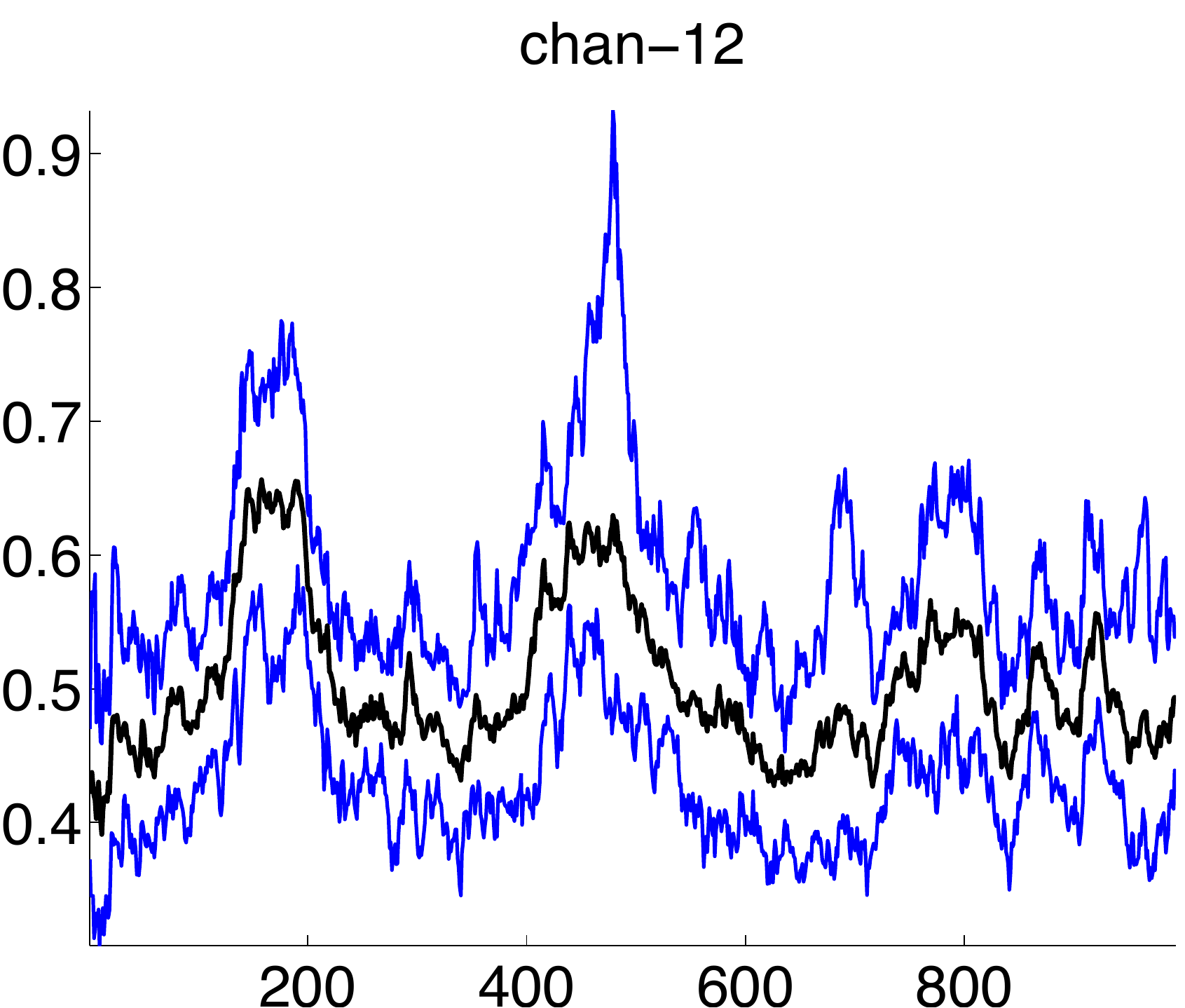} & \hspace{-0.1in}
		\includegraphics[width = 1.65in]{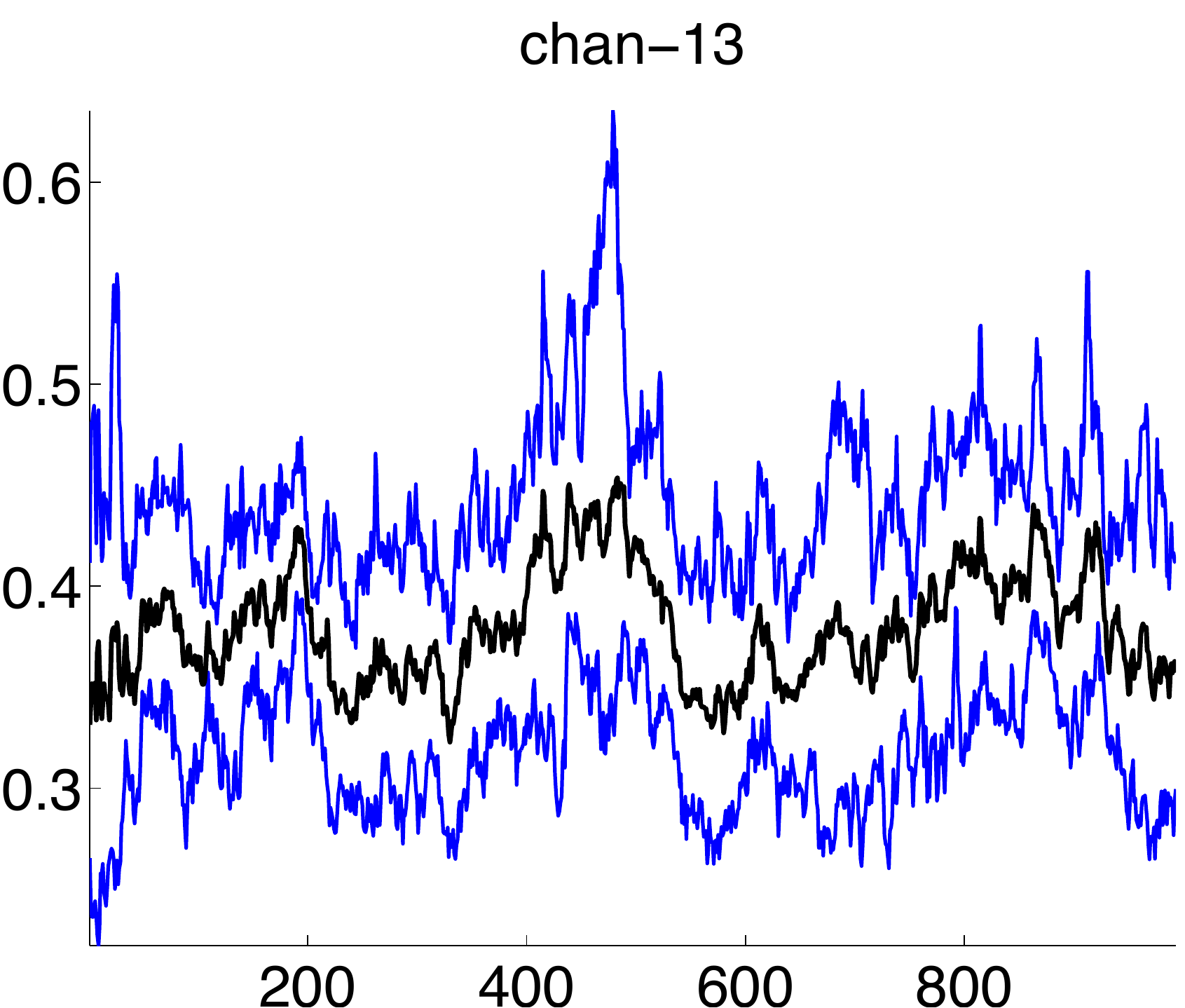} \\
		\hspace{-0.2in}
		\includegraphics[width = 1.65in]{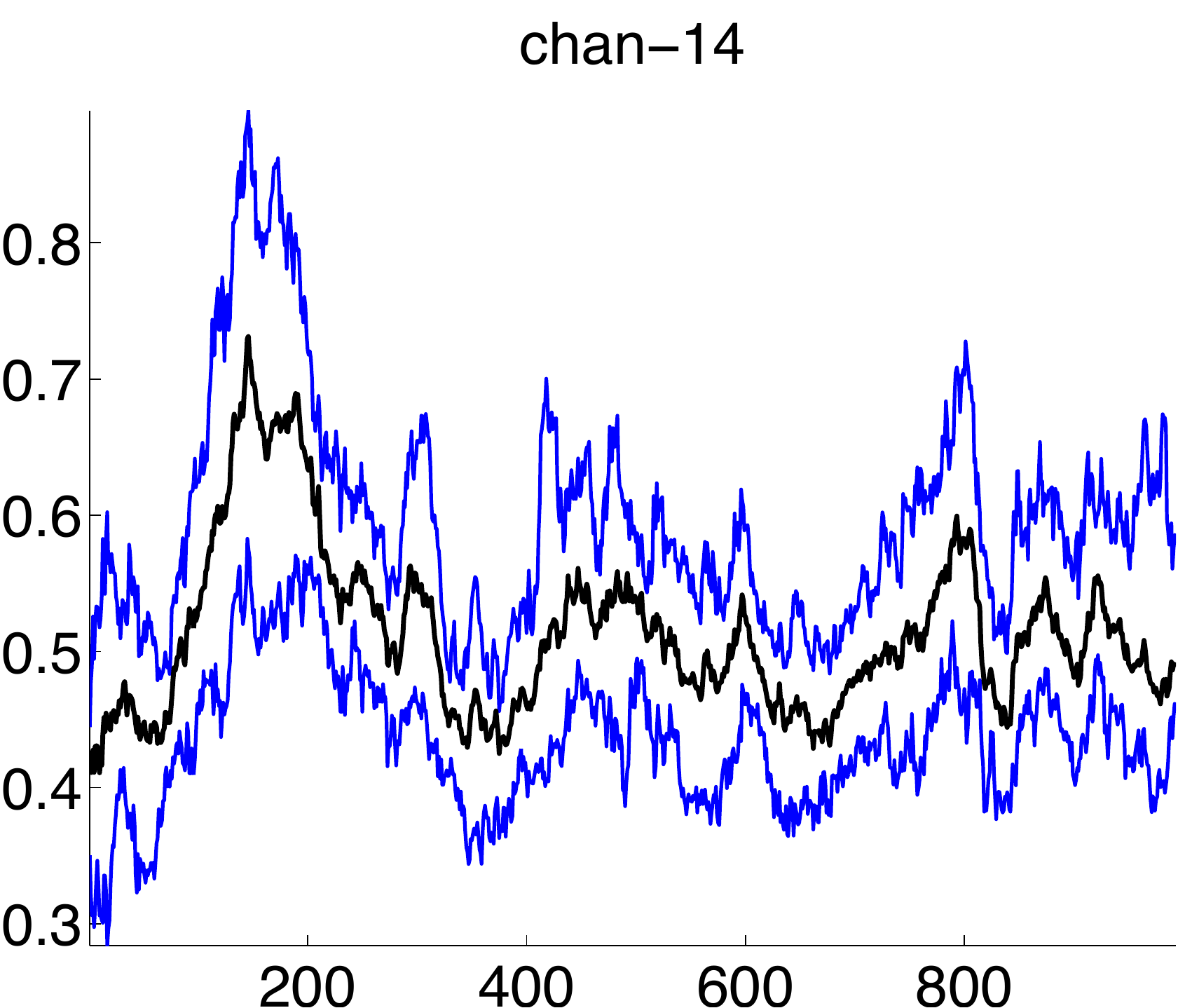} & \hspace{-0.1in}
		\includegraphics[width = 1.65in]{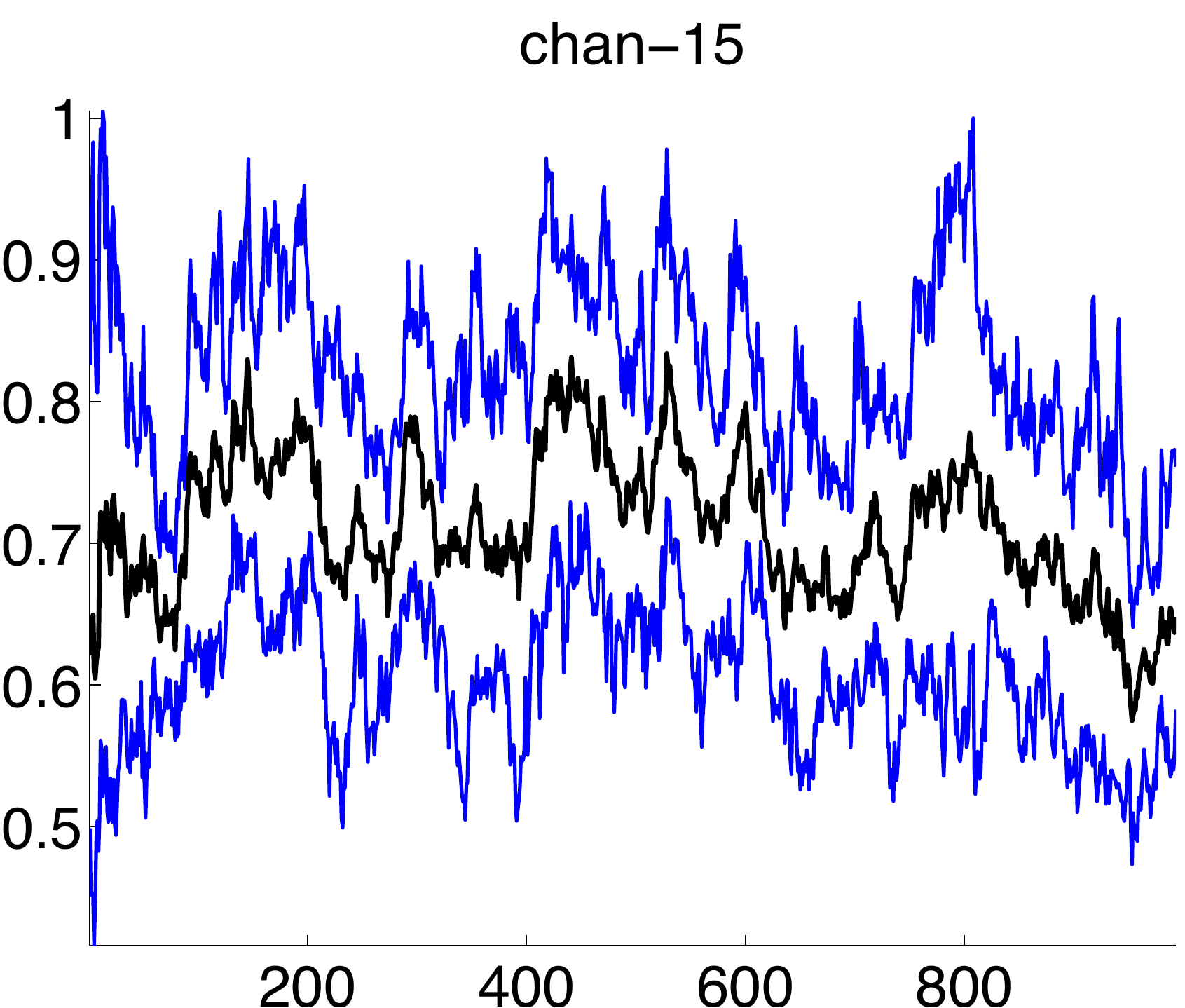} & \hspace{-0.1in}
		\includegraphics[width = 1.65in]{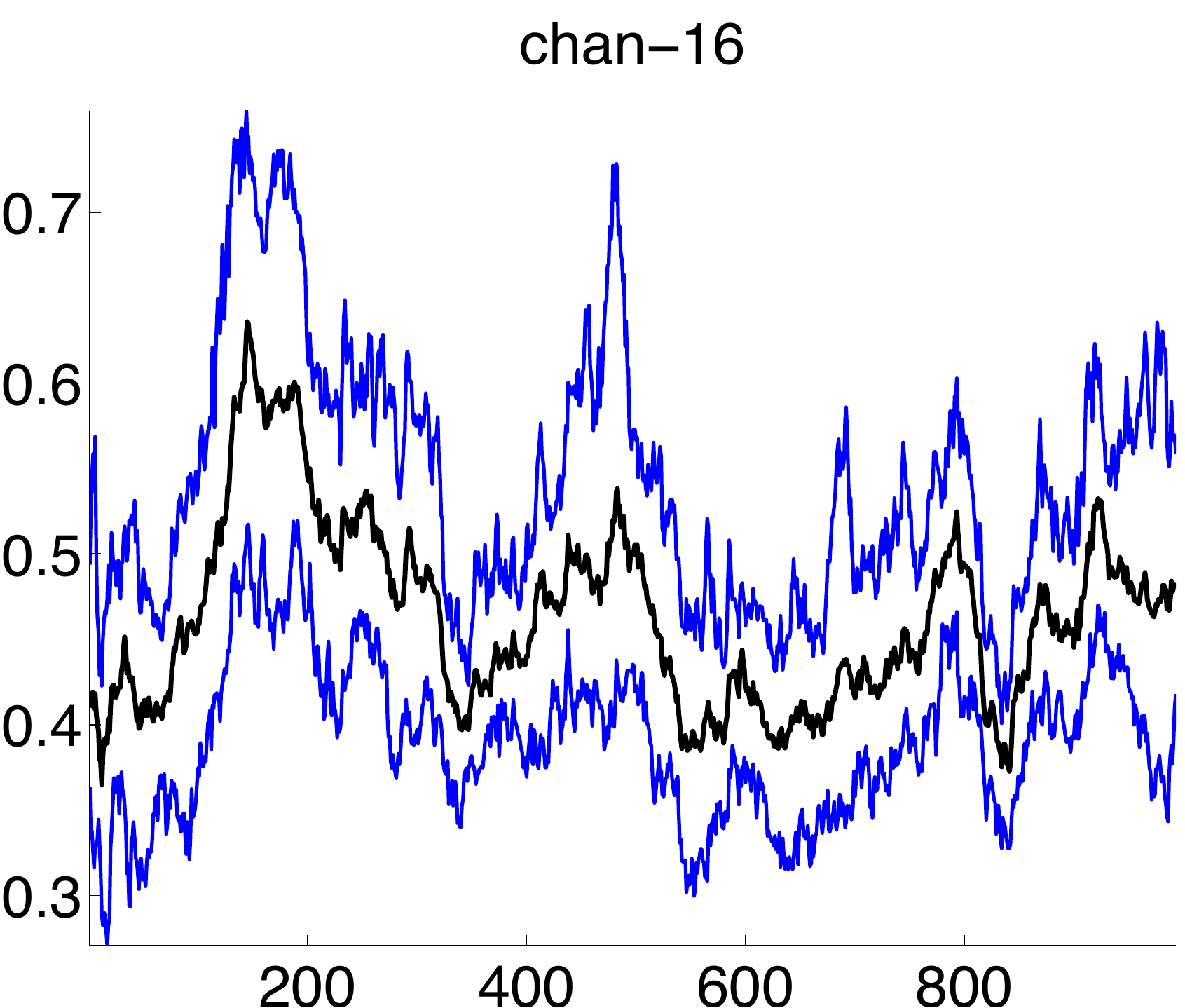}
	\end{tabular}
	\caption{EEG time series from 2 of the 10 channels (first two frames) followed by estimated trajectories of volatilities  $\Sigma_{ii,t}^{1/2}$ for each of $i=1:10$ time series representing EEG channels 7-16 in the original data set.  
	The IW-AR posterior mean, computed based on averaging over 5 chains from iterations $[1000:10:5000]$, is shown in black.  The point-wise 95\% highest posterior density intervals are indicated in blue.} \label{fig:eeg_volatility} \postcap \vspace{0.1in}
\end{figure}

\begin{figure}[htbp!]
	\centering
	\begin{tabular}{ccc}
		\hspace{-0.2in}
		\includegraphics[width = 1.65in]{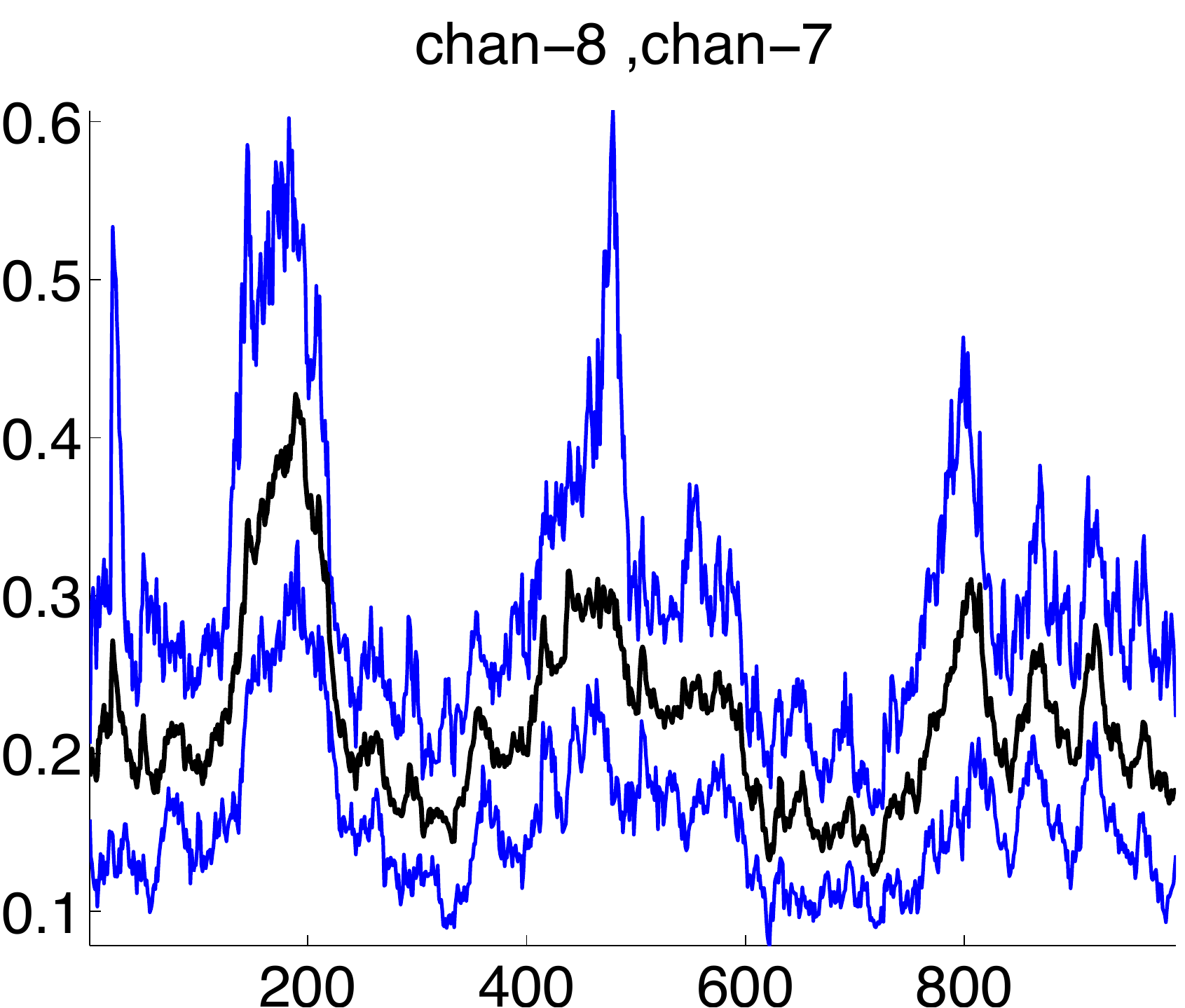} & \hspace{-0.1in}
		\includegraphics[width = 1.65in]{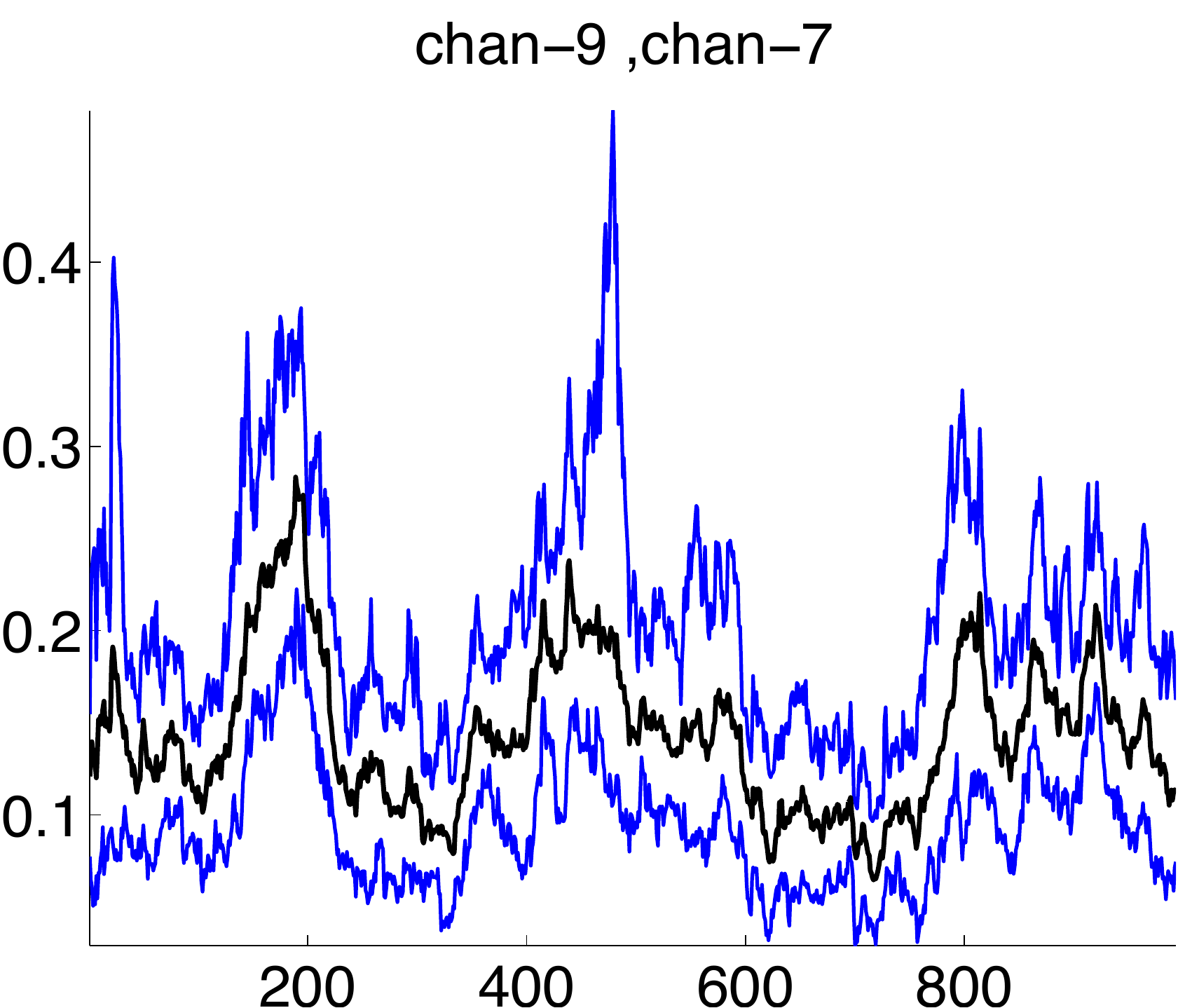} & \hspace{-0.1in}
		\includegraphics[width = 1.65in]{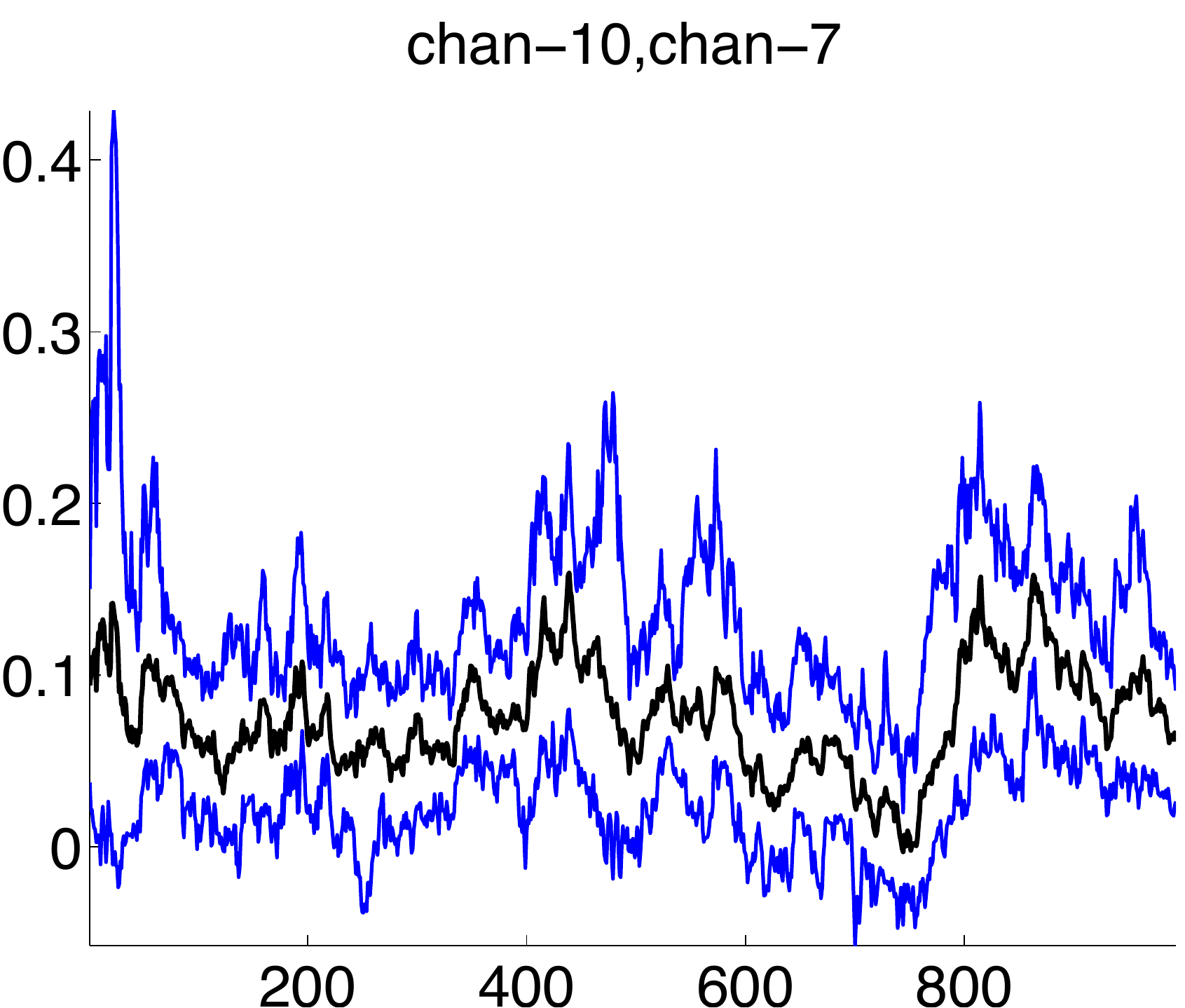} \\
		\hspace{-0.2in}
		\includegraphics[width = 1.65in]{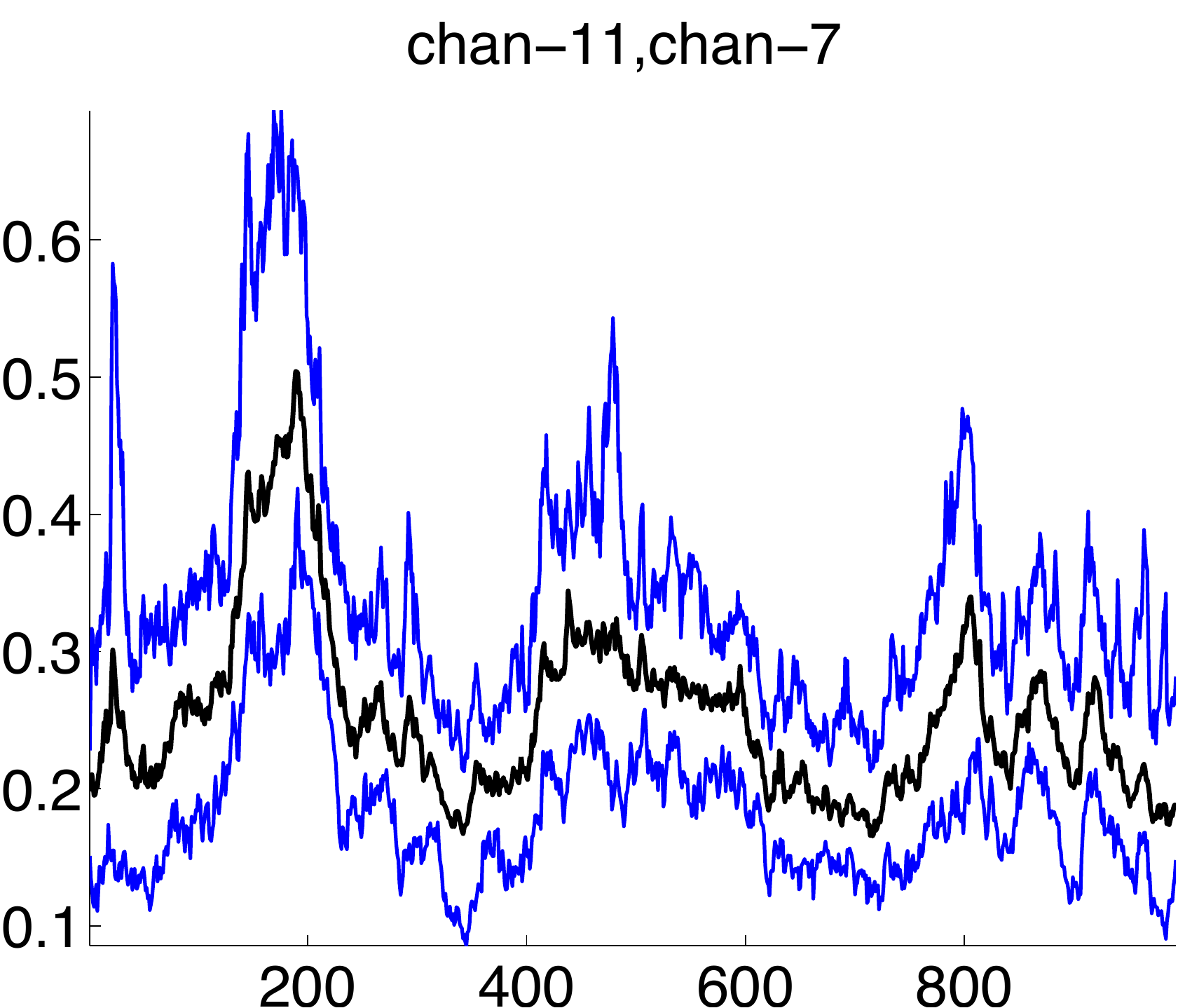} & \hspace{-0.1in}
		\includegraphics[width = 1.65in]{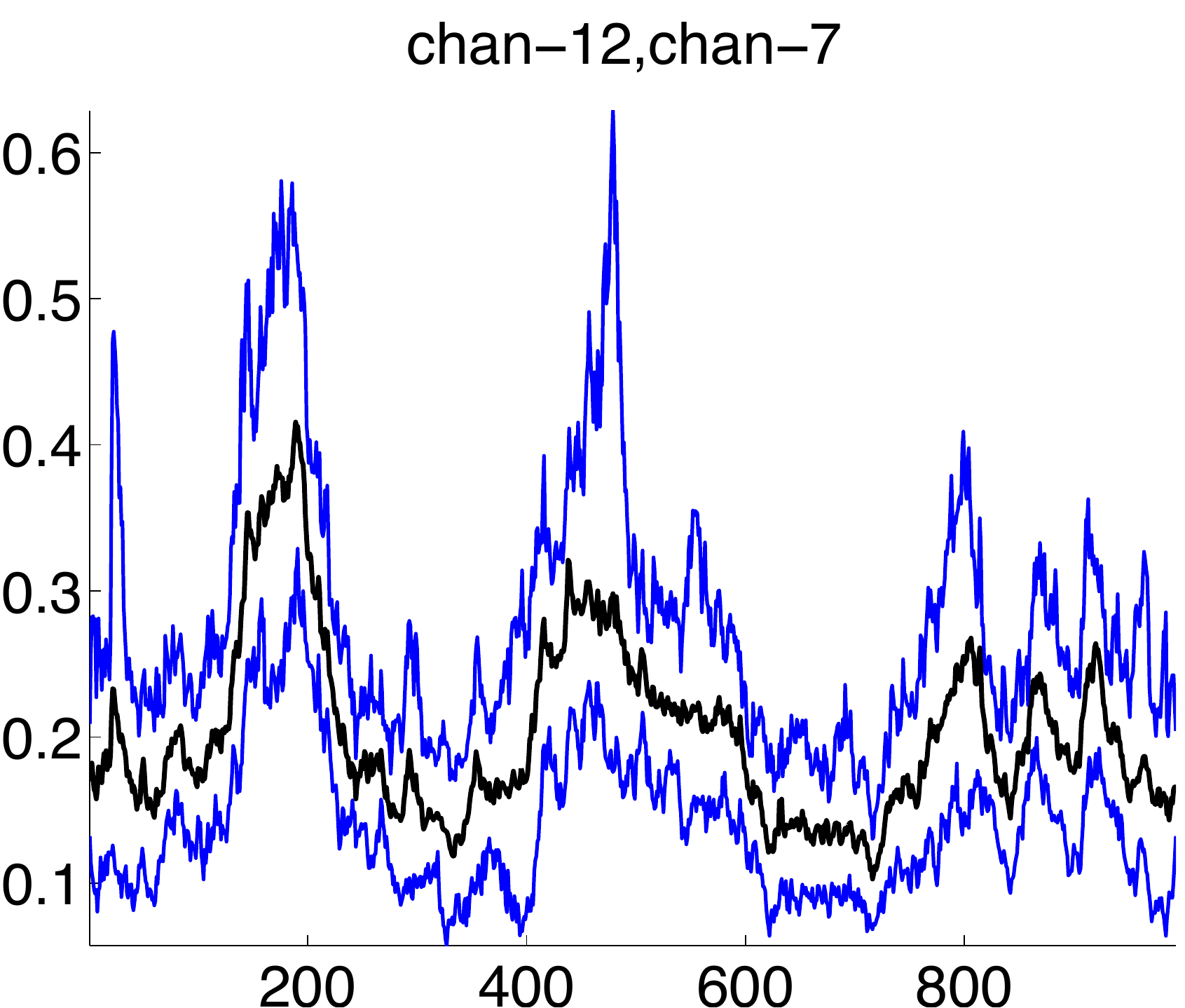} & \hspace{-0.1in}
		\includegraphics[width = 1.65in]{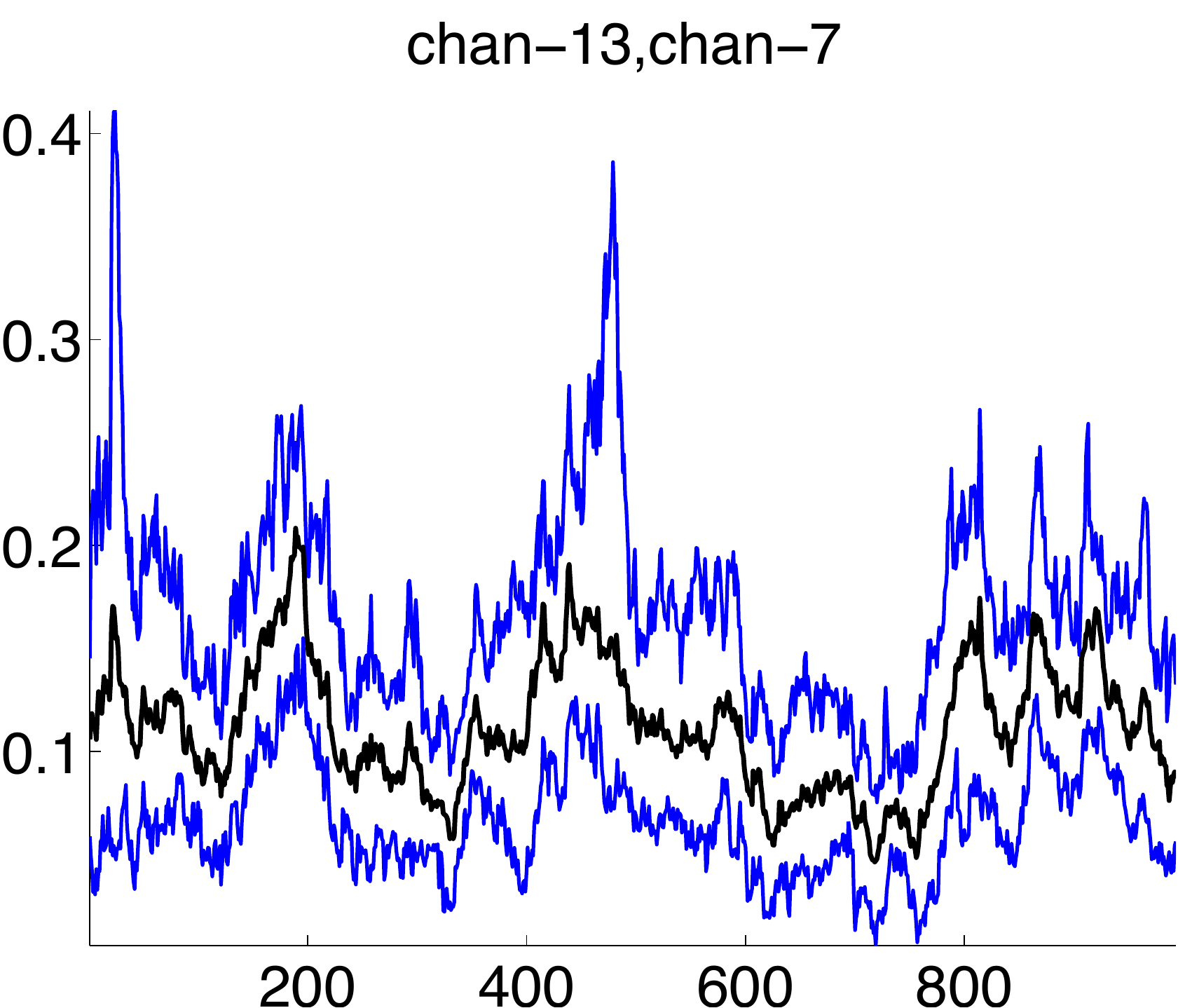} \\
		\hspace{-0.2in}
		\includegraphics[width = 1.65in]{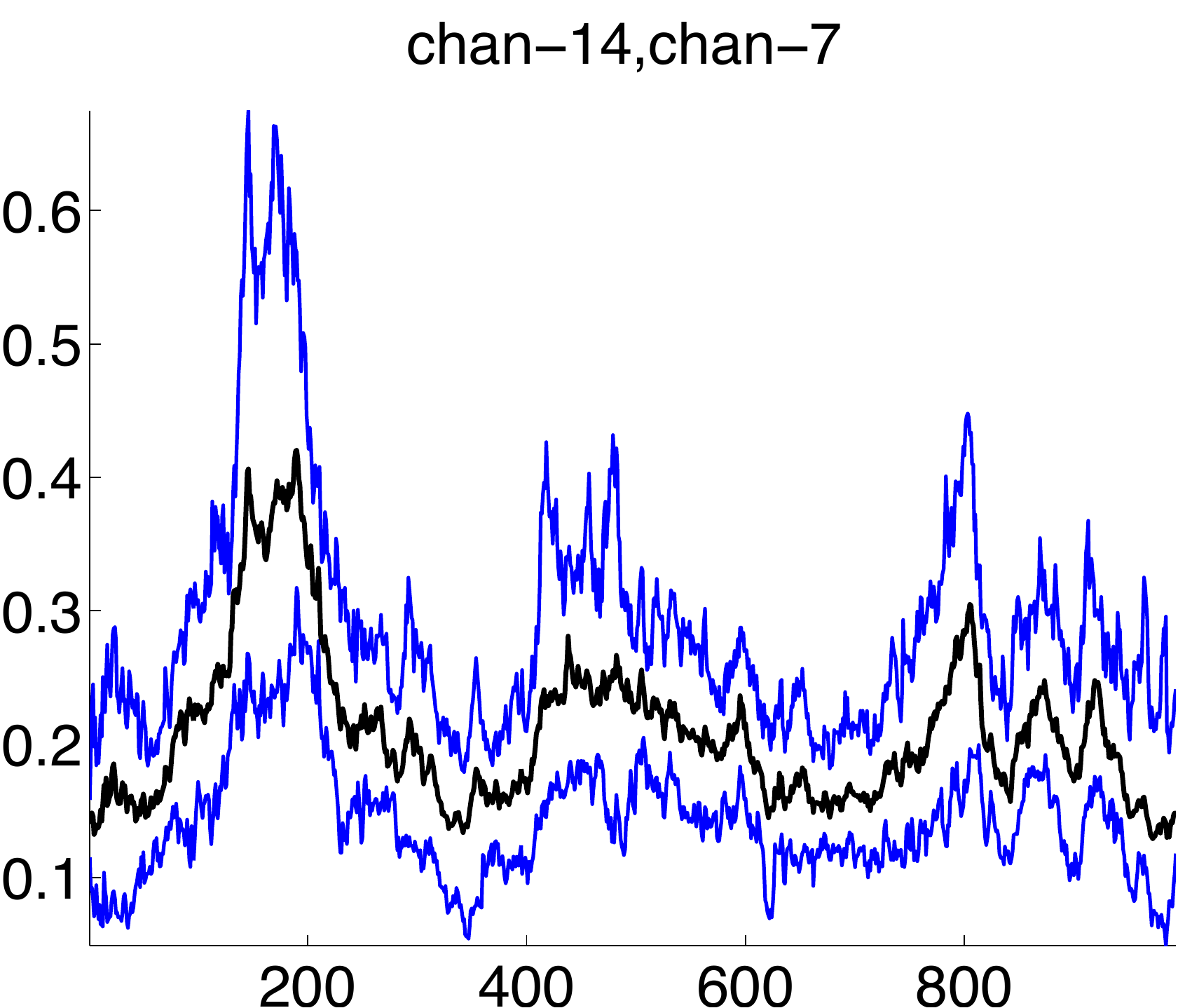} & \hspace{-0.1in}
		\includegraphics[width = 1.65in]{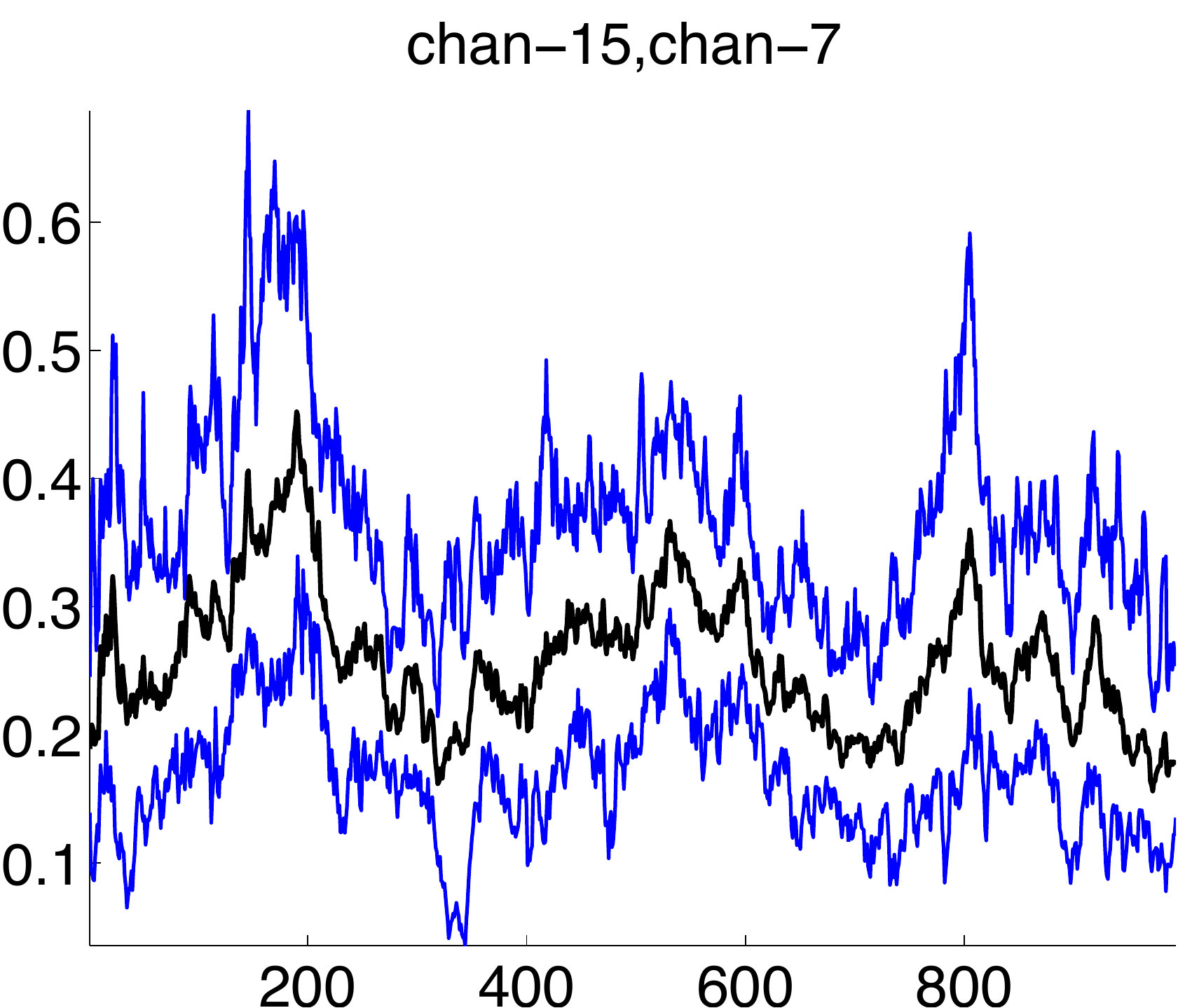} & \hspace{-0.1in}
		\includegraphics[width = 1.65in]{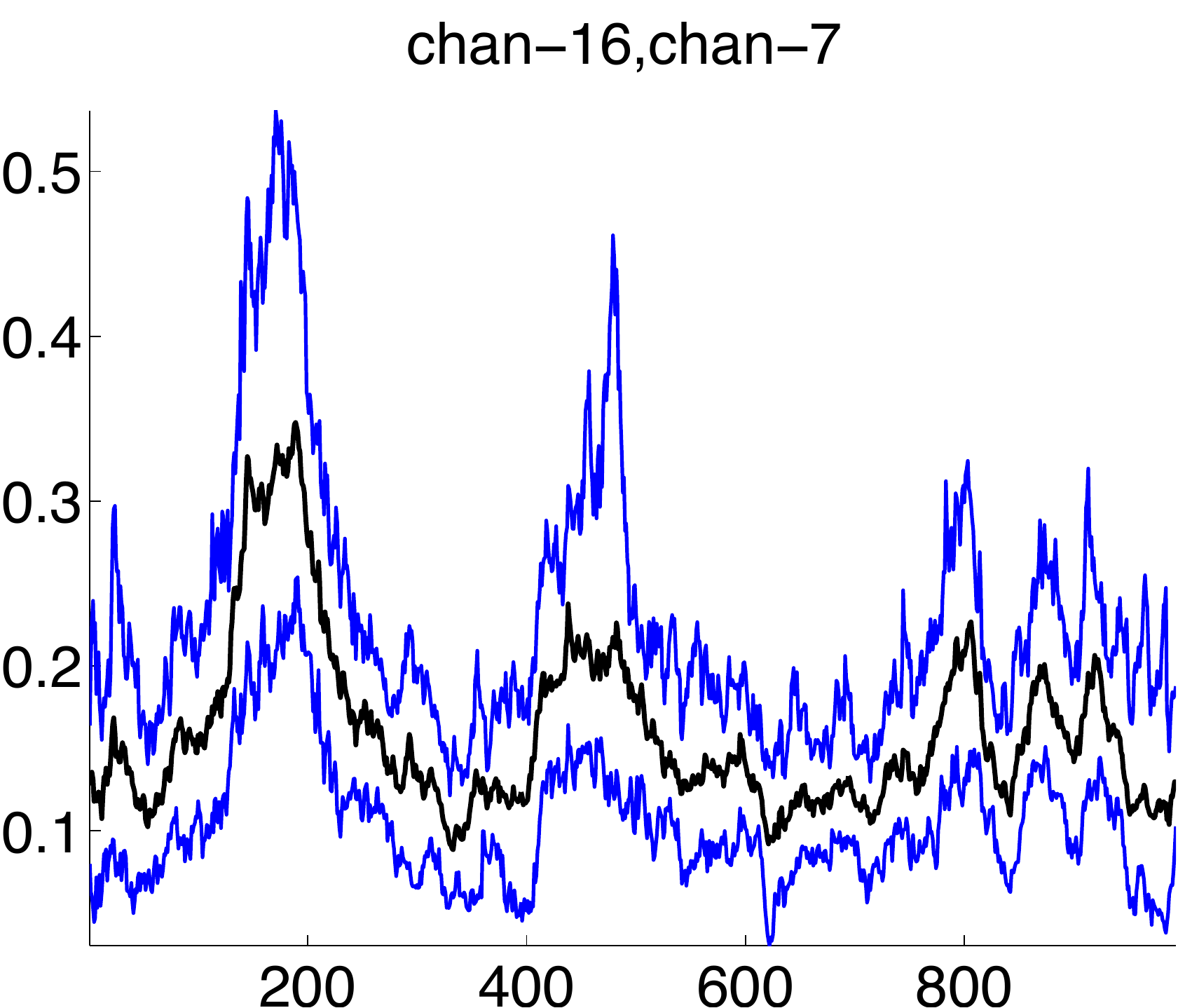}
	\end{tabular}
	\caption{Estimated trajectories of covariance terms $\Sigma_{ij,t}$ for $i\neq j$ for $j=1$ (EEG channel 7) colored as in Figure~\ref{fig:eeg_volatility}.} \label{fig:eeg_cov} \postcap \vspace{0.1in}
\end{figure}

\begin{figure}[htbp!]
	\centering
	\begin{tabular}{ccc}
		\hspace{-0.2in}
		\includegraphics[width = 1.65in]{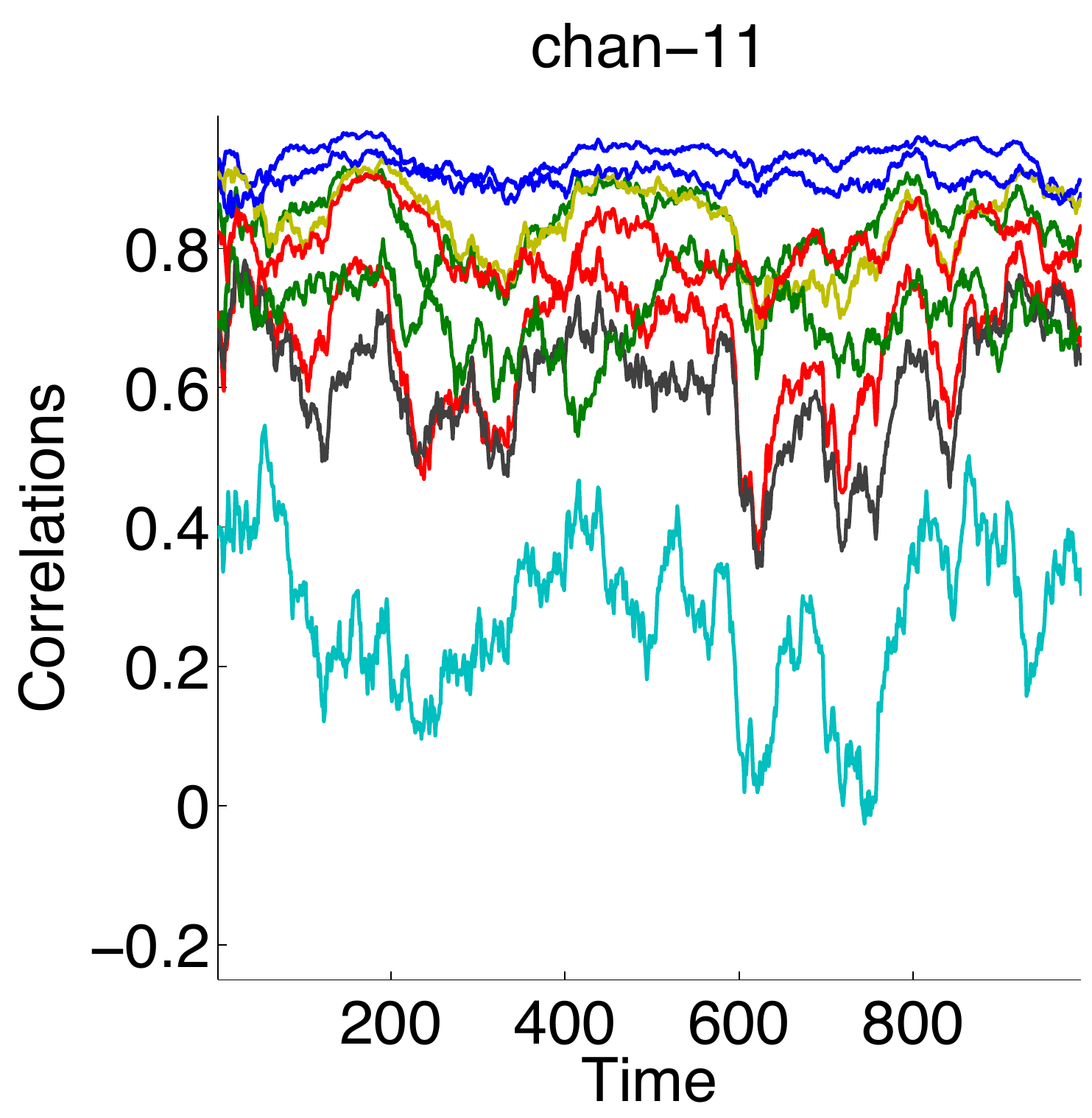} & \hspace{-0.1in}
		\includegraphics[width = 1.65in]{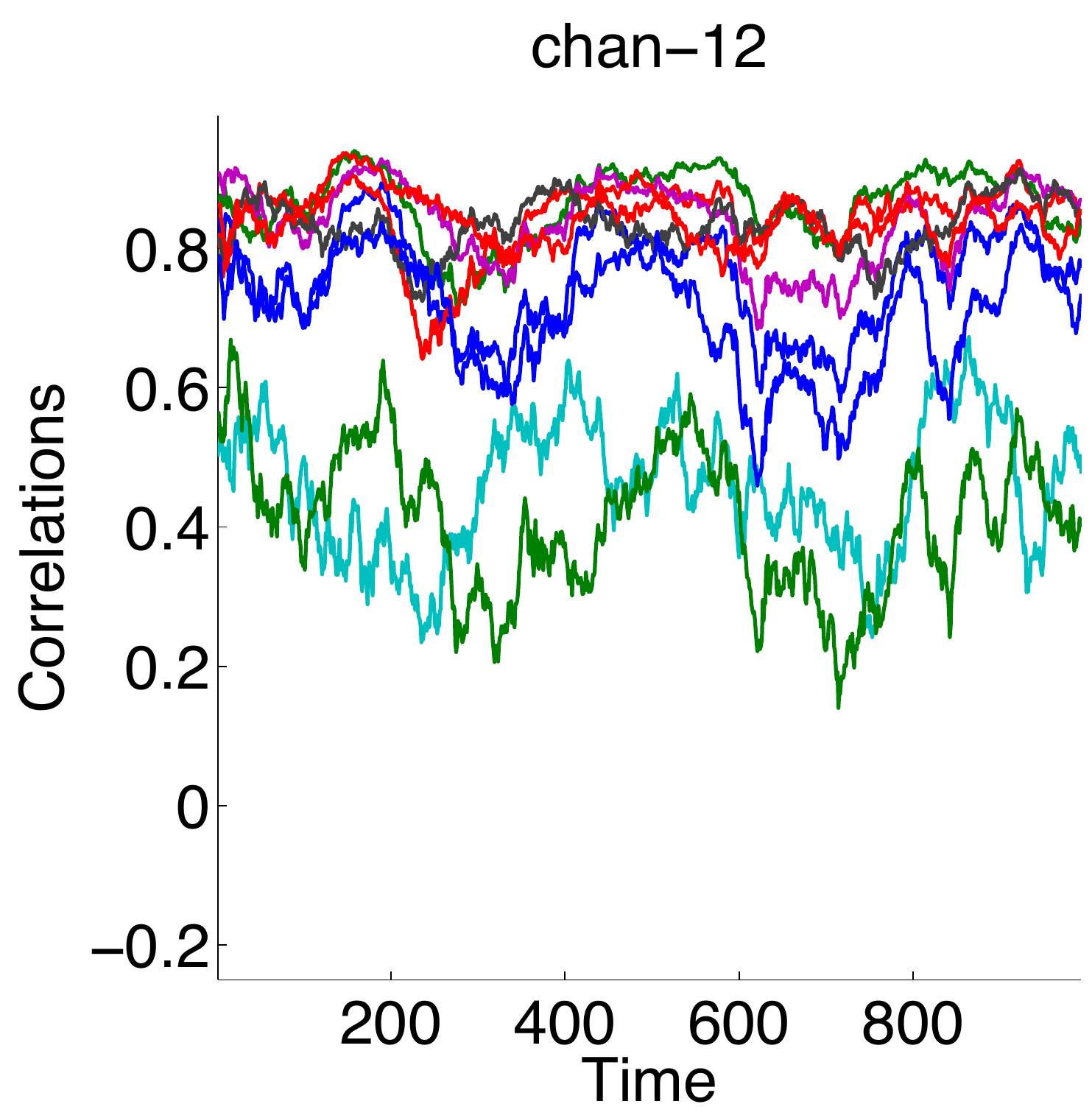} & \hspace{-0.1in}
		\includegraphics[width = 1.65in]{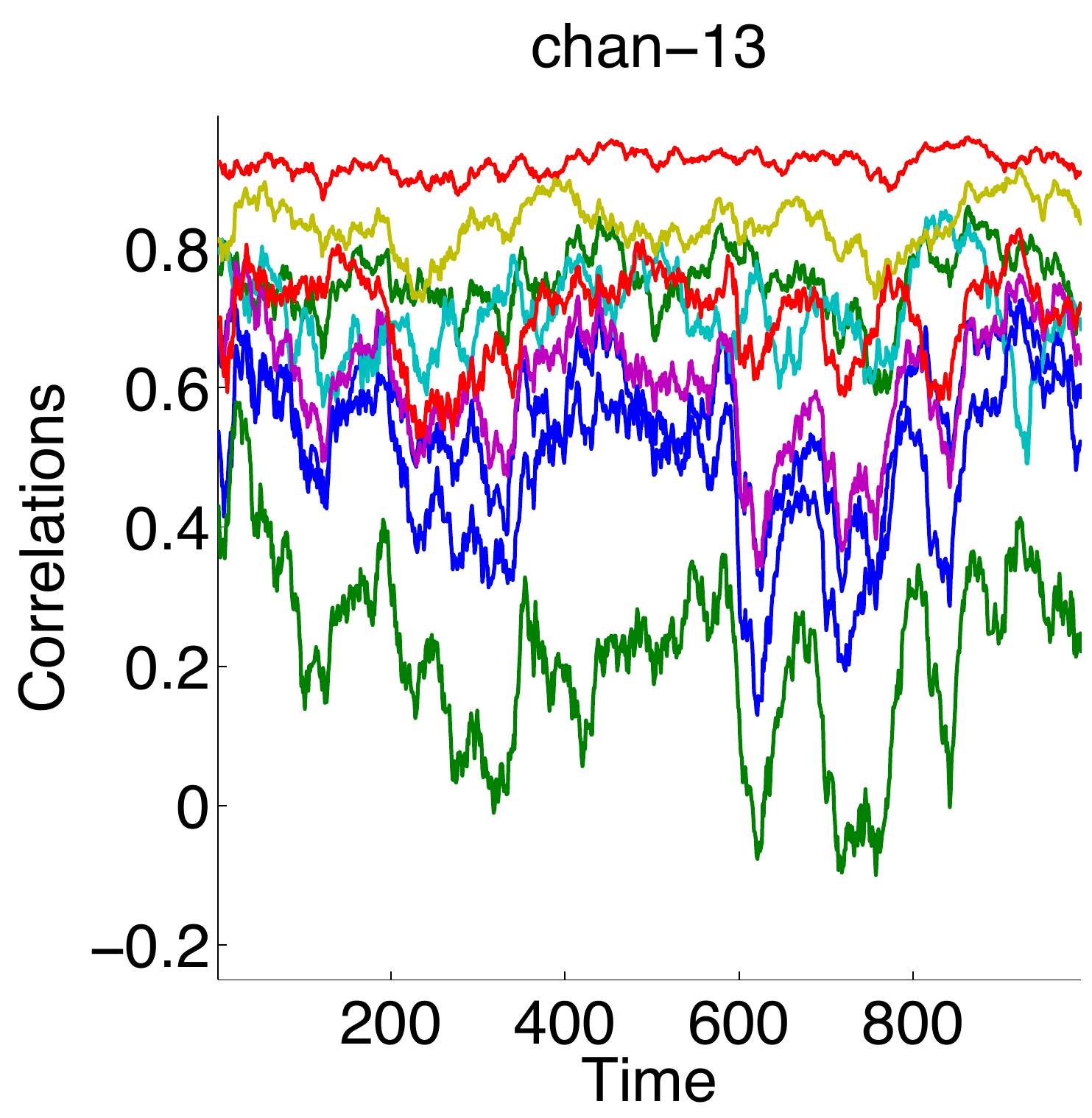} \\
		\hspace{-0.2in}
		\includegraphics[width = 1.65in]{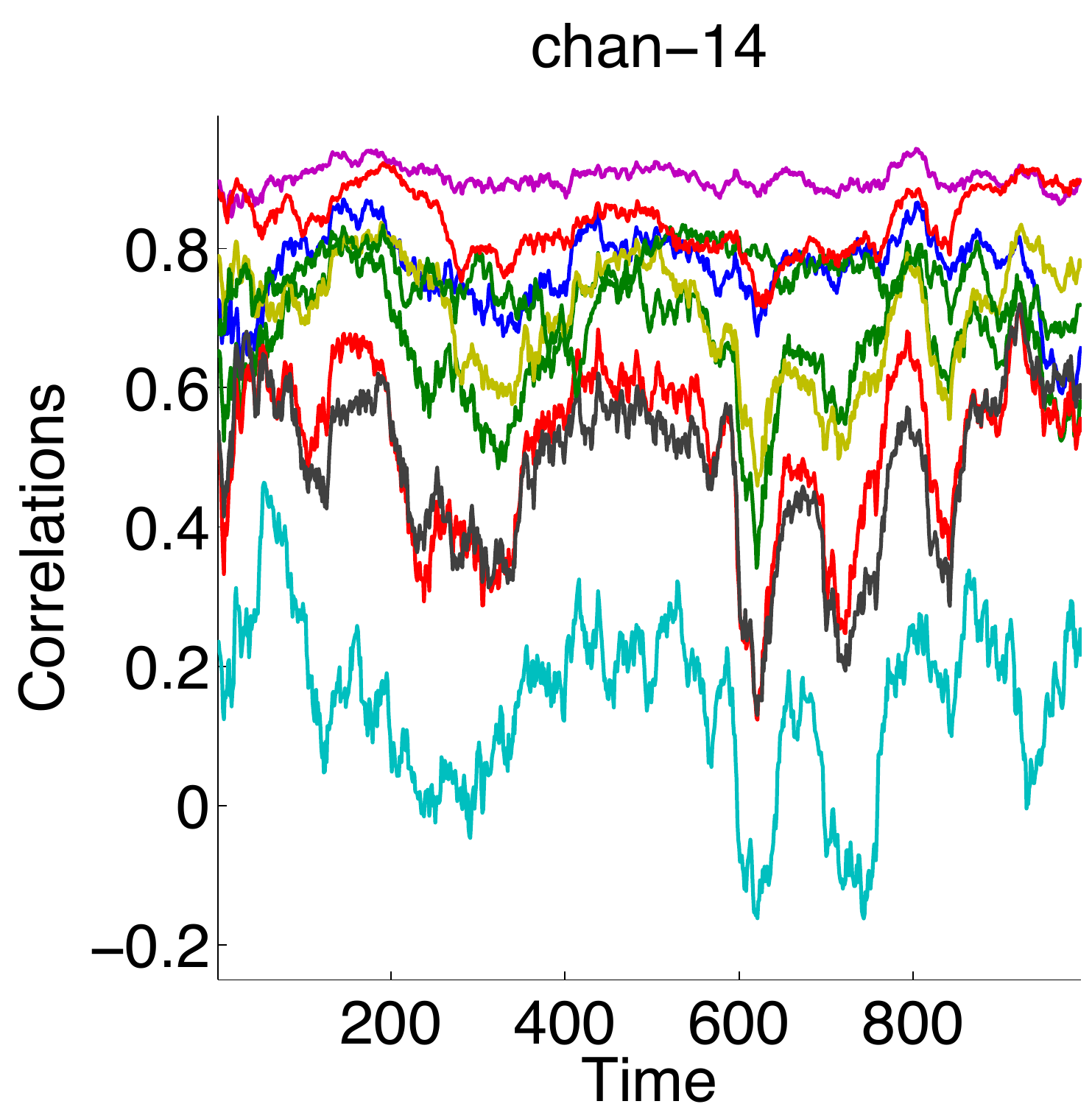} & \hspace{-0.1in}
		\includegraphics[width = 1.65in]{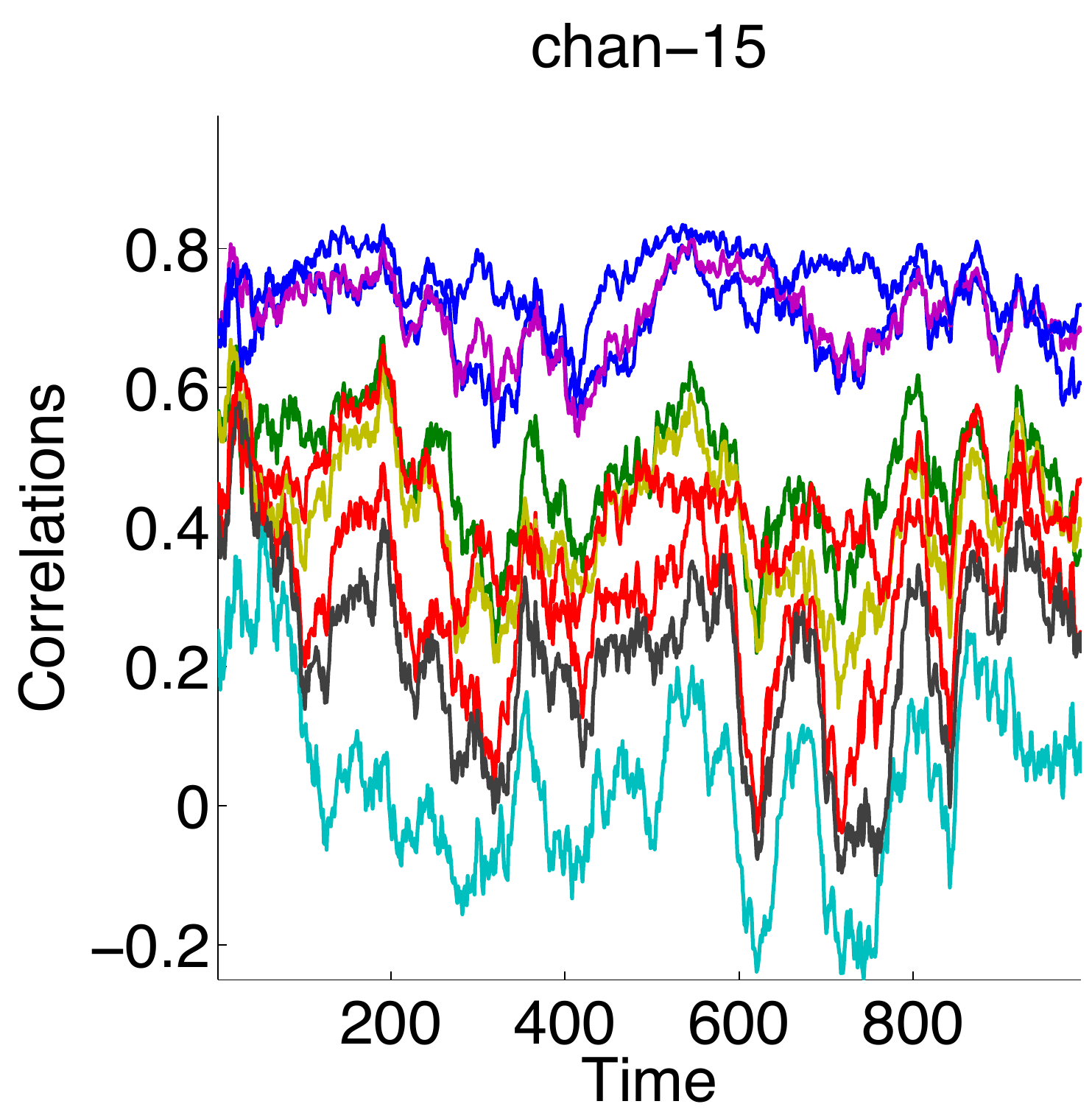} & \hspace{-0.1in}
		\includegraphics[width = 1.65in]{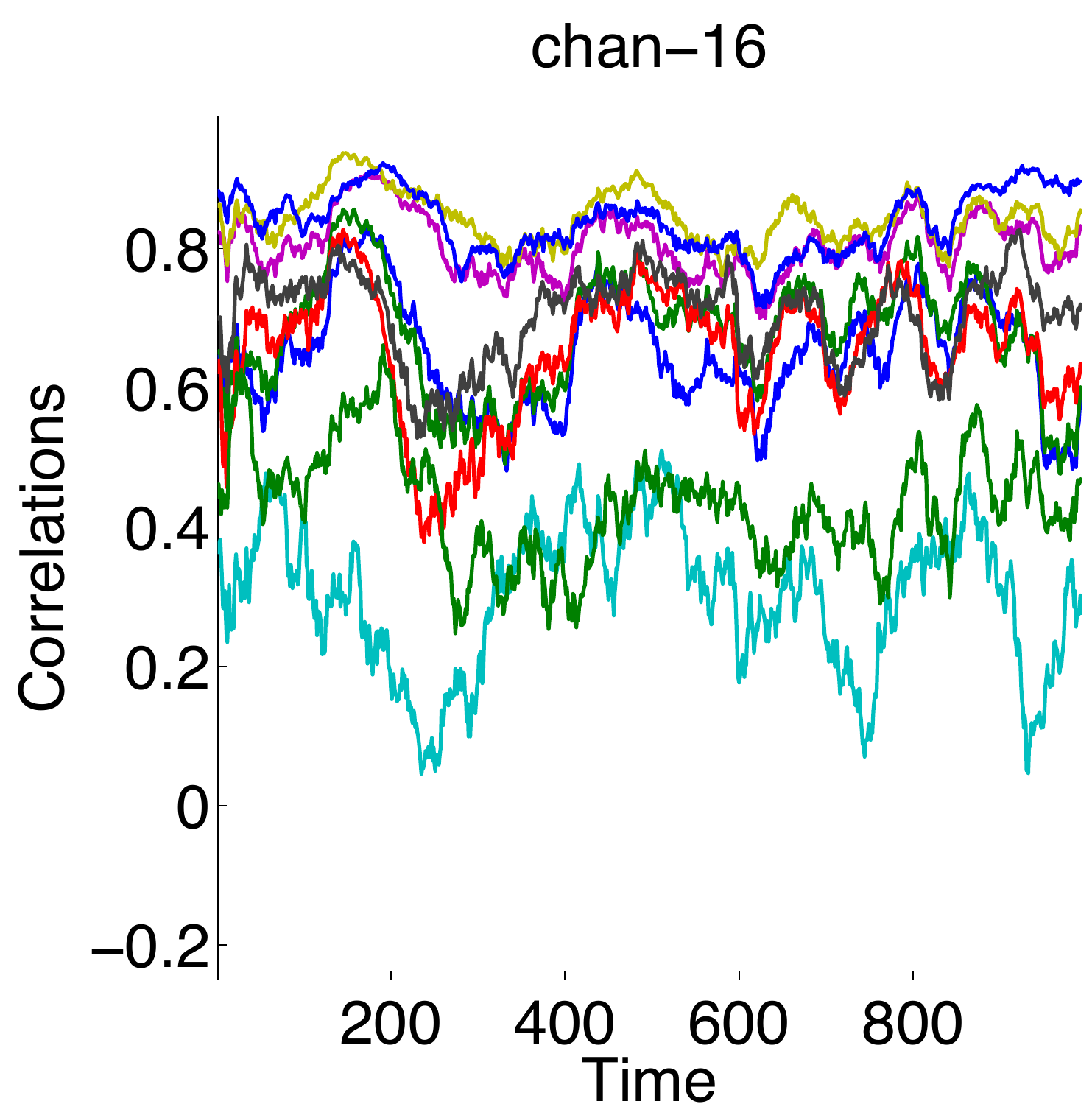}
		%\hspace{-0.2in}
		%\includegraphics[width = 1.2in]{\figdir/eeg_corr_chan-16_corr} & \hspace{-0.1in}
		%\includegraphics[width = 1.2in]{\figdir/eeg_corr_chan-17_corr} & &
	\end{tabular}
	\caption{Estimated trajectories of correlations between each of 6 channels and all other channels as a function of time.  The correlations are computed based on posterior means of $\Sigma_t$ using MCMC samples $[1000:10:5000]$ from 5 chains.} \label{fig:eeg_corr} \postcap \vspace{0.1in}
\end{figure}

\begin{figure}[htbp!]
	\centering
	\begin{tabular}{cc}
		\hspace{-0.2in}
		\includegraphics[width = 2in]{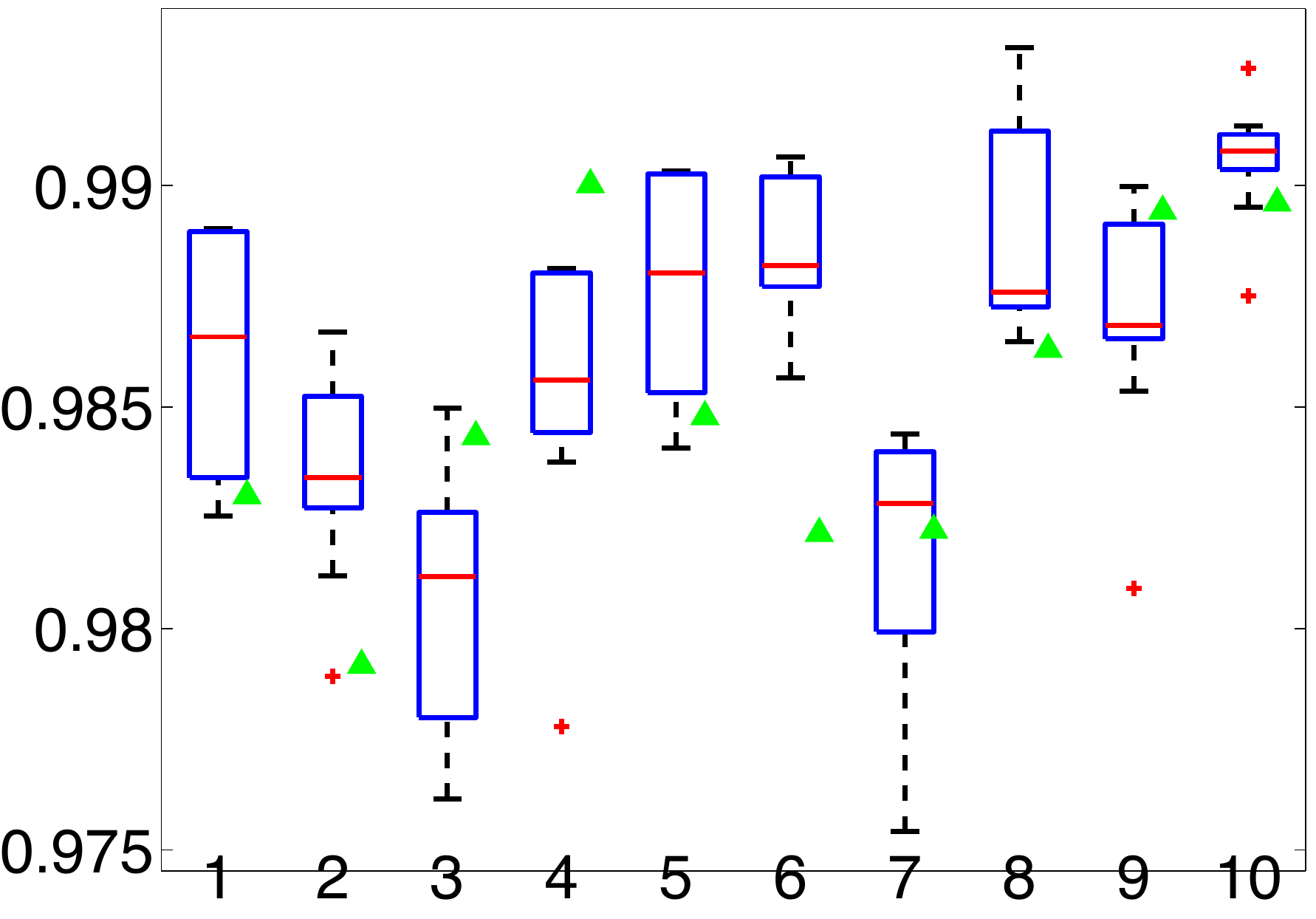} &
		\includegraphics[width = 2in]{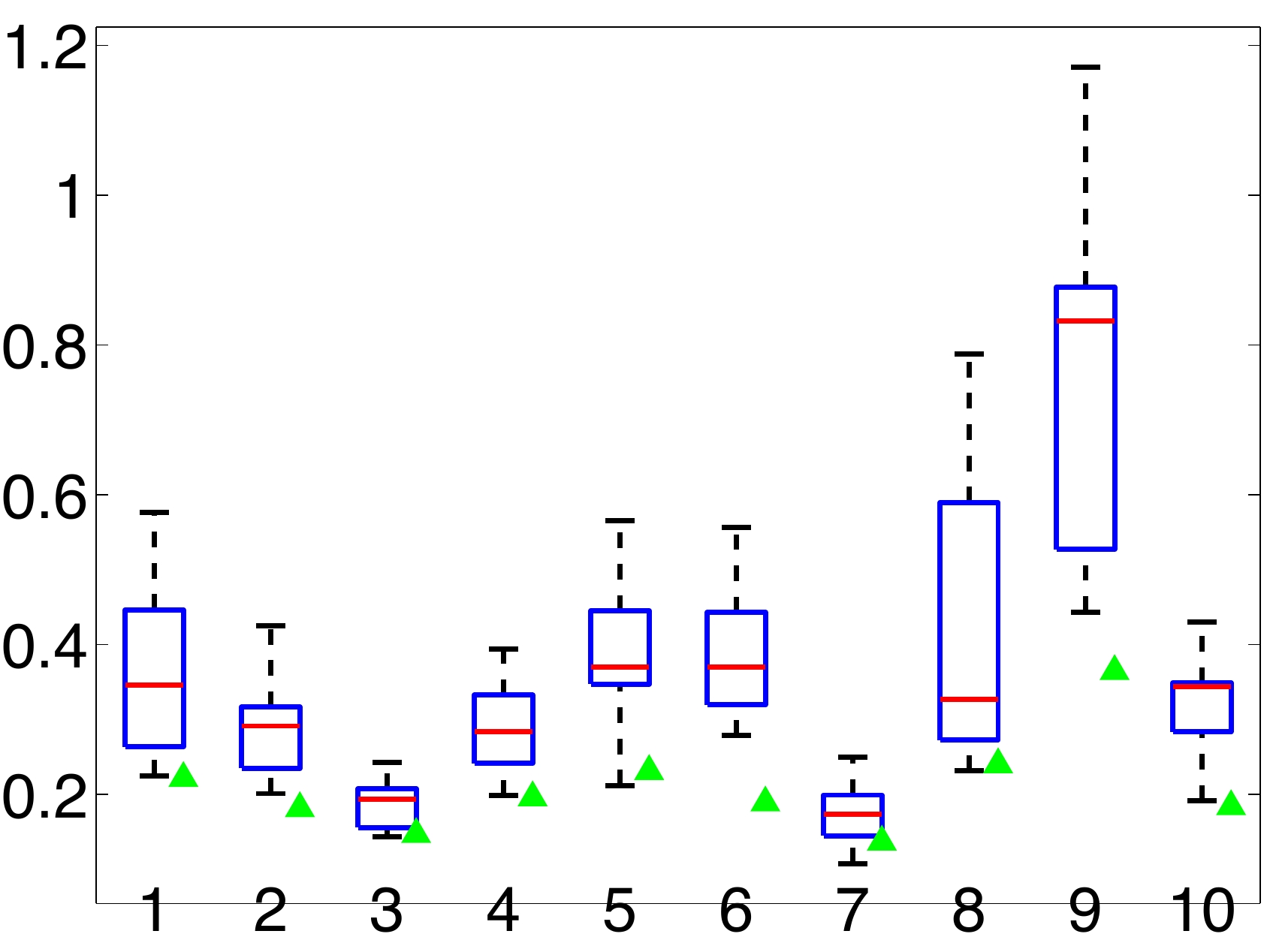}
	\end{tabular}
	\caption{Box plots of the posterior samples of $\rho_i$ and $S_{ii}$, respectively, based on the same MCMC samples as in Figure~\ref{fig:eeg_volatility}.  The prior means are marked by green triangles.} \label{fig:eeg_hypers} \postcap \vspace{0.1in}
\end{figure}

Figure~\ref{fig:eeg_volatility} displays volatility trajectories for each of the 10 examined EEG channels showing
clear changes in volatility over the 8 seconds of data, while related temporal structure in cross-series
covariances is evident in Figure~\ref{fig:eeg_cov}.  These changes are also captured in Figure~\ref{fig:eeg_corr}, which display the time-varying correlations between the EEG channel AR innovations.  For the model parameters, Figure~\ref{fig:eeg_hypers} 
shows clear evidence of learning via changes from prior to posterior summaries for the $\rho_i$ and $S_{ii}$ elements; this
figure also highlights the high dependence in the IW-AR(1) model and heterogeneity across EEG channels. 

\section{IW-AR(2) and Higher Order Models}
\label{sec:higherOrder}
The constructive approach for IW-AR(1) models extends to higher orders. This can be done in a
number of ways, as follows. 

\subsection{Direct Extension} 
For any order $p\ge 1,$
transition distributions of IW-AR($p$) processes can be defined by the conditionals of $q\times q$ diagonal blocks 
of underlying  inverse Wishart distributions for $(p+1)q\times (p+1)q$ matrices.  This involves a direct 
extension of the basic idea underlying the IW-AR(1) model construction. We develop this here for the case of $p=2.$ 

With $p=2,$ begin with 
\beq\label{eqn:IW3q}
\bpmat \De{t-1}             & \De{t-1}\Ga{t}' \\
       \Ga{t}\De{t-1}    & \Sig{t} \\
       \epmat
=
\bpmat \Sig{t-2}             & \phi_{t-1}'     & \gamma_{t}' \\
	   \phi_{t-1}     		& \Sig{t-1}              & \phi_t' \\
       \gamma_{t}                & \phi_t       & \Sig{t} \\
       \epmat
\sim  IW_{3q}(n+2, nS_3),
\eeq
where
\beq
S_3=\bpmat S & G' & H' \\
            G & S & G'\\
     		H & G & S \epmat.
\eeq
Then, for all $t$,
\begin{align}
\De{t} &\sim IW_{2q}(n+2,nS_2)\quad\textrm{with}\quad \quad S_2 = \bpmat S & G'\\ G & S \epmat, \\
\Sig{t} &\sim IW_{q}(n+2,nS).
\end{align}
This then constructively defines a stationary order 2 process with common bivariate and univariate margins.
In contrast with the IW-AR(1) construction of Section~\ref{sec:model}, $\{\phi_t,\gamma_t\}$ in eqn.~\eqref{eqn:IW3q} are not independent over time.  Rather, if $\Ga{t}=[\Ga{1,t}, \Ga{2,t}]$, then we have defined an autoregressive process on the augmented variance elements:
\begin{align}
\gamma_t &= \Ga{t1}\Sig{t-2} + \Ga{t2}\phi_{t-1}\\
\phi_t &= \Ga{t2}\Sig{t-1} + \Ga{t1}\phi_{t-1}'.
\end{align}
The \lq\lq memory" induced by these off-diagonal elements is evident as the full conditional distribution for 
$\Sig{t}$ is $p(\Sig{t}\mid \De{t-1}),$ whereas the IW-AR(2) observation model is 
$p(\Sig{t}\mid \Sig{t-1:t-2})$ which involves marginalization over the relevant conditional for the off-diagonal matrices. 

As in the case of the IW-AR(1), the construction of eqn.~\eqref{eqn:IW3q} implies that
\beq
\Sigma_t = \Omega_t + \Gamma_t\Delta_{t-1}\Gamma_t',
\eeq
with time $t$ innovation matrices $\Ga{t}$ $(q\times 2q)$ and $\Om{t}$ $(q\times q)$ independent of $\De{t-1}$ and distributed as
\begin{align}
\Omega_t &\sim IW_{q}\left(n+2+2q,n\left(S-\bbmat H & G \ebmat S_2^{-1} \bbmat H & G \ebmat'\right)\right)\\
\Gamma_t \mid \Omega_t &\sim N\left(\bbmat H & G \ebmat S_2^{-1},\Omega_t,(nS_2)^{-1}\right).
\end{align}
If we assume that $H= GS^{-1}G$ such that
\beq
S_3=\bpmat S & G' & G'S^{-1}G' \\
            G & S & G'\\
     		GS^{-1}G & G & S \epmat = \bpmat S_2 & S_2\bbmat 0 & GS^{-1} \ebmat'  \\
     	   \bbmat 0 & GS^{-1} \ebmat S_2 & S \epmat,
\eeq
then
\begin{align}
\Omega_t &\sim IW_{q}\left(n+2+2q,n\left(S-GS^{-1}G'\right)\right)\\
\Gamma_t \mid \Omega_t &\sim N\left(\bbmat 0 & GS^{-1} \ebmat,\Omega_t,(nS_2)^{-1}\right).
\end{align}
Furthermore, taking $G = FS$ leads to 
\beq
S_3=\bpmat S & SF' & SF^{2'} \\
            FS & S & SF'\\
     		F^2S & FS & S \epmat = \bpmat S_2 & S_2\bbmat 0 & F \ebmat'  \\
     	   \bbmat 0 & F \ebmat S_2 & S \epmat
\eeq
and
\begin{align}
\Omega_t &\sim IW_{q}\left(n+2+2q,n\left(S-FSF'\right)\right)\\
\Gamma_t \mid \Omega_t &\sim N\left(\bbmat 0 & F \ebmat,\Omega_t,(nS_2)^{-1}\right).
\end{align}
Note that for the specified $IW_{3q}$ to be a valid distribution, we need $S_3$ positive definite.  As before, this is equivalent to $S_2$ and the Schur complement of $S_3$ being positive definite.  Taking $G=FS$ and $H = F^2S$, the Schur complement is simply $S-FSF'$ and
\beq
|S_3| = |S_2||S-FSF'| = |S||S-FSF'|^2.
\eeq
So, just as in the IW-AR(1), we require $S$ and $V = S-FSF'$ to be positive definite; the conditions for a valid process and  stationarity have not changed in this extension to the IW-AR(2) process based on the chosen parameterization.

%The conditional mean of this IW-AR(2) process is as follows; a derivation appears in the Appendix:
%%
%\begin{multline}
%E[\Sig{t} \mid \De{t-1}] = F\Sig{t-1}F' + \frac{V}{n+2q}tr(\Sig{t-1}V^{-1})\\
% + \frac{V}{n+2q}tr(\Sig{t-2}(S^{-1}+F'V^{-1}F)) - 2\frac{V}{n+2q}tr(\phi_{t-1}V^{-1}F).
%\label{eqn:IWAR2_condMean1}
%\end{multline}
%%
%As $n \rightarrow \infty$, the process concentrates around an $AR(1)$ process with
%%
%\beq
%E[\Sig{t} \mid \De{t-1}] = S + F(\Sig{t-1}-S)F'.
%\eeq
%%
%Note that with $H$ defined more generically than $H = F^2S$, one could induce a stronger dependency on $\Sig{t-2}$ (e.g., a term analogous to $F\Sig{t-1}F'$) though with a more complicated condition for stationarity and expression of the conditional mean.

More general IW-AR$(p)$ follow from the obvious extension of this constructive approach. Note that the ancillary off-diagonal blocks of the extended $IW_{(p+1)q}$ matrix defining the IW-AR$(p)$ transition distributions are latent variables that will feature in Bayesian fitting.

\subsection{A Second Constructive Approach to Higher-Order Models}
\label{sec:IWARp}
A related, alternative and novel approach is defined by coupling AR components to generate higher order AR structures.  Specifically, take $\Sig{t} = \Ps{t} + \Up{t}\Sig{t-1}\Up{t}'$ with $p(\Up{t},\Ps{t})$ having a marginal matrix normal, inverse Wishart form as in the IW-AR$(1)$ model of Section~\ref{sec:model}.  Denote the hyperparameters of this IW-AR(1) by $\mu_1 = \{ n,F,S\}$.

Now introduce Markovian dependence into the $\Ps{t}$ sequence while maintaining the same conditional independence of $(\Up{t}|\Ps{t})$ on the history of the $\Sig{t}$ process.  Specifically, take an IW-AR$(1)$ model for $\Ps{t}$ so that for each $t$
\beq
\Ps{t} =  \Xi_t + \Ph{t}\Ps{t-1}\Ph{t}',
\eeq
with time $t$ innovations $\{\Ph{t},\Xi_t\}$ having independent matrix normal, inverse Wishart distributions with defining parameters $\mu_2 = \{ n+q,H,V\}$, where $V=S-FSF'$.

This induces a second-order Markov model
\beq
\Sig{t} = \Up{t}\Sig{t-1}\Up{t}' + \Ph{t}\Sig{t-1}\Ph{t}' - \Ph{t}\Up{t-1}\Sig{t-2}\Up{t-1}'\Ph{t}' + \Xi_t.
\eeq

Stationarity of this IW-AR(2) process is implied by simply ensuring the stationarity of the IW-AR(1) $\Sig{t}$ process and the embedded IW-AR(1) $\Psi_{t}$ process: each of $S, V$ and now 
 $W = V-HVH'$ must be positive definite.

Conditional on  $\De{t-1}$ earlier defined, we have
\begin{multline}
E[\Sig{t} \mid \De{t-1}] = F\Sig{t-1}F' + \\ \left\{H\Psi_{t-1}H' + c_{n,2q}W + c_{n,2q}W \tr(\Psi_{t-1}(nV)^{-1})\right\}\left(1 + \tr(\Sig{t-1}(nS)^{-1})\right),
\label{eqn:IWAR2_condMean2}
\end{multline}
where $c_{n,2q}=n/(n+2q)$  and 
$\Psi_{t-1} = \Sigma_{t-1} - \Upsilon_{t-1}\Sigma_{t-2}\Upsilon_{t-1}'$ is a deterministic function of the elements of $\De{t-1}$.  In the limit as $n\rightarrow \infty$,
\beq
\begin{aligned}
E[\Sig{t} \mid \De{t-1}] &= W + F\Sig{t-1}F' + H\Psi_{t-1}H' \\
						 &= S + F(\Sig{t-1} - S)F' +\nonumber\\
						&\hspace{1in} H((\Sigma_{t-1} - S) - (\Upsilon_{t-1}\Sigma_{t-2}\Upsilon_{t-1}'-FSF'))H'.
\end{aligned}
\eeq
Derivations of these conditional moments are provided in the Appendix.

The new structure of joint distributions of the innovations $\{ \Up{t},\Ps{t} \}$ is to be explored, as are extensions of 
the MCMC for model fitting.

\section{Related Models}
\label{sec:related}
\paragraph{A Markov Latent Variable Construction}
As discussed in Section~\ref{sec:q1}, our IW-AR(1) model in $q=1$ dimensions relates closely to the 
univariate model arising via a latent variable construction introduced by~\cite{Pitt:02,Pitt:05}.  We can extend the 
univariate model of that reference to the multivariate case, as follows.  The $\Sig{1:T}$ process is coupled with a latent $q\times q$ variance matrix process $\La{1:T}$ via time $t$ conditionals:  $(\Sig{t}|\La{t})\sim IW_q(n+2+a,nS+aB\La{t}B')$ with $(\La{t}|\Sig{t-1})\sim W_q(a,A\Sig{t-1}A'/a)$ a Wishart conditional for some $a>0$ and non-singular $q\times q$ matrix $A = B^{-1}$.  It can be shown that this latent variable construction defines a valid joint distribution with margin $\Sig{t}\sim IW_q(n+2,nS)$ for all $t$.  This leads to an AR(1) transition model $p(\Sig{t}|\Sig{t-1})$ in closed form and appears to be the most general AR(1) construction based on the latent variable/process idea of~\cite{Pitt:05}.

Although producing identical margins to the IW-AR(1), the proposed multivariate extension of the~\cite{Pitt:05} construction is limited.  Such models are always reversible. Most critically, the construction
implies $E(\Sig{t}|\Sig{t-1})=S+w(\Sig{t-1}-S)$, where $w=a/(n+a)$ is scalar. So, in contrast to the $F$ matrix of the IW-AR(1), there is no notion of multiple autoregressive coefficients for flexible autocorrelation structures on the elements of $\Sig{t}$.  
%In addition, $p(\Sig{t},\Sig{t-1}) = \int p(\Sig{t} \mid \La{t})p(\La{t}\mid \Sig{t-1})p(\Sig{t-1})d\La{t}$ does not have an analytic form. 
Finally, it is not clear how to extend  to higher-order autoregressive models.

\paragraph{Direct Specification of Transition Distributions}
The interesting class of models of~\cite{PhilipovGlickman:06} specifies the \emph{transition} distribution for $\Sig{t}$ given $\Sig{t-1}$ as inverse Wishart, discounting information from the previous matrix.  Specifically, the conditional mean is given by $cF\Sig{t-1}^dF'$ for some $c,d>0$ and matrix $F$.  The Markov construction generates models with stationary structure.  Scaling to higher dimensions, the authors apply the proposed stationary Wishart models to the variance matrix of a lower-dimensional latent factor in a latent factor volatility model \citep{PhilipovGlickman:06b}, extending prior approaches based on dynamic latent factor models~\citep{AguilarWest00,West:1999}.  Based on the specification of Wishart Markov transition kernels, the proposed models do not yield a clear marginal structure and extensions to higher dimensions appear challenging.  Furthermore, the proposed sampling-based model fitting strategy yields low acceptance rates in moderate dimensions (e.g., $q=12$).

Related approaches in~\cite{Gourieroux:09} define Wishart (and non-central Wishart) processes via functions of sample variance matrices of a collection of latent vector autoregressive processes.  Specifically, when the degree of freedom $n$ is integer, $\Sigma_t = \sum_{k=1}^n x_{kt}x_{kt}'$, with each $x_{k}$ independently defined via $x_{k,t} = Mx_{k,t-1} + e_{k,t}$ and $e_{k,t} \sim N(0,\Sigma_0)$.  For stationary autoregressions, one can analyze the marginal distribution of $\Sig{t}$. Extensions to higher order processes are also presented.  For model fitting, the authors rely on a (non-asymptotically efficient) method of moments assuming that a sequence of observed \emph{volatility/co-volatility matrices} are available.  Extensions to embedding the proposed Wishart autoregressive process within a standard stochastic volatility framework is computationally complicated: a mean model can be estimated based on nonlinear filtering approximations of latent volatilities.  Within Bayesian analysis of such a setup, the non-central Wishart does not yield an analytic posterior distribution and is challenging to sample.  One might be able to exploit latent process constructions, but the analysis is not straightforward.

%Neither of these approaches seems at all conducive to extensions that embed in more structure graphical models
%noted above as one of our motivating research goals. Fundamentally that will rely on probabilistic specifications
%via inverse Wisharts to link with the hyper-inverse Wisharts of graphical models.

\paragraph{AR models for Cholesky elements}
Several recent works use linear, normal AR(1) models for off-diagonal elements of the Cholesky of $\Sig{t}$ and for the log-diagonal elements~\citep{CogleySargent05,Primiceri05,Lopes2010c,Nakajima2011a}, building on the Cholesky-based heteroscedastic model of~\cite{Pourahmadi:99}, and a 
natural parallel of Bayesian factor models for multivariate volatility~\citep{AguilarWest00,West:1999}.  However, each autoregression has an interpretation as the time-varying regression parameters in a model in which the \emph{ordering} of the elements of the observation vector is required and plays a key role in model formulation.  For models in which this is not the case, the parameters employed in the autoregressions are less interpretable.  We can cast our IW-AR within a similar framework.  The inverse Wishart margins for $\Sig{t}$ and $\Sig{t-1}$ translate to Wishart margins for the precision matrices $\Sig{t}^{-1}$ and $\Sig{t-1}^{-1}$.  Since each $q\times q$ Wishart matrix can be equivalently described via an outer product of a collection of $q\times q$ identically distributed normal random variables, our IW-AR implicitly arises from a first-order Markov process on the normal random variables and thus defines a Gaussian autoregression, though possibly of a nonlinear form.  Note that there are a few key differences between the IW-AR induced element-wise autoregressions and the Cholesky component AR models: (i) the IW-AR autoregressions are on elements of the \emph{precision} matrix and (ii) these elements comprise a matrix square root, but not the Cholesky square root.  The issue of implicitly defining an ordering of observations when using a Cholesky decomposition is not present in the matrix square root considered in the IW-AR case.

\section{Final Comments} \label{sec:final}
The structure of the proposed IW-AR processes immediately open possibilities for examining alternative computational methods and extensions to parsimonious modeling of higher-dimensional time series.

The inherent state-space structure of the IW-AR also suggests opportunity to develop more effective computational methods using some variant of particle filtering and particle learning~\citep{Lopes2010a,Lopes2010b}. Among the main challenges here is that 
of including the fixed parameters -- or expanded state variables that include approximate sufficient statistics for these parameters -- in particulate representations of filtering distributions~\citep{Liu2001}.  One possible approach is to harness ideas from particle MCMC~\citep{Andrieu:10}.  Otherwise, the new IW-AR model class is inherently well-suited to the most effective {\em reweight/resample} strategies of particle learning for sequential Monte Carlo.

The inverse Wishart distribution also has extensions to \emph{hyper-inverse Wishart} (HIW) distributions for variance matrices constrained by specified graphical models~\citep{Dawid93,Carvalho:07b}.  Graphical models provide scalable structuring for higher-dimensional problems, and it would be interesting to consider extensions of the IW-AR to \emph{HIW-AR} processes that evolve maintaining the sparsity structure (of the precision matrix) specified by a graphical model.

\appendix

\section{Proofs of Theoretical Properties}
\subsection{Proof of Theorem~\ref{thm:condMean}}
Let $\upsilon_{t_k}$ denote the $k$th row of $\Upsilon_t$ and $f_k$ the $k$th column of $F$. Then, for the IW-AR(1) we can write the $(i,j)$ element of $\Sigma_t$ as 
\begin{align}
	\Sigma_{t,ij} = \Psi_{t,ij} + \upsilon_{t_i}\Sigma_{t-1}\upsilon_{t_j}'.
\end{align}
Taking the expectation conditioned on $\Sigma_{t-1}$,
\begin{align}
	E\left[\Sigma_{t,ij}\mid \Sigma_{t-1}\right] &= \frac{nV_{ij}}{n+q} + \sum_{k=1}^q\sum_{\ell=1}^q \Sigma_{t-1,k\ell}E[\upsilon_{t_{ik}}\upsilon_{t_{j\ell}}]\nonumber\\
		&= \frac{nV_{ij}}{n+q} + \sum_{k=1}^q\sum_{\ell=1}^q\Sigma_{t-1,k\ell}\left\{\left[\frac{nV_{ij}}{n+q}(nS)^{-1}\right]_{k\ell} + F_{ik}F_{j\ell}\right\}\nonumber\\
		&= \frac{nV_{ij}}{n+q} + \tr\left(\Sigma_{t-1}\frac{nV_{ij}}{n+q}(nS)^{-1}\right) + f_i \Sigma_{t-1} f_j'
\end{align}
where we have used the fact that $E[\Psi_t] = nV/(n+q)$.  In matrix form, we have
\begin{align}
	E[\Sigma_t \mid \Sigma_{t-1}] = \frac{(1+\tr(\Sigma_{t-1}(nS)^{-1}))}{n+q}nV + F\Sigma_{t-1}F'.
\end{align}
\subsection{Proof of Theorem~\ref{thm:principalIWAR}}
	For the case of $F=ERE'$ and $S=EQE'$, we can write eqn.~\eqref{eqn:SigmaJoint} as	
	\begin{align}
		\left(\begin{array}{cc}
			\Sigma_{t-1} & \Sigma_{t-1}\Upsilon_t'\\ \Upsilon_t\Sigma_{t-1} & \Sigma_t
		\end{array}\right)
		\sim \mbox{IW}_{2q}\left(n+2,n
		\left(\begin{array}{cc}
		E & 0\\ 0 & E
		\end{array}\right)\left(\begin{array}{cc}
		Q & QR\\ QR & Q
		\end{array}\right)\left(\begin{array}{cc}
		E' & 0\\ 0 & E'
		\end{array}\right)\right).
	\end{align}
	%

	%%
	%implying that
	%%
	%\begin{align}
%		\left(\begin{array}{cc}
%		E' & 0\\ 0 & E'
%		\end{array}\right)\left(\begin{array}{cc}
%			\Sigma_{t-1} & \Sigma_{t-1}\Upsilon_t'\\ \Upsilon_t\Sigma_{t-1} & \Sigma_t
%		\end{array}\right)^{-1}	\left(\begin{array}{cc}
%			E & 0\\ 0 & E
%			\end{array}\right)
%		\sim \mbox{W}_{2q}\left(n+2,\frac{1}{n}
%		\left(\begin{array}{cc}
%		Q & QR\\ QR & Q
%		\end{array}\right)^{-1}\right).
%	\end{align}
%	%
	Standard theory implies that 
	\begin{align}
		\left(\begin{array}{cc}
		E' & 0\\ 0 & E'
		\end{array}\right)\left(\begin{array}{cc}
			\Sigma_{t-1} & \Sigma_{t-1}\Upsilon_t'\\ \Upsilon_t\Sigma_{t-1} & \Sigma_t
		\end{array}\right)	\left(\begin{array}{cc}
			E & 0\\ 0 & E
			\end{array}\right)
		\sim \mbox{IW}_{2q}\left(n+2,n
		\left(\begin{array}{cc}
		Q & QR\\ QR & Q
		\end{array}\right)\right).
	\end{align}
	The derivation of the conditional mean is exactly as in the general IW-AR case,  noting that $\tr(\hat{\Sigma}_{t-1}(nQ)^{-1}) = \sum_i \hat{\Sigma}_{t-1,ii}/(n\xi_i).$

\subsection{Proof of Theorem~\ref{thm:SigmaCondMean}}
Assume that $||S||_{\infty} \leq \lambda$, $||S^{-1}||_{\infty} \leq \lambda$, $||F||_{\infty} \leq \lambda$, $||\Sigma_0||_{\infty} \leq \lambda$, and for some $t-1$
\begin{align}
	E[\Sigma_{t-1}\mid \Sigma_0] = F^{t-1}\Sigma_0 F^{t-1'} + \frac{n}{n+q}\left(S- F^{t-1} S F^{t-1'}\right) + O\left(\frac{1\cdot 1'}{n}\right).
	\label{eqn:baseCase}
\end{align}
To prove that eqn.~\eqref{eqn:baseCase} holds for general $t$, we apply iterated expectations to the conditional expectation of eqn.~\eqref{eqn:SigmaCondMean}:
\begin{align}
	E[\Sigma_t \mid \Sigma_0] &= FE[\Sigma_{t-1}\mid \Sigma_0]F' + \frac{n}{n+q}V + \frac{V}{n+q} \tr\left(E[\Sigma_{t-1}\mid \Sigma_0] S^{-1}\right)\\
%		&= F^{t}\Sigma_0 F^{'t} + \frac{n}{n+q}(FSF'-F^tSF^{'t}) + \frac{n}{n+q}V + \frac{V}{n+q} \tr\left(E[\Sigma_{t-1}\mid \Sigma_0] S^{-1}\right)\\
		&= F^{t}\Sigma_0 F^{'t} + \frac{n}{n+q}\left(S - F^{t} S F^{'t}\right) + \frac{V}{n+q} \tr\left(E[\Sigma_{t-1}\mid \Sigma_0] S^{-1}\right),
\end{align}
where we have used the definition $V = S-FSF'$.  Since
\begin{multline}
	\frac{V}{n+q} tr\left(E[\Sigma_{t-1}\mid \Sigma_0] S^{-1}\right)\\
	\leq \frac{V}{n+q} \tr\left\{\left(\lambda^{4}1\cdot 1' + \frac{n}{n+q}(1+\lambda^{4})1\cdot 1' + \lambda O\left(\frac{1\cdot 1'}{n}\right)\right) 1\cdot 1'\right\} = O\left(\frac{1\cdot 1'}{n}\right),
\end{multline}
we conclude that, indeed,
\begin{align}
	E[\Sigma_{t}\mid \Sigma_0] = F^{t}\Sigma_0 F^{'t} + \frac{n}{n+q}\left(S- F^{t} S F^{'t}\right) + O\left(\frac{1\cdot 1'}{n}\right).
\end{align}
Then also 
\begin{align}
	\lim_{n\rightarrow \infty} E[\Sigma_{t}\mid \Sigma_0] = S + F^{t}(\Sigma_0-S) F^{'t}.
\end{align}
\subsection{Proof of Theorem~\ref{thm:eigCondMean}}
Let $\Lambda = Q(I-R^2)$ such that $V = E\Lambda E'$. According to eqn.~\eqref{eqn:SigmaCondMean}, $E[\Sigma_1 \mid \Sigma_0]$ then shares the same eigenspace as $\Sigma_0$ (i.e., the eigenvectors are given by $E$), and by induction, so does $E[\Sigma_t \mid \Sigma_0]$ for all $t$.  Let $\Theta_{t|0}$ denote the diagonal matrix of eigenvalues of $E[\Sigma_t \mid \Sigma_0]$.  Since $\tr(\Sigma S^{-1}) = \tr(\Theta Q^{-1})$, eqn.~\eqref{eqn:SigmaCondMean} can be rewritten solely in terms of the eigenvalues:
\begin{align}
	\Theta_{t|0} = \frac{\Lambda}{n+q}\tr\left(\Theta_{t-1|0}Q^{-1}\right) + \frac{n}{n+q}\Lambda + R\Theta_{t-1|0}R'.
\end{align}
In terms of the vectors of eigenvalues $\theta_{t|0} = \mbox{diag}\left(\Theta_{t|0}\right)$, $\xi = \mbox{diag}(Q)$ and $\xi_{-1} = \mbox{diag}\left(Q^{-1}\right)$ we have
\begin{align}
	\theta_{t|0} %&= \frac{\mbox{diag}(Q)}{n+q}\mbox{diag}\left(\Theta_{t-1}\right)'\mbox{diag}\left(\Lambda^{-1}\right) + \frac{n}{n+q}\mbox{diag}(Q) + R^2\mbox{diag}\left(\Theta_{t-1}\right)\\
	&= \frac{n}{n+q}(I-R^2)\xi + \left[\frac{1}{n+q}(I-R^2)\xi\xi_{-1}' + R^2\right]\theta_{t-1|0}.
\end{align}
Letting $\alpha = \frac{n}{n+q}(I-R^2)\xi$ and $B = \frac{1}{n+q}(I-R^2)\xi\xi_{-1}' + R^2$, we conclude that
\begin{align}
	\theta_{t|0} &= B^t\theta_0 + \sum_{\tau =0}^{t-1}B^\tau \alpha.
\end{align}
Since $B$ represents a matrix (strictly) convex combination of $\frac{\xi\xi_{-1}'}{n+q}$ and $I$, the maximum eigenvalue of $B$ is bounded by
\begin{align}
	\left|\left|(I-R^2)\max\left\{\mbox{eig}\left(\frac{\xi\xi_{-1}'}{n+q}\right)\right\}\cdot 1 + R^2 \cdot 1 \right|\right|_0.
\end{align}
Here, $\max\{\mbox{eig}(A)\}$ denotes the maximum eigenvalue of $A$. The term $\xi\xi_{-1}'$ is a rank 1 matrix implying that the only non-zero eigenvalue is equal to $\tr(\xi\xi_{-1}') = q$.  Thus, regardless of $n$, $B$ has eigenvalues with modulus strictly less than 1 since $\frac{\xi\xi_{-1}'}{n+q}$ has $q-1$ eigenvalues equal to 0 and one equal to $\frac{q}{n+q} < 1$.  This implies that the conditional mean of the process forgets the initial condition $\Sigma_0$ exponentially fast regardless of $n$.  Furthermore, since the eigenvalues of $B$ have modulus less than 1,
\begin{align}
	\theta_{t|0} &= B^t\theta_0 + (I-B)^{-1}(I-B^t)\alpha,
\end{align}
implying that, as expected, the eigenvalues of the limiting conditional mean are exactly those of the marginal mean $S$:
\begin{align}
	\lim_{t\rightarrow \infty} \theta_{t|0} &= (I-B)^{-1}\alpha\\
	&= \left[\left(I-\frac{\xi\xi_{-1}'}{n+q}\right)^{-1}(I-R^2)^{-1}\right]\frac{n}{n+q}(I-R^2)\xi\\
	&= \frac{n}{n+q}\left(I-\frac{\xi\xi_{-1}'}{n+q}\right)^{-1}\xi\\
	&= \xi.
\end{align}
The last equality follows from matrix inversion and the fact that $\xi_{-1}'\xi = q$.

\section{Derivation of Forward Filtering Backward Sampling Algorithm}
\subsection{Approximate Forward Filtering}
The inverse Wishart prior on $\Delta_1$ can be analytically updated to an inverse Wishart posterior conditioned on $y_1$:
\begin{align}
p(\Delta_1 \mid y_1) = \mbox{IW}_{2q}\left(n+3,n
\left(\begin{array}{cc}
S & SF'\\ FS & S
\end{array}\right) + y_1y_1'\right).
\label{eqn:up1}
\end{align}
To propagate to $t=2$, we use the Chapman-Kolmogorov equation, integrating over $\Delta_1$:
\begin{align}
p(\Delta_2 \mid y_1) &\propto \int p(\Delta_2 \mid \Delta_1)p(\Delta_1 \mid y_1) d\Delta_1\nonumber\\
&\propto \int \delta_{\Sigma_1 = \Psi_1 + \Upsilon_1\Sigma_0\Upsilon_1'} p(\Psi_2)p(\Upsilon_2 \mid \Psi_2)p(\Delta_1\mid y_1) d\Upsilon_1 d\Psi_1 d\Sigma_0\nonumber\\
&\propto \mbox{IW}_q(\Psi_2\mid  n+q+2,n\tilde{V})N(\Upsilon_2\mid \tilde{F},\tilde{\Psi}_2,(nS)^{-1})\nonumber\\
&\hspace{2in}\mbox{IW}_q(\Sigma_1\mid n+3,nS+x_1x_1').
\label{eqn:prop1}
\end{align}
Here, we have used the fact that the transition kernel $p(\Delta_2\mid \Delta_1)$ simply involves independent innovations $\{\Upsilon_2,\Psi_2\}$ and deterministically computing $\Sigma_1$.  Integrating the elements used to compute $\Sigma_1$ (a component of $\Delta_2$), the marginal posterior can be derived from the joint posterior of the augmented variance matrix at time $t$ given in eqn.~\eqref{eqn:up1}. Although an independent normal-inverse Wishart set of random variables can be combined with a $q$-dimensional inverse Wishart matrix to form a $2q$-dimensional inverse Wishart, as discussed in Section~\ref{sec:model}, there are restrictions on the parameterizations of these respective distributions.  The set of distributions specified in eqn.~\eqref{eqn:prop1} do not satisfy these constraints, and thus do not combine to form a $2q$-dimensional inverse Wishart distribution on $\Delta_2$.  Namely, the $\Psi_2$ prior and $\Sigma_1$ posterior degrees of freedom do not match, nor do the $\Sigma_1$ posterior scale matrix and the $\Upsilon_2$ prior variance term $(nS)^{-1}$.

Since  exact, analytic forward filtering is not possible, we instead approximate the propagate step with a moment-matched inverse Wishart distribution.  That is,
\begin{align}
p(\Delta_2 \mid y_1) &\approx \mbox{IW}(n+2, nE[\Delta_2 \mid y_1])\nonumber\\
&\triangleq g_{2|1}(\Delta_2 \mid y_1).
\end{align}
Based on the approximations made in propagating, the subsequent update step is exact due to the conjugacy of the Gaussian observation and inverse Wishart predictive distribution.

In general, we can choose an arbitrary degree of freedom parameter in our approximate forward filtering.  Assume that at time $t$ we use $r_t$ degrees of freedom for the moment-matched approximation $g_{t|t-1}(\Delta_t \mid y_{1:t-1})$ to the predictive distribution $p(\Delta_t \mid y_{1:t-1})$.  We use $g_{t|t}(\Delta_t \mid y_{1:t})$ to denote the resulting approximation to the updated posterior $p(\Delta_t \mid y_{1:t})$.

%Also, let use define
%%
%\begin{align}
%	U_t = E_{g_{t-1|t-1}}\left[ \left.\left(\begin{array}{cc}
%		\Sigma_{t-1} & \Sigma_{t-1}\Upsilon'_t\\
%		 \Upsilon_t\Sigma_{t-1} & \Sigma_t
%	\end{array}\right) \right| y_{1:t-1}  \right],
%\end{align}
%%
%where $g_{t|t}(\Delta_t \mid y_{1:t})$ is the approximation to the updated posterior $p(\Delta_t \mid y_{1:t})$.

We initialize at $t=1$ with $r_1=n+2$ and
\begin{equation}
\begin{aligned}
   g_{1|0}(\Delta_1) &= p(\Delta_1) = \mbox{IW}(r_1,(r_1-2)E[\Delta_1]), \hspace{0.2in}	E[\Delta_1] = \left(\begin{array}{cc}
		S & SF'\\
		 FS & S
	\end{array}\right),\\
	g_{1|1}(\Delta_1\mid y_1) &= p(\Delta_1 \mid y_1) =  \mbox{IW}(r_1+1,(r_1-2)E[\Delta_1]+y_1y_1').
\end{aligned}
\end{equation}
Propagating forward,
\begin{equation}
\begin{aligned}
   	g_{2|1}(\Delta_2\mid y_1) &= \mbox{IW}(r_2,(r_2-2)E_{g_{1|1}}[\Delta_2\mid y_1]),\\
	g_{2|2}(\Delta_2\mid y_{1:2}) &= \mbox{IW}(r_2+1,(r_2-2)E_{g_{1|1}}[\Delta_2\mid y_1] + y_2y_2').
\end{aligned}
\end{equation}
Here, the predictive mean is derived as
\begin{align}
	E_{g_{1|1}}[\Delta_2\mid y_1] = \left(\begin{array}{cc}
			S_1 & S_1F'\\ FS_1 & FS_1F' + \frac{nV}{n+q}(1+\tr(S_1(nS)^{-1}))
		\end{array}\right),
\end{align}
with $S_1 = E_{g_{1|1}}[\Sigma_1 \mid y_1] = \frac{(r_1-2)S + x_1x_1'}{r_1-1}$.  The term $E_{g_{1|1}}[\Sigma_2\mid y_1]$ is derived via iterated expectations, namely $E_{g_{1|1}}[E[\Sigma_2\mid \Sigma_1,y_1]]$, and employing eqn.~\eqref{eqn:SigmaCondMean} noting that $\Sigma_2$ is conditionally independent of $y_1$ given $\Sigma_1$.

The forward filter then recursively defines
\begin{equation}
\begin{aligned}
	g_{t|t-1}(\Delta_t \mid y_{1:t-1}) &= \mbox{IW}\left(r_t,(r_t-2)E_{g_{t-1|t-1}}[\Delta_t\mid y_{1:t-1}] \right),\\
	g_{t|t}(\Delta_t \mid y_{1:t}) &= \mbox{IW}\left(r_t+1,(r_t-2)E_{g_{t-1|t-1}}[\Delta_t\mid y_{1:t-1}] + y_ty_t'\right),
\end{aligned}
\end{equation}
with
\begin{multline}
	E_{g_{t-1|t-1}}[\Delta_t\mid y_{1:t-1}]\\ = \left(\begin{array}{cc}
			S_{t-1} & S_{t-1}F'\\ FS_{t-1} & FS_{t-1}F' + \frac{nV}{n+q}(1+\tr(S_{t-1}(nS)^{-1}))
		\end{array}\right)
\end{multline}
and
\begin{align}
	S_t &= \frac{(r_t-2)\left(FS_{t-1}F' + \frac{nV}{n+q}(1+\tr(S_{t-1}(nS)^{-1}))\right) + x_tx_t'}{r_t-1}.
\end{align}
\subsection{Backward Sampling}
The density required for backward sampling is the posterior of $\{\Sigma_{t-1},\Upsilon_t,\Psi_t\}$ conditioned on $\Sigma_t$ and $y_{1:T}$, which can be written as
\begin{align}
	p(\Sigma_{t-1},\Upsilon_t,\Psi_t \mid \Sigma_t,y_{1:T}) &=p(\Sigma_{t-1},\Upsilon_t,\Psi_t \mid \Sigma_t,y_{1:t})\\
	 	&\propto p(\Sigma_{t-1},\Upsilon_t,\Psi_t \mid y_{1:t})\delta_{\Sigma_t = \Psi_t + \Upsilon_t \Sigma_{t-1}\Upsilon_t'} \\
		&\propto g_{t|t}(\Delta_t \mid y_{1:t})\delta_{\Sigma_t = \Psi_t + \Upsilon_t \Sigma_{t-1}\Upsilon_t'}.
\end{align}
Thus, sampling $\{\Sigma_{t-1},\Upsilon_t,\Psi_t\}$ from this conditional posterior is equivalent to fixing $\Sigma_t$ in the $\Delta_t$ matrix and sampling a valid $\{\Sigma_{t-1},\Upsilon_t,\Psi_t\}$ conditioned on $\Sigma_t$ based on the forward filtering distribution $g_{t|t}(\Delta_t \mid y_{1:t})$.  By \emph{valid}, we mean a value that corresponds to $\Sigma_t = \Psi_t + \Upsilon_t \Sigma_{t-1}\Upsilon_t'$.

Based on eqn.~\eqref{eqn:updatedProposal},
%%
%\begin{align}
%\{\Sigma_{t-1},\Upsilon_t,\Psi_t\} &\sim p(\Sigma_{t-1},\Upsilon_t,\Psi_t \mid \Sigma_t,y_{1:t})\nonumber\\
%&= \delta_{\Sigma_t = \Psi_t + \Upsilon_t\Sigma_{t-1}\Upsilon_t'} p(\Sigma_{t-1},\Upsilon_t,\Psi_t \mid y_{1:t}) \nonumber\\
%&\approx \delta_{\Sigma_t = \Psi_t + \Upsilon_t\Sigma_{t-1}\Upsilon_t'}g_{t|t}(\Delta_t|y_{1:t})
%\end{align}
%%
\begin{align}
g_{t|t}\left(\left(\left.\begin{array}{cc}
		\Sigma_{t-1} & \Sigma_{t-1}\Upsilon_t'\\ \Upsilon_t\Sigma_{t-1} & \Sigma_t
	\end{array}\right)\right|y_{1:t}\right) =
	\mbox{IW}\left(r_t+1,
	\left(\begin{array}{cc}
	G_t^{11} & G_t^{21'}\\ G_t^{21} & G_t^{22}
	\end{array}\right)\right),
	\label{eqn:gt1}
\end{align}
implying that
\begin{align}
	g_{t|t}\left(\left(\left.\begin{array}{cc}
			\Sigma_t  & \Upsilon_t\Sigma_{t-1}\\ \Sigma_{t-1}\Upsilon_t' & \Sigma_{t-1}
		\end{array}\right)\right|y_{1:t}\right) =
		\mbox{IW}\left(r_t+1,
		\left(\begin{array}{cc}
		G_t^{22} & G_t^{21}\\ G_t^{21'} & G_t^{11}
		\end{array}\right)\right).
		\label{eqn:gt2}
\end{align}		
Here, $G_t$ is the forward filtering term defined in eqn.~\eqref{eqn:Gt}, with $G_t^{11}$, $G_t^{21}$, $G_t^{22}$ denoting the three unique $q\times q$ sub-blocks ($G_t^{12} = G_t^{21'}$).  The form of eqn.~\eqref{eqn:gt2} allows us to use the previously discussed properties of the inverse Wishart distribution to sample $\{\Sigma_{t-1},\Upsilon_t,\Psi_t\}$ conditioned on $\Sigma_t$ and $y_{1:t}$.  Specifically, as discussed in Section~\ref{sec:model}, there exists a $\{\tilde{\Upsilon}_t,\tilde{\Psi}_t\}$ such that $\tilde{\Upsilon}_t\Sigma_t = \Sigma_{t-1}\Upsilon_t'$ and
\begin{equation}
\begin{aligned}		
\tilde{\Psi}_t\mid y_{1:t} &\sim \mbox{IW}(r_t+1+q,G_t^{11} - G_t^{21'}(G_t^{22})^{-1}G_t^{21}),\\
\tilde{\Upsilon}_t &\mid \tilde{\Psi}_t,y_{1:t} \sim N(G_t^{21'}(G_t^{22})^{-1},\tilde{\Psi}_t,(G_t^{22})^{-1})
\end{aligned}
\label{eqn:tildeInnovationsAppendix}
\end{equation}
with
\begin{align}
\Sigma_{t-1} = \tilde{\Psi}_t + \tilde{\Upsilon}_t\Sigma_{t}\tilde{\Upsilon}_t'.
\label{eqn:tildeSigmaAppendix}
\end{align}
Thus, to sample $\Sigma_{t-1}$ conditioned on $\Sigma_t$ from the approximation to $p(\Sigma_{t-1} \mid \Sigma_t,y_{1:t})$, we first sample $\{\tilde{\Upsilon}_t,\tilde{\Psi}_t\}$ as specified in eqn.~\eqref{eqn:tildeInnovationsAppendix} and then compute $\Sigma_{t-1}$ based on eqn.~\eqref{eqn:tildeSigmaAppendix}.

%\section{Higher Order Models}
%%
%\subsection{Conditional Mean of IW-AR(2) - Construction 1}
%%
%For the first constructive definition of an IW-AR(2) in Section~\ref{sec:higherOrder}, in direct analogy to the derivation of the conditional mean in the IW-AR(1) case we derive eqn.~\eqref{eqn:IWAR2_condMean1} as
%%
%\begin{align}
%E[\Sig{t} \mid \De{t-1}] &= \bpmat 0 & F \epmat \De{t-1} \bpmat 0 & F \epmat' + \frac{nV}{n+2q}\left(1 + \tr\left(\De{t-1}(nS_2)^{-1}\right)\right)\\
%						 &= F \Sig{t-1} F' + \frac{nV}{n+2q}\nonumber\\
%						&\hspace{0.5in} + \frac{V}{n+2q}tr\left(\bpmat \Sig{t-2} & \phi_{t-1}'\\ \phi_{t-1} & \Sig{t-1} \epmat \bpmat S^{-1} + F'V^{-1}F & -V^{-1}F\\ -F'V^{-1} & V^{-1} \epmat\right)\\
%						% &= F \Sig{t-1} F' + \frac{nV}{n+2q}\nonumber\\
%						%&\hspace{1in} + \frac{V}{n+2q}tr\left(\bpmat \Sig{t-2}(S^{-1} + F'V^{-1}F) - \phi_{t-1}'F'V^{-1} & \cdot\\ \cdot & -\phi_{t-1}V^{-1}F + \Sig{t-1}V^{-1} \epmat\right)\\
%						 &= F\Sig{t-1}F' + \frac{V}{n+2q}\tr(\Sig{t-1}V^{-1})\nonumber\\
%						&\hspace{0.5in} + \frac{V}{n+2q}\tr(\Sig{t-2}(S^{-1}+F'V^{-1}F)) - 2\frac{V}{n+2q}\tr(\phi_{t-1}V^{-1}F).
%\end{align}
%%

\section{Multivariate t Distribution}

The $q$-dimensional multivariate t distribution with $\nu$ degrees of freedom and 
parameters $\mu$ and $\Sigma$ has density 
\begin{align}
	t_{\nu}(x\mid \mu,\Sigma/\nu) = a_{\nu,q} |\Sigma|^{-1/2} \left(1 + \frac{1}{\nu}(x-\mu)'\Sigma^{-1}(x-\mu)\right)^{-(\nu + q)/2}
\end{align}
where $$a_{q,\nu} = \frac{\Gamma((\nu + q)/2)}{\Gamma(\nu/2) \nu^{q/2} \pi^{q/2}}.$$
For the proposed IW-AR, since $\Upsilon_t \mid \Psi_t \sim N(F,\Psi_t,(nS)^{-1})$ standard theory
gives $\Upsilon_t z_t \mid \Psi_t,z_t \sim N(Fz_t,\Psi_t(z_t'(nS)^{-1}z_t))$.  Marginalizing over $\Upsilon_t$ in eqn.~\eqref{eqn:xzreg}, we have
\begin{align}	
	p(x_t \mid z_t,\Psi_t) = N(Fz_t,\Psi_t(1+z_t'(nS)^{-1}z_t)).
	\label{eqn:xzreg_marg1}
\end{align}
Now, $\Psi_t \sim IW(n+q+2,nV)$ implies $\Psi_t(1+z_t'(nS)^{-1}z_t) \mid z_t \sim IW(n+q+2,nV(1+z_t'(nS)^{-1}z_t))$   Marginalizing $\Psi_t$ from the distribution of eqn.~\eqref{eqn:xzreg_marg1} yields the t distribution with density
\begin{align}	
	p(x_t \mid z_t) = t_{n+q+2}\left(Fz_t,(1+z_t'(nS)^{-1}z_t)\frac{nV}{n+q+2}\right).
	\label{eqn:xzreg_marg2}
\end{align}
The marginal likelihood of $F$ and $S$ given $z_{1:T}$ follows immediately.

\section{Conditional Mean of IW-AR(2) Model}
For the IW-AR(2) process in Section~\ref{sec:IWARp} we have
\begin{align}
E[\Psi_{t} \mid \Psi_{t-1}] &= H\Psi_{t-1}H' + \frac{nW}{n+2q}\left(1+\tr(\Psi_{t-1}(nV)^{-1})\right).
\label{eqn:Psi_condMean}
\end{align}
Since $E[\Sig{t} \mid \De{t-1}] = E[\Psi_t \mid \De{t-1}] + E[\Upsilon_t\Sigma_{t-1}\Upsilon_t'\mid \De{t-1}]$, in deriving the conditional mean of $\Sig{t}$ given $\De{t-1}$ we first need
\begin{equation}
\begin{aligned}
E[[\Upsilon_t\Sigma_{t-1}\Upsilon_t']_{ij}\mid \De{t-1}] &= E[E[[\Upsilon_t\Sigma_{t-1}\Upsilon_t']_{ij}\mid \Psi_t] \mid \De{t-1}]\\
						 &= E[\sum_{k,\ell}\Sigma_{t-1,k\ell}E[\Upsilon_{t,ik}\Upsilon_{t,j\ell}\mid \Psi_t] \mid \De{t-1}]\\
						&= E[\sum_{k,\ell}\Sigma_{t-1,k\ell}\left([\Psi_{t,ij}(nS)^{-1}]_{k\ell} + F_{ik}F_{j\ell}\right) \mid \De{t-1}]\\
						&= E[\tr(\Sigma_{t-1}(nS)^{-1})\Psi_{t,ij} + F_{i\cdot}\Sigma_{t-1}F_{j\cdot}' \mid \De{t-1}]\\
						&= \tr(\Sigma_{t-1}(nS)^{-1})E[\Psi_{t,ij}\mid \De{t-1}] + F_{i\cdot}\Sigma_{t-1}F_{j\cdot}',
\end{aligned}
\end{equation}
implying
\begin{align}
E[\Upsilon_t\Sigma_{t-1}\Upsilon_t' \mid \De{t-1}] &= \tr(\Sigma_{t-1}(nS)^{-1})E[\Psi_{t}\mid \De{t-1}] + F\Sigma_{t-1}F'.
\end{align}

Noting that $E[\Psi_t \mid \De{t-1}] = E[E[\Psi_t\mid \Psi_{t-1}] \mid \De{t-1}]$ and $E[\Psi_{t-1} \mid \De{t-1}] = \Psi_{t-1}$ since $\Psi_{t-1}$ is a deterministic function of the elements of $\De{t-1}$, eqn.~\eqref{eqn:IWAR2_condMean2} follows directly.  That is, 
$$E[\Sig{t} \mid \De{t-1}] = F\Sigma_{t-1}F' + E[\Psi_t \mid \Psi_{t-1}](1+ \tr(\Sigma_{t-1}(nS)^{-1}))$$ with $E[\Psi_t \mid \Psi_{t-1}]$ as in eqn.~\eqref{eqn:Psi_condMean}.

In the limit as $n\rightarrow \infty$, we have
\begin{equation}
\begin{aligned}
E[\Sig{t} \mid \De{t-1}] &= W + F\Sig{t-1}F' + H\Psi_{t-1}H' \\
	&= V-HVH' + F\Sig{t-1}F' + H(\Sigma_{t-1} - H\Upsilon_{t-1}\Sigma_{t-2}\Upsilon_{t-1}')H'\\
	&= S-FSF' + F\Sig{t-1}F'\nonumber\\
	&\hspace{0.75in} + H((\Sigma_{t-1} - \Upsilon_{t-1}\Sigma_{t-2}\Upsilon_{t-1}') - (S-FSF'))H'\\
	&= S + F(\Sig{t-1} - S)F'\nonumber\\
	&\hspace{0.75in} + H((\Sigma_{t-1}-S) - (\Upsilon_{t-1}\Sigma_{t-2}\Upsilon_{t-1}' - FSF'))H'.
\end{aligned}
\end{equation}

\section{Sampling for Stochastic Volatility in Time Series Models with IW-AR(1) Components}

In eqn.~\eqref{eqn:ARobs}, the conditional posterior of the autoregressive parameters $A$ (i.e., \emph{Step 0} of the sampler) is given as follows.

\paragraph{Step 0}

Sample the observation autoregressive parameter $A$ given $\Delta_{1:T}$ and $\xi_{1-r:T}$.  Assume $A_i$ diagonal defined by the $q$-vector $a_i=\mbox{diag}(A_i)$.  The autoregressive process of eqn.~\eqref{eqn:ARobs} can be equivalently represented as
\begin{equation}
\begin{aligned}
	\xi_t &= \begin{bmatrix}\mbox{diag}(\xi_{t-1}) & \cdots & \mbox{diag}(\xi_{t-r})\end{bmatrix}\begin{bmatrix}a_1' & \cdots & a_r'\end{bmatrix}'+x_t.
\end{aligned}
\end{equation}
Under a prior $\begin{bmatrix}a_1' & \cdots & a_r'\end{bmatrix}' \sim N(\mu_a,\Sigma_a)$, standard theory yields the conditional for 
$\begin{bmatrix}a_1' & \cdots & a_r'\end{bmatrix}' \mid \Sigma_{0:T},\xi_{1-r:T}$ as multivariate normal
with easily computed moments. 
%
%\begin{align}
%	N^{-1}\left(\Sigma_{\phi}^{-1}\mu_{\phi} + \sum_t \bar{Y}_t\Sigma_t^{-1}\xi_t, \Sigma_{\phi}^{-1} + \sum_t \bar{Y}_t'\Sigma_t^{-1}\bar{Y}_t\right),
%\end{align}
%%
%where $N^{-1}(\Sigma^{-1}\mu,\Sigma^{-1})$ denotes a Gaussian $N(\mu,\Sigma)$ in information form.  
For $t=1,\dots,T$, set $x_t = \xi_t - \sum_{i=1}^r A_i \xi_{t-i}$.

\bibliographystyle{plainnat}
%\bibliography{stats,../../Bibliography/Bibliography}
\bibliography{stats,Bibliography}

\end{document}